\documentclass[hyper]{JHEP3} % 10pt is ignored!

\JHEPspecialurl{http://jhep.sissa.it/JOURNAL/JHEP3.tar.gz}

\usepackage{epsfig,multicol,amsmath}

\def\beq{\begin{equation}}
\def\eeq{\end{equation}}
\def\bea{\begin{eqnarray}}
\def\eea{\end{eqnarray}}
\def\eq#1{{Eq.~(\ref{#1})}}
\def\fig#1{{Fig.~\ref{#1}}}

\newcommand{\Lb}{\left(}
\newcommand{\Rb}{\right)}
\parskip=\itemsep               %?

\renewcommand{\thefootnote}{\fnsymbol{footnote}}

\newcommand{\FFr}{\frac}

\makeatletter \@addtoreset{equation}{section}
\makeatother \makeatletter \@addtoreset{equation}{section}

\def\thefootnote{\fnsymbol{footnote}}

\title{\centering\LARGE \bf Higgs Boson Mass predicted by the Four Color Theorem
}
\author{\centering \large
Ashay Dharwadker$^{}$\thanks{Email: ashay@dharwadker.org}\,\,\,
and\,\,
Vladimir Khachatryan$^{}$\thanks{Email: vkhachatryan@dharwadker.org} \\\

{\footnotesize{\it Institute of Mathematics, H-501 Palam Vihar, District Gurgaon, Haryana  122017, India  \\
}}}

\abstract{We show that the mathematical proof of the four color theorem yields a perfect \mbox{interpretation} of the Standard Model of particle physics. The steps of the proof enable us to construct the \textit{t}-Riemann surface and particle frame which forms the gauge. We specify well-defined rules to match the Standard Model in a one-to-one correspondence with the topological and algebraic structure of the particle frame. This correspondence is exact - it only allows the particles and force fields to have the observable properties of the Standard Model, giving us a Grand Unified Theory. In this paper, we concentrate on explicitly specifying the quarks, gauge vector bosons, the Standard Model scalar Higgs $H^{0}$ boson and the weak force field. Using all the specifications of our mathematical model, we show how to calculate the values of the Weinberg and Cabibbo angles on the particle frame. Finally, we present our prediction of the Higgs $H^{0}$ boson mass \,$M_{{\!}_{H^{0}}}\,=\,125.992\,\simeq 126\,GeV$,\, as a direct consequence of the proof of the four color theorem.}


\begin{document}

\def\thefootnote{\arabic{footnote}}
\section{Introduction}

The famous four color theorem \footnote{The four color theorem was first conjectured by M\"{o}bius in 1840, later by DeMorgan and the Guthrie brothers in 1852, and again by Cayley in 1878 \cite{Cour}. It remained one of the most celebrated and long outstanding conjectures in mathematics after many mathematicians tried to prove it for over a century. The conjecture was finally verified using an extensive computer search for potential counter-examples by Appel and Haken in the 1970's \cite{ApHak2}. However, this computer verification cannot be checked by humans, even in principle. On the other hand, Dharwadker's proof \cite{Dhar1} is a standard mathematical proof and can be verified by hand.} was proved mathematically for the first time in 2000, with a standard mathematical proof using algebraic and topological methods \cite{Dhar1}. The corresponding physical interpretation leads inexorably to a grand unified theory of particle physics that has been developed in detail by Dharwadker over the last decade \cite{Dhar2}. In this paper with Khachatryan, we present the mathematical foundations and further development of the theory primarily for physicists. The main focus is on applying the proof of the four color theorem to calculate and predict
the mass of the missing Higgs particle of the Standard Model (SM).

The four color theorem arises as a fundamental problem in topology when the surface of a sphere or plane is partitioned into finitely many contiguous regions called a map. Two regions in the map are considered adjacent if they share a whole segment of their boundaries in common. The theorem states that the regions of any such map can always be colored by using at most four different colors, so that no two adjacent regions have the same color. The theorem may also be rephrased in terms of graph theory \cite{DharPir1,DharPir2,Di}. A planar graph can be obtained
from a map by substituting every region of the map by a vertex in the graph, and connecting two vertices by an edge whenever two regions share a common boundary segment. Then the four color theorem states that the vertices of every planar graph can be colored with at most four colors, so that no two adjacent vertices receive the same color.

The mathematical proof of the four color theorem \cite{Dhar1} has a rich topological and algebraic
structure and here we propose its most important and fundamental application: a physical interpretation of the mathematical
proof which directly implies the existence of the SM \cite{Dhar2}. Conversely, the success of the theory in explaining the physically observable properties of the SM shows that nature uses the algebraic and topological structure of the proof of the four color theorem at the most fundamental level. We describe the whole picture consistently, unifying all the main features of \cite{Dhar1} and \cite{Dhar2} in the framework of one paper. We review some results of \cite{Dhar2} and
perform new calculations using the most recent experimental data. The principal aim of this paper is to present a prediction for the mass of the missing Higgs boson, which could possibly be detected at Tevatron and/or LHC in the near future, given the recent promising developments \cite{Aal,CDF_D0_1,CDF_D0_2,Ber,Tev}.

The Higgs boson is the only SM particle that has not yet been observed. Experimental detection of the Higgs boson would help
explain the origin of mass. For example, it could explain the difference between the massless photon, which mediates
electromagnetism, and the massive $W^{\pm}$ and $Z$ weak gauge bosons, which mediate the weak force. Theoretically, it is
known that the SM Higgs boson is one neutral quantum component of the Higgs field, along with another neutral and two
charged components acting as Goldstone bosons \cite{Gold1,Gold2}. This field has a non-zero vacuum expectation value, or stated
otherwise, an amplitude different from zero. The existence of this non-zero vacuum expectation spontaneously breaks
electroweak gauge symmetry which itself is the Higgs-Kibble mechanism \cite{Hig1,Hig2,Hig3,Hig4,Kib,Kib1,Djo1,Bed}, and which is responsible for giving masses to the other particles and to the Higgs particle as well. In more technical terms, the appearance of this
non-zero vacuum expectation value selects a preferred direction in weak isospin-hypercharge space spontaneously breaking the
\,$SU(2)_{L} \times U(1)_{Y}$\, symmetry. The vacuum everywhere can emit or absorb the neutral colorless quantum of the
Higgs field that carries weak isospin and hypercharge. The gauge bosons and fermions that couple to the field acquire masses,
meanwhile, photons and gluons that cannot couple to it remain massless. The current experimental constraint on the existence
of the Higgs boson is found to be in the energy realm from $114$ to $160\,GeV$. There is also a small possibility of its
existence in the $170-185\,GeV$ range, however, the lower energies are far more probable. Some MSSM (minimal
supersymmetric extension of the Standard Model \cite{Noj,Mar,Mar1,Mar2}) models predict that the upper limit of the mass of the lightest MSSM Higgs boson $H_{1}^{0}$ (of several) and the lower limit of the mass of the heaviest Higgs boson $H_{2}^{0}$ will have
values at around \,$100 \sim 130\,GeV$\, \cite{Djo2,Djo3,CaHa,Low,HaHe}. The lowest constraint for the mass value
of $H_{1}^{0}$ is \,$M_{{\!}_{H_{1}^{0}}} > 92.8\,GeV$\, \cite{PDG}.

To calculate the Higgs boson mass, we use the  strategy outlined in the following construction of the paper. In Section $2$ we introduce the four color theorem as described in \cite{Dhar1}. We present all the steps of the proof, stating all the lemmas which logically lead to the formulation of the theorem. For the complete and detailed proof of each lemma we refer to \cite{Dhar1}. For our stated goal, it is important to carefully define the construction of a specific mathematical model, the labeled \textit{t}-Riemann surface and particle frame, that will form the base onto which the SM can be matched in a one-to-one correspondence. We  define Schr\"{o}dinger disks centered at the origin of the complex plane from which the labeled \textit{t}-Riemann surfaces and particle frames are constructed. We build the SM of particles and force fields from copies of the oriented Schr\"{o}dinger disks, arranged in a certain way on the particle frame, as dictated by the mathematical proof of the four color theorem. In Section $3$ we specify the general procedure by which all particles of the SM can be defined, together with their basic physical properties: spin, charge and mass. In this context the labeled \textit{t}-Riemann surface is called a particle frame and forms the gauge. We work with the particle frame structure that corresponds to the present epoch in the cosmological timeline or equivalent energy scales and formulate the precise rules for the definition of particles and antiparticles that make up the SM. The matching of the SM onto the particle frame allows us to calculate the value of the Higgs boson mass, given the experimental values of the masses of $Z^{0}$ and $W^{\pm}$ bosons. In Section $4$ we explicitly define fermions, bosons and their antiparticles on the particle frame. We also define the weak force field which allows us to calculate the Weinberg and Cabibbo angles. In Section $5$ we show
how to calculate the value of the Higgs boson mass in our model,  given the experimental values of the masses of $Z^{0}$ and $W^{\pm}$ bosons. We also compare some mass constraints (existing
in the literature) of the SM and MSSM Higgs bosons with our prediction.

\section{The Four Color Theorem developed for the Standard Model}
\label{sec:Theorem}

In this Section, we present the steps of the proof of the four color theorem developed for the SM, following \cite{Dhar1}. We state all the lemmas that lead to the formulation of the theorem, referring the reader to \cite{Dhar1} for the complete and detailed proof of each lemma.

\subsection{Formulation of the Theorem}
The statement of the four color theorem is formulated as follows:

\begin{quote}
{\textit{For any subdivision of the plane or the surface of a sphere into non-overlapping regions, it is always possible
to mark each of the regions with one of the colors \,$0, 1, 2, 3$\, in such a way that no two adjacent regions receive the
same color.}}
\end{quote}

\subsubsection{Map Coloring and Steiner Systems}
A \emph{map} on the sphere is a subdivision of the spherical surface into finitely many regions. The map is regarded as
\emph{properly colored} if each region receives a color and no two regions having a whole segment of their boundaries in
common receive the same color. Since deformations of the regions and their boundary lines do not affect the proper coloring
of the map, we confine ourselves to maps whose regions are bounded by simple closed polygons. For purposes of the proper
coloring it is equivalent to consider maps drawn on the plane. Any map on the sphere may be represented on the plane by
boring a small hole through the interior of one of the regions and deforming the resulting surface until it is flat.
Conversely, by a reversal of this process, any map on the plane may be represented on the sphere. Furthermore, it suffices
to consider \emph{$3$-regular} maps, i.e. maps with exactly three edges meeting at each vertex, by the following argument.
Replace each vertex, at which more than three edges meet, by a small circle and join the interior of each such circle to one
of the regions meeting at the vertex. A new map is obtained which is $3$-regular. If this new map can be properly colored by
using at most $n$ colors then by shrinking the circles down to points, the desired coloring of the original map using at
most $n$ colors is obtained. From Euler's polyhedron formula \cite{Stil} we have

\begin{itemize}
\item[]LEMMA $1$. {\textit{Any map on the sphere can be properly colored using at most six colors.}}
\end{itemize}

By LEMMA $1$, the minimal number of colors required to properly color any map from the class of all maps on the sphere is a
well-defined natural number. We may now make the following basic definitions.

\begin{quote}
$\bullet$ Define $N$ to be the minimal number of colors required to properly color any map from the class of all maps on the
sphere. That is, given any map on the sphere, no more than $N$ colors are required to properly color it and there exists a
map on the sphere which requires no fewer than $N$ colors to be properly colored. \\
$\bullet$ Based on the definition of $N$, select a specific map $\textit{\textbf{m}(N)}$ on the sphere which requires no
fewer than $N$ colors to be properly colored. \\
$\bullet$ Based on the definition of the map $\textit{\textbf{m}(N)}$, select a proper coloring of the regions of the map
$\textit{\textbf{m}(N)}$ using the $N$ colors \,$0, 1, ..., N\!-\!1$.
\end{quote}

The natural number $N$, the map $\textit{\textbf{m}(N)}$ and the proper coloring of the regions of $\textit{\textbf{m}(N)}$
are fixed for all future references.

\begin{center}
\FIGURE[h]{
\centerline{\epsfig{file=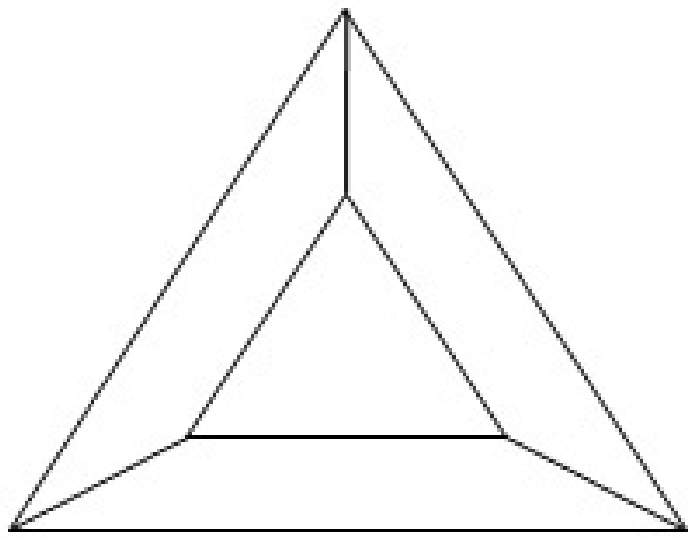,width=65mm,height=45mm}}
\caption{A map that requires four colors to be properly colored.}
\label{FCT_fig1}}
\end{center}

The map shown in \fig{FCT_fig1} has four regions that are adjacent to each other, hence requires four colors to be properly colored. There are infinitely many maps that require four colors to be properly colored but for the following argument we need to demonstrate the existence of only one such map.  Thus, by the map shown in \fig{FCT_fig1} and LEMMA $1$, we have proved that \,$4 \leq N \leq 6$.\,

The goal in Section $2$ is to show that \,$N \leq 4$\,, which together with the above inequality would imply $N = 4$ to prove the four color theorem \footnote{In our mathematical model for the SM, the natural number $N$ will represent the number of unsigned electric charges (see the Electric Charge Rule in Section $3.2$). Thus $N = 4$ is also a necessary condition for applying our mathematical model to the SM.}. To achieve this goal, we first need to define Steiner systems \cite{Ass, Dhar3}. A \emph{Steiner system} with parameters $t$, $k$,
$\nu$, written as $S(t, k, \nu)$, is an $\nu$-element set $\textit{\textbf{P}}$ of \emph{points} together with a set
$\textit{\textbf{B}}$ of $k$-element subsets of $\textit{\textbf{P}}$ (called \emph{blocks}) with a property that each
$t$-element subset of $\textit{\textbf{P}}$ is contained in exactly one block. Thus, in the Steiner system $S(t, k, \nu)$
\begin{quote}
$\bullet$ There are $\nu$ points. \\
$\bullet$ Each block consists of $k$ points. \\
$\bullet$ Every set of $t$ points is contained in a unique block.
\end{quote}
Note that by definition $t$, $k$, $\nu$ are nonnegative integers with \,$t\leq k \leq \nu$.\, The Steiner systems with
\,$\nu = k$\, (only one block that contains all the points) or \,$k = t$\, (every $k$-element subset of points is a block)
are called trivial. A particularly interesting nontrivial Steiner system is \,S(5, 8, 24),\, known as the
\emph{Witt Design} (see \cite{Dhar3} for an explicit construction), whose blocks are also known as Golay codewords
of weight eight \cite{Dhar3, Ass}. The automorphism group of \,S(5, 8, 24)\, (permutations of points that transform blocks to blocks) is the largest of the Mathieu groups, $M_{24}$ \cite{Ass}. The following inequality \cite{Tits} gives an important constraint on the parameters of any nontrivial Steiner system.

\begin{itemize}
\item[]LEMMA $2$. {\textit{If there exists a nontrivial Steiner system \,$S(t, k, \nu)$\,, then \,$\nu \geq (t + 1)(k - t + 1)$\,
}}.
\end{itemize}

From the map $\textit{\textbf{m}(N)}$ and its fixed proper coloring, we shall construct a Steiner system \,$S(N\!+\!1, 2N, 6N)$\,
by defining the points and blocks in a certain way. Using LEMMA $2$, the next lemma shows that this construction would force
\,$N \leq 4$.

\begin{itemize}
\item[]LEMMA $3$. {\textit{Referring to the definition of $N$, if there exists the Steiner system \,$S(N\!+\!1, 2N, 6N)$\,
then \,$N \leq 4$.}}
\end{itemize}

If such a Steiner system \,$S(N\!+\!1, 2N, 6N)$\, can be shown to exist, then since \,$4 \leq N \leq 6$,\, LEMMA $3$ would imply that \,$N = 4$\,. This would also imply that the Steiner system \,$S(N\!+\!1, 2N, 6N)$\, is actually just the Witt design \,$S(5, 8, 24)$\,. Thus, from now on our goal is to demonstrate the existence of the Steiner system \,$S(N\!+\!1, 2N, 6N)$\, based upon the
definition of the map $\textit{\textbf{m}(N)}$.

\subsubsection{Eilenberg Modules and Hall Matchings}
In this section we build the essential algebraic structure \footnote{This will also provide us with the algebraic structure that we need to define our mathematical model of the SM.} that is required to construct the Steiner system \,$S(N\!+\!1, 2N, 6N)$\,. The reader may refer to \cite{Lan} for elementary definitions about groups, rings and modules.

Let $\textit{G}$ be a group \footnote{A \emph{monoid} $G$ is a set together with a binary operation that satisfies the associative law $x(yz) = (xy)z$ and has an identity element $e$ such that $ex = x = xe$ for all $x$, $y$, $z$ in $G$ . We usually write the binary operation as addition, with identity element $0$, if it satisfies the abelian commutative law $x + y = y + x$ and as multiplication (juxtaposition), with identity element $1$ or $e$, in general. A $group$ is a monoid in which each element $x$ has an inverse $x^{-1}$ such that $xx^{-1} = e = x^{-1}x$.} with identity element $\textit{e}$\, and let $\textit{\textbf{Z}}$ denote the ring \footnote{A \emph{ring} $R$ is a set together with two binary operations of addition and multiplication such that $R$ is an abelian group with respect to addition, a monoid with respect to multiplication and satisfies the distributive laws $(x+y)z = xz+yz$ and $z(x+y) = zx+zy$ for all $x$, $y$, $z$ in $R$. If the subset of nonzero elements also form an abelian group under multiplication then $R$ is called a \emph{field}.} of integers. The
\emph{integral group algebra} \,$(\textit{\textbf{Z}}G,+,\cdot)$\, is a ring whose elements are formal sums
$$\sum_{g\,\in\,G}n_{g}\,g\,,$$ with $g$ in $G$ and $n_{g}$ in $\textit{\textbf{Z}}$ such that \,$n_{g} = 0$\, for all but
a finite number of $g$. Addition and multiplication in $\textit{\textbf{Z}}G$ are defined by
\beq \label{Theor1}
\sum_{g\,\in\,G}n_{g}\,g\,+\,\sum_{g\,\in\,G}m_{g}\,g\,=\,\sum_{g\,\in\,G}(n_{g}\,+\,m_{g})\,g
\eeq
and
\beq \label{Theor2}
\Lb \sum_{g\,\in\,G}n_{g}\,g \Rb\!\cdot\!\Lb \sum_{g\,\in\,G}m_{g}\,g \Rb\,=\,\sum_{g\,\in\,G} \Lb \sum_{h\,\in\,G}
n_{gh^{-1}}\,m_{h} \Rb g\,.
\eeq
Note that $h^{-1}$ is the inverse of $h$ in the group $G$ and the product $gh^{-1}$ is an element of the group $G$.
Thus \,$n_{gh^{-1}}$\, is an element of the group algebra $\textit{\textbf{Z}}G$ indexed by the element $gh^{-1}$.
The element $n$ of $\textit{\textbf{Z}}$ is identified with the element \,$n \cdot e$\, of $\textit{\textbf{Z}}G$
and the element $g$ of $G$ is identified with the element \,$1 \cdot g$\, of $\textit{\textbf{Z}}G$, so that
$\textit{\textbf{Z}}$ and $G$ are to be regarded as subsets of $\textit{\textbf{Z}}G$. The underlying additive abelian group
\,$(\textit{\textbf{Z}}G,+)$\, is the direct sum of copies of the integers $\textit{\textbf{Z}}$ indexed by elements of $G$.
If $H$ is a subgroup of $G$ then $\textit{\textbf{Z}}H$ is a subring of $\textit{\textbf{Z}}G$ in a natural way.

For each element $g$ of $G$, the right multiplication \,\,$R(g)\!: G \rightarrow G; \,x \rightarrow xg$\,\, and the left
multiplication \,\,$L(g)\!: G \rightarrow G; \,x \rightarrow gx$\,\, are permutations of the set $G$. Denote the group of
all permutations of the set $G$ by $\textit{Sym}(G)$. Then
{\setlength\arraycolsep{2pt}
\bea \label{Theor3}
& & R\!:\,G\,\rightarrow\,Sym(G);\,\,\,g\,\rightarrow\,R(g)\,,
\nonumber\\
& &
\!\!\!\!\!\!L^{-1}\!:\,G\,\rightarrow\,Sym(G);\,\,\,g\,\rightarrow\,L^{-1}(g)\,=\,L(g^{-1})\,
\eea}are embeddings of the group $G$ in $\textit{Sym}(G)$. The images $R(G)$, $L^{-1}(G)$ are called the \emph{Cayley right and
left regular representations of $G$}, respectively.

A \emph{representation} \cite{Lan,DharSmi} of a group $G$ is defined as a module \footnote{A \emph{module} $M$ is an abelian group together with scalar multiplication by elements of a ring $R$ that satisfies $(a+b)x = ax+bx$ and $a(x+y) = ax+ay$ for $a$, $b$ in $R$ and $x$, $y$ in $M$. In particular, modules \,$(M,+)$\ over the integers $\textit{\textbf{Z}}$\ are just abelian groups and modules \,$(M,+)$\ over a field $F$ are vector spaces.} \,$(M,+)$\,, for which there is a homomorphism \,$T\!:\,G \rightarrow Aut(M, +)$\, showing how $G$ acts \footnote{A group $G$ \emph{acts} on a set $S$ (by right action) if (1) $sg$ is a well-defined element of $S$ for all $g$ in $G$ and $s$ in $S$, (2) $s(gh) = (sg)h$ for all $g$, $h$ in $G$ and $s$ in $S$, and (3) $se = s$ for all $s$ in $S$.} as a group of automorphisms of the module. The ordinary representation theory is obtained by considering modules \,$(M,+)$\, over the complex field $\textit{\textbf{C}}$\, i.e. complex vector spaces. An equivalent approach to representation theory due to Eilenberg \cite{Eilen} views a module $M$ for the group $G$ as follows. The set \,$M\!\times\!G$\, equipped with multiplication
\beq \label{Theor5}
(m_{{}_{1}},g_{{}_{1}})(m_{{}_{2}},g_{{}_{2}})\,=\,(m_{{}_{1}}\,+\,m_{{}_{2}}T(g_{{}_{1}}),g_{{}_{1}}g_{{}_{2}})
\eeq
becomes a group \,$M]G$\, known as the \emph{split extension} of $M$ by $G$, \,for \,$m_{{}_{1}}$, $m_{{}_{2}}$\, in
$M$ and \,$g_{{}_{1}}$, $g_{{}_{2}}$ in $G$.\, There is an exact sequence of groups
\beq \label{Theor6}
1\,\rightarrow\,M\,\,\,{}^{{}^{\iota_0}}\!\!\!\!\!\!\rightarrow\,M]G\,\,\,{}^{{}^{\pi_0}}\!\!\!\!\!\!
\rightarrow\,G\,\rightarrow\,1\,,
\eeq
with the injection function \,$\iota_0$: $M \rightarrow M]G$; $m \rightarrow (m,e)$\, and the projection function \,$\pi_0$: $M]G \rightarrow G$; $(m,g) \rightarrow g$\, split by the zero function \,$0$: $G \rightarrow M]G$; $g \rightarrow (0,g)$. The group action $T$ is recovered from the split extension \,$M]G$\, by \,$m\,T(g)\,\iota_0\, = m\,\iota_0 \,R(0,g)\,L^{-1}(0,g)$\, for $m$ in $M$ and $g$ in $G$. In this context, $M$ is called an \emph{Eilenberg module for the group $G$}. For example, the trivial representation for the group $G$ is obtained by defining \,\,$T\!:\,G \rightarrow Aut(M, +);\,g \rightarrow {\bf{1}}_{M}$,\,\, where ${\bf{1}}_{M}$ denotes the identity automorphism of $(M,+)$ and the corresponding split extension is the group direct product \,$M\!\times\!G$.\, The Cayley
right regular representation for the group $G$ is obtained by defining
\beq \label{Theor7}
T\!:\,\,G\,\rightarrow\,Aut(\textit{\textbf{Z}}G,+);\,\,\,g\,\rightarrow\,
\Lb \sum_{h\,\in\,G}n_{h}h\,\rightarrow\,\sum_{h\,\in\,G}n_{h}h\,R(g) \Rb\,.
\eeq
Here \,$T(g) = R(g)L(g^{-1})$\, with $L(g^{-1})$ acting trivially on the module elements and $R(g)$ acting as the usual
right multiplication. The split extension \,$\textit{\textbf{Z}}G]G$\, has multiplication given by
\beq \label{Theor8}
(m_{{}_{1}},g_{{}_{1}})(m_{{}_{2}},g_{{}_{2}})\,=\,(m_{{}_{1}}\,+\,m_{{}_{2}}R(g_{{}_{1}}),g_{{}_{1}}g_{{}_{2}})\,,
\eeq
for $m_{{}_{1}}$, $m_{{}_{2}}$ in $\textit{\textbf{Z}}G$ and $g_{{}_{1}}$, $g_{{}_{2}}$ in $G$.

Referring to the definitions in Section $2.1.1$, $N$ is the minimal number of colors required to properly color any
map from the class of all maps on the sphere and $\textit{\textbf{m}(N)}$ is a specific map that requires all of $N$ colors
to be properly colored. Note that $\textit{\textbf{m}(N)}$ has been properly colored by using the $N$ colors $0, 1, ..., N\!-\!1$,
and this proper coloring is fixed. The set of regions of $\textit{\textbf{m}(N)}$ is then partitioned into subsets
$\underline{0}, \underline{1}, ..., \underline{N\!-\!1}$ where the subset $\underline{m}$ consists of all regions which
receive the color $m$.  Note that the subsets $\underline{0}, \underline{1}, ..., \underline{N\!-\!1}$ are each nonempty
(since $\textit{\textbf{m}(N)}$ requires all of $N$ colors to be properly colored) and form a partition of the set of
regions of $\textit{\textbf{m}(N)}$ (by virtue of proper coloring). Identify the set
$\left\{\underline{0}, \underline{1}, ..., \underline{N\!-\!1}\right\}$ with the underlying set of the $N$-element cyclic
group $\textit{\textbf{Z}}_{N}$ under addition modulo $N$. Let $S_{3}$ denote the symmetric group on three letters \footnote{The symmetric group $S_3$ is embedded in the algebraic structure in various ways, and will play a crucial role in our mathematical model of the SM. The six elements of $S_3$ will represent the unsigned electromagnetic, weak and strong charges in Section 3.} ,
identified with the dihedral group \cite{Lan, Rot} of order six generated by $\rho$ and $\sigma$, where
\,$|\rho| = 3$\, and \,$|\sigma| = 2$\,:
\begin{displaymath}
S_{3}\,=\,< \sigma,\,\rho >\,=\,\left\{1,\,\rho,\,\rho^{2},\,\sigma\rho^{2},\,\sigma\rho,\,\sigma\right\}\,.
\end{displaymath}
\begin{itemize}
\item[]LEMMA $4$. {\textit{$(\textit{\textbf{Z}}_{N},+)$ is an Eilenberg module for the group $S_{3}$ with the trivial
homomorphism
\beq \label{Theor9}
T_{1}\!:\,S_{3} \rightarrow Aut(\textit{\textbf{Z}}_{N}, +);\,\alpha \rightarrow {\bf{1}}_{\textit{\textbf{Z}}_{N}}\,,
\eeq
where ${\bf{1}}_{\textit{\textbf{Z}}_{N}}$ denotes the identity automorphism of $(\textit{\textbf{Z}}_{N},+)$.
The corresponding split extension \,$\textit{\textbf{Z}}_{N}]S_{3}$\, has multiplication given by
\beq \label{Theor10}
(\underline{m}_{{}_{\,1}},\alpha_{{}_{1}})\cdot(\underline{m}_{{}_{\,2}},\alpha_{{}_{2}})\,=\,(\underline{m}_{{}_{\,1}}\,+\,
\underline{m}_{{}_{\,2}},\alpha_{{}_{1}}\alpha_{{}_{2}})
\eeq
and is a group isomorphic to the direct product \,$\textit{\textbf{Z}}_{N}\!\times\!S_{3}$.\,
}}
\end{itemize}

Recall from Section $2.1.1$ that our goal is to construct the Steiner system \,$S(N\!+\!1, 2N, 6N)$\,. We shall define the set of points of the Steiner system to be the underlying set of the split extension $\textit{\textbf{Z}}_{N}]S_{3}$. The following lemma gives us the essential algebraic structure that will be used to define the blocks of the Steiner system.

\begin{itemize}
\item[]LEMMA $5$. {\textit{Let \,$\Lb \textit{\textbf{Z}}(\textit{\textbf{Z}}_{N}]S_{3}),+\Rb$\, and
\,$(\textit{\textbf{Z}}S_{3},+)$\, denote the underlying additive
groups of the integral group algebras
\,$\textit{\textbf{Z}}(\textit{\textbf{Z}}_{N}]S_{3})$\, and
$\textit{\textbf{Z}}S_{3}$, respectively. Then \,$\Lb
\textit{\textbf{Z}}(\textit{\textbf{Z}}_{N}]S_{3}),+\Rb$\, is an
Eilenberg module for the group \,$(\textit{\textbf{Z}}S_{3},+)$
with the trivial homomorphism
\beq \label{Theor11}
T_{2}\!:\,(\textit{\textbf{Z}}S_{3},+) \rightarrow
Aut(\textit{\textbf{Z}}(\textit{\textbf{Z}}_{N}]S_{3}),+);\,
\sum_{\alpha\,\in\,S_{3}}n_{\alpha}\alpha \rightarrow
{\bf{1}}_{\textit{\textbf{Z}}(\textit{\textbf{Z}}_{N}]S_{3})}\,,
\eeq where
\,${\bf{1}}_{\textit{\textbf{Z}}(\textit{\textbf{Z}}_{N}]S_{3})}$\,
denotes the identity automorphism of \,$\Lb
\textit{\textbf{Z}}(\textit{\textbf{Z}}_{N}]S_{3}),+\Rb$.\, The
corresponding split extension
\,$\textit{\textbf{Z}}(\textit{\textbf{Z}}_{N}]S_{3})]\textit{\textbf{Z}}S_{3}$\,
has multiplication given by {\setlength\arraycolsep{2pt} \bea
\label{Theor12} & & \Lb
\sum_{(\underline{m},\beta)\,\in\,\textit{\textbf{Z}}_{N}]S_{3}}n_{(\underline{m},\beta)}(\underline{m},\beta),
\,\sum_{\alpha\,\in\,S_{3}}n_{\alpha}\alpha \Rb \Lb
\sum_{(\underline{m},\beta)\,\in\,\textit{\textbf{Z}}_{N}]S_{3}}n^{'}_{(\underline{m},\beta)}(\underline{m},\beta),
\,\sum_{\alpha\,\in\,S_{3}}n^{'}_{\alpha}\alpha  \Rb\,=
\nonumber\\
& &
\,\,\,\,\,\,\,\,\,\,\,\,\,\,\,\,\,\,\,\,\,\,\,\,\,\,\,\,\,\,=\,\Lb
\sum_{(\underline{m},\beta)\,\in\,\textit{\textbf{Z}}_{N}]S_{3}}\Lb
n_{(\underline{m},\beta)}(\underline{m},\beta)\,+\,n^{'}_{(\underline{m},\beta)}(\underline{m},\beta) \Rb
,\,\sum_{\alpha\,\in\,S_{3}}(n_{\alpha}+n^{'}_{\alpha})\alpha \Rb
\eea}
and is a group isomorphic to the direct product
\,$\Lb \textit{\textbf{Z}}(\textit{\textbf{Z}}_{N}]S_{3})\!\times\!\textit{\textbf{Z}}S_{3}, + \Rb$.
}}
\end{itemize}

We require the concept of Hall Matchings \cite{Hall} to obtain a common system of coset representatives in LEMMA $8$ below. Let $\Gamma$ be a \emph{bipartite graph} \cite{Ass} with a vertex set \,$V = X\cup\,Y$\, and an edge set $E$ (every edge has one end in $X$ and the other end in $Y$). A \emph{matching} from $X$ to $Y$ in $\Gamma$ is a subset $M$ of $E$ such that no vertex is incident with more than one
edge in $M$. The matching from $X$ to $Y$ in $\Gamma$ is called \emph{complete} if every vertex in $X$ is incident with an edge
in $M$. If $A$ is a subset of $V$ then let $adj(A)$ denote the set of all vertices adjacent to a vertex in $A$.

\begin{itemize}
\item[]LEMMA $6$. {\textit{If \,$|adj(A)| \geq |A|$\, for every subset $A$ of $X$ then there exists a complete matching from
$X$ to $Y$ in $\Gamma$.
}}
\end{itemize}

\begin{itemize}
\item[]LEMMA $7$. {\textit{Referring to \eq{Theor3}, let
\,$Sym(\textit{\textbf{Z}}_{N}]S_{3})$\, denote the group of all permutations of the underlying set of the split extension
$\textit{\textbf{Z}}_{N}]S_{3}$ of LEMMA $4$. Then $S_{3}$ is embedded in \,$Sym(\textit{\textbf{Z}}_{N}]S_{3})$\, via the
Cayley right regular representation.
}}
\end{itemize}

\begin{itemize}
\item[]LEMMA $8$. {\textit{By LEMMA $7$, regard $S_{3}$ as a subgroup of \,$Sym(\textit{\textbf{Z}}_{N}]S_{3})$.\,There exists
a common system of coset representatives \,$\varphi_{1},...,\varphi_{k}$\, such that
\,$\left\{\varphi_{1}S_{3},...,\varphi_{k}S_{3}\right\}$\, is the family of the left cosets of $S_{3}$ in
\,$Sym(\textit{\textbf{Z}}_{N}]S_{3})$\, and \,$\left\{S_{3}\varphi_{1},...,S_{3}\varphi_{k}\right\}$\, is the family of
the right cosets of $S_{3}$ in \,$Sym(\textit{\textbf{Z}}_{N}]S_{3})$.
}}
\end{itemize}

\subsubsection{Riemann Surfaces}
In this section we build the essential topological structure \footnote{This will also provide us with the topological structure that we need to define our mathematical model of the SM.} that is required to construct the Steiner system \,$S(N\!+\!1, 2N, 6N)$\,. The reader may refer to \cite{Ahl} for elementary definitions about the topology of the complex plane.

Let $\textit{\textbf{C}}$\, denote the complex plane. Consider the function $\textit{\textbf{C}} \rightarrow \textit{\textbf{C}}$\,;
\,\,$z \rightarrow w = z^{n}$,\, where $n \geq 2$. There is a one-to-one correspondence between each sector
\beq \label{Theor13}
\left\{\,z \mid (k\,-\,1)(2\pi/n)\,<\,arg\,z\,<\,k(2\pi/n)\right\} (k\,=\,1,...,n)
\eeq
and the whole $w$-plane except for the positive real axis. The image of each sector is obtained by performing a cut along
the positive real axis; this cut has an upper and a lower edge. Corresponding to $n$ sectors in the $z$-plane, take $n$
identical copies of the $w$-plane with the cut. These will be the sheets of the \emph{Riemann surface} \cite{Ri} and are distinguished
by a label $k$ which serves to identify the corresponding sector. For \,$k = 1,...,n-1$,\, attach the lower edge of the sheet labeled by $k$ with the upper edge of the sheet labeled by $k+1$. To complete the cycle, attach the lower edge of the sheet labeled by $n$ with the upper edge of the sheet labeled by $1$. In a physical sense, this is not possible without self-intersection but the idealized model shall be free of this discrepancy.

\begin{center}
\FIGURE[h]{
\centerline{\epsfig{file=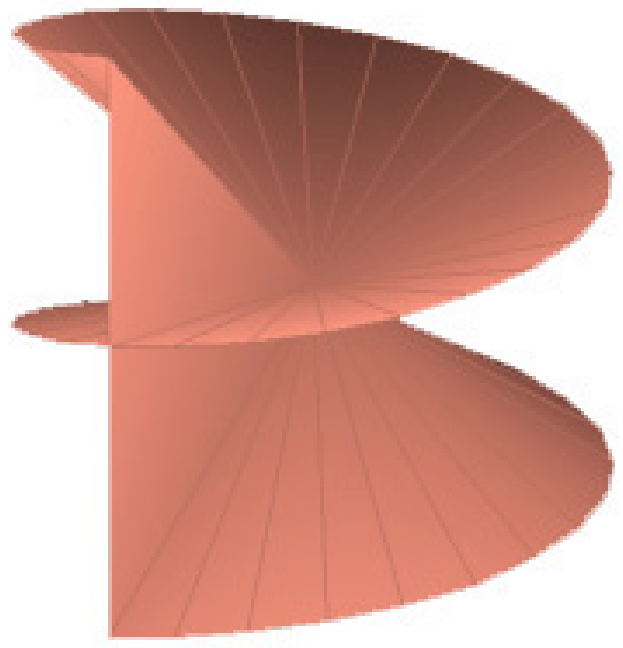,width=50mm,height=50mm}}
\caption{An example of a Riemann surface.}
\label{FCT_fig2}}
\end{center}

The result of the construction is the $w$-Riemann surface whose points are in one-to-one correspondence with the points of the $z$-plane. This correspondence is continuous in the following sense. When $z$ moves in its plane, the corresponding point $w$ is free to move on the $w$-Riemann surface. The point \,$w = 0$\, connects all the sheets and is called a \emph{branch point}. A curve must wind $n$ times around the branch point before it closes. Now consider the $n$-valued relation
\beq \label{Theor14}
z\,=\,\sqrt[n]{w}\,.
\eeq
To each \,$w \neq 0$,\, there correspond $n$ values of $z$. If the $w$-plane is replaced by the $w$-Riemann surface just
constructed, then each complex number \,$w \neq 0$\, is represented by $n$ points of the Riemann surface at superposed
positions. Let the point on the uppermost sheet represent the principal value and the other \,$n - 1$\, points represent
the other values. Then \,$z$\, in \eq{Theor14} becomes a single-valued, continuous, one-to-one correspondence of the points
of the $w$-Riemann surface with the points of the $z$-plane.

Recall the definition of the map $\textit{\textbf{m}(N)}$ from Section $2.1.1$. The map $\textit{\textbf{m}(N)}$ is on
the sphere. Place the sphere on the complex plane with its south pole touching the origin $0$ of $\textit{\textbf{C}}$. Then every straight line from the north pole of the sphere to the complex plane will intersect the sphere in exactly one point $s$ (other than the north pole) and the complex plane in exactly one point $c$. The north pole itself is a special point and corresponds (by all lines through it parallel to the complex plane $\textit{\textbf{C}}$\,) to the horizon or $\infty$ of the extended complex plane $\textit{\textbf{C}}$. The function that takes all such points $s$ (including the north pole) on the sphere to the points $c$ on the extended complex plane $\textit{\textbf{C}}$\, (including $\infty$) is called the \emph{stereographic projection}. Pick a region and deform the map so that both the south pole ($0$) and the north pole ($\infty$) are two distinct points inside this region when the sphere is regarded as an extended complex plane.  Using the stereographic projection one obtains the map
$\textit{\textbf{m}(N)}$ on the complex plane $\textit{\textbf{C}}$\, with the region containing 0 and $\infty$ forming
a ``sea'' surrounding the other regions which form an ``island'', in analogy with ordinary geographical maps. Put this copy of the complex plane with the map on each of the $n$ sheets of the $w$-Riemann surface corresponding to \,$w = z^{n}$.\, The $n$ ``islands'' will be superposed on the $w$-Riemann surface and the branch point lies in the $n$ superposed ``seas''.\, The inverse function \,$z$\, in \eq{Theor14} results in $n$ copies of the map $\textit{\textbf{m}(N)}$ on the $z$-plane in the $n$ sectors defined in \eq{Theor13}. The origin of the $z$-plane lies in the $n$ ``seas''. \fig{FCT_fig3} shows an example of the construction for $n$ = 4.

\begin{center}
\FIGURE[h]{
\centerline{\epsfig{file=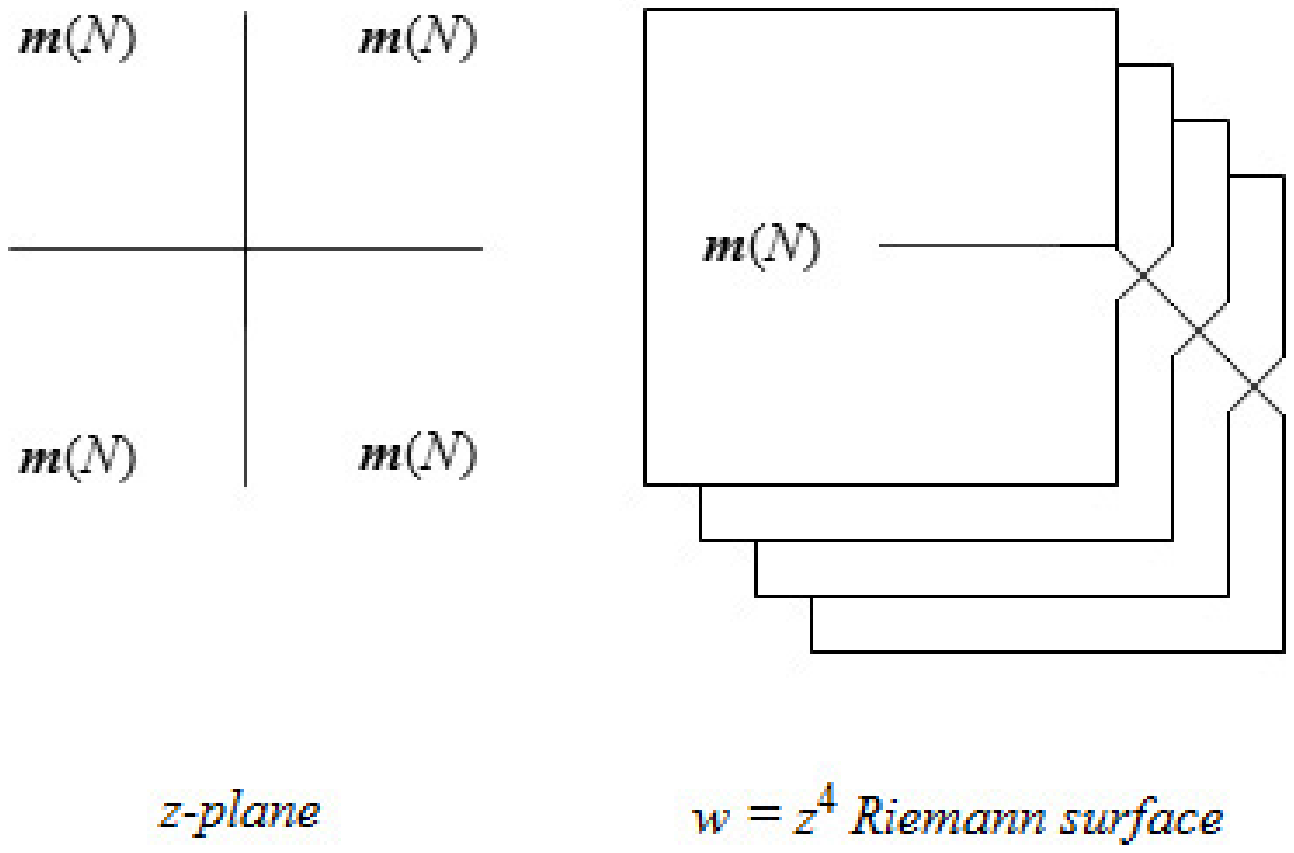,width=90mm,height=55mm}}
\caption{An example of a $w$-Riemann surface for \,$n = 4$.}
\label{FCT_fig3}}
\end{center}

We shall now construct the $\textit{t}$-Riemann surface. Consider the composition of the functions
\beq \label{Theor15}
\textit{\textbf{C}}\,\rightarrow\,\textit{\textbf{C}}\,;\,\,\,\,\,z\,\rightarrow\,t\,=\,z^{2}\,\,\,\,\,\,\,\,\,\mbox{and}
\,\,\,\,\,\,\,\,\,\textit{\textbf{C}}\,\rightarrow\,\textit{\textbf{C}}\,;\,\,\,\,\,t\,\rightarrow\,w\,=\,t^{12}\,.
\eeq
The composite is given by the assignment
\beq \label{Theor16}
z\,\rightarrow\,t\,=\,z^{2}\,\,\rightarrow\,\,w\,=\,t^{12}\,=\,z^{24}\,.
\eeq
There are twenty-four superposed copies of the map $\textit{\textbf{m}(N)}$ on the $w$-Riemann surface corresponding to
the twenty-four sectors
\beq \label{Theor17}
\left\{\,z \mid (k\,-\,1)(2\pi/24)\,<\,arg\,z\,<\,k(2\pi/24)\right\} (k\,=\,1,...,24)
\eeq
on the $z$-plane. These are divided into two sets. The first set consists of twelve superposed copies of the map
$\textit{\textbf{m}(N)}$ corresponding to the sectors
\beq \label{Theor18}
\left\{\,z \mid (k\,-\,1)(2\pi/24)\,<\,arg\,z\,<\,k(2\pi/24)\right\} (k\,=\,1,...,12)
\eeq
of the upper half of the $z$-plane which comprise the upper sheet of the $t$-Riemann surface. The second set consists
of twelve superposed copies of the map $\textit{\textbf{m}(N)}$ corresponding to the sectors
\beq \label{Theor19}
\left\{\,z \mid (k\,-\,1)(2\pi/24)\,<\,arg\,z\,<\,k(2\pi/24)\right\} (k\,=\,13,...,24)
\eeq
of the lower half of the $z$-plane which comprise the lower sheet of the $t$-Riemann surface.

\begin{center}
\FIGURE[h]{
\centerline{\epsfig{file=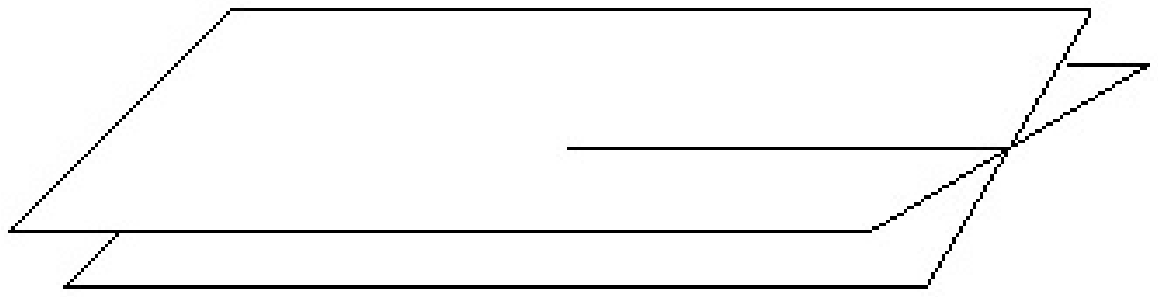,width=75mm,height=27mm}}
\caption{The two sheets of the $t$-Riemann surface.}
\label{FCT_fig4}}
\end{center}

Referring to Section $2.1.2$, the full symmetric group \,$Sym(\textit{\textbf{Z}}_{N}]S_{3})$\, acts faithfully \footnote{A group $G$ acts \emph{faithfully} on a set $S$ (faithful right action) if $sg = sh$ for all $s$ in $S$ implies that $g = h$ in $G$.} on the set
\,$\textit{\textbf{Z}}_{N}]S_{3}$.\, The action of an element $\psi$ of \,$Sym(\textit{\textbf{Z}}_{N}]S_{3})$\, on an
element \,$(\underline{m},\alpha)$\, of \,$\textit{\textbf{Z}}_{N}]S_{3}$\, will be written as \,$(\underline{m},\alpha)\psi$.\,
This action extends to the integral group algebra \,$\textit{\textbf{Z}}(\textit{\textbf{Z}}_{N}]S_{3})$\, by linearity
\beq \label{Theor20}
\Lb \sum_{(\underline{m},\alpha)\,\in\,\textit{\textbf{Z}}_{N}]S_{3}}n_{(\underline{m},\alpha)}(\underline{m},\alpha)\Rb\!
\psi\,=\,\sum_{(\underline{m},\alpha)\,\in\,\textit{\textbf{Z}}_{N}]S_{3}}n_{(\underline{m},\alpha)}\Lb (\underline{m},\alpha)
\psi\Rb\,.
\eeq
Referring to LEMMA 8, fix a common coset representative $\varphi_{i}$ of $S_{3}$ in
\,$Sym(\textit{\textbf{Z}}_{N}]S_{3})$\, and fix a pair \,$(\beta, \gamma) \in S_{3}\!\times\!S_{3}$\,.\,

We shall now specify the labeling scheme for the $t$-Riemann surface. Referring to Section $2.1.2$, the regions of the map $\textit{\textbf{m}(N)}$ have been partitioned into disjoint, nonempty
equivalence classes \,$\underline{0}, \underline{1},...,\underline{N - 1}$,\, and this set of equivalence classes forms the
underlying set of the cyclic group $\textit{\textbf{Z}}_{N}$. Hence there are twelve copies of $\textit{\textbf{Z}}_{N}$ on
the upper sheet and twelve copies of $\textit{\textbf{Z}}_{N}$ on the lower sheet of the $t$-Riemann surface. The copies of
$\textit{\textbf{Z}}_{N}$ are indexed by the elements of
\,$\textit{\textbf{Z}}(\textit{\textbf{Z}}_{N}]S_{3})]\textit{\textbf{Z}}S_{3}$\, which label the sectors on a particular
sheet. The branch point of the $t$-Riemann surface is labeled by the element \,$(0,\,\beta + \gamma)$\, of
\,$\textit{\textbf{Z}}(\textit{\textbf{Z}}_{N}]S_{3})]\textit{\textbf{Z}}S_{3}$,\, where $0$ denotes the zero element
of \,$\textit{\textbf{Z}}(\textit{\textbf{Z}}_{N}]S_{3})$.\, The labeling schemes for the twelve sectors of the upper sheet and the twelve sectors of the lower sheet of the $t$-Riemann surface by
the elements of \,$\textit{\textbf{Z}}(\textit{\textbf{Z}}_{N}]S_{3})]\textit{\textbf{Z}}S_{3}$\, are shown in \fig{FCT_fig5}
and \fig{FCT_fig6}, respectively. There are two cases depending on whether \,$\beta \neq \gamma$\, or whether \,$\beta = \gamma$.\,

\begin{center}
\FIGURE[h]{
\centerline{\epsfig{file=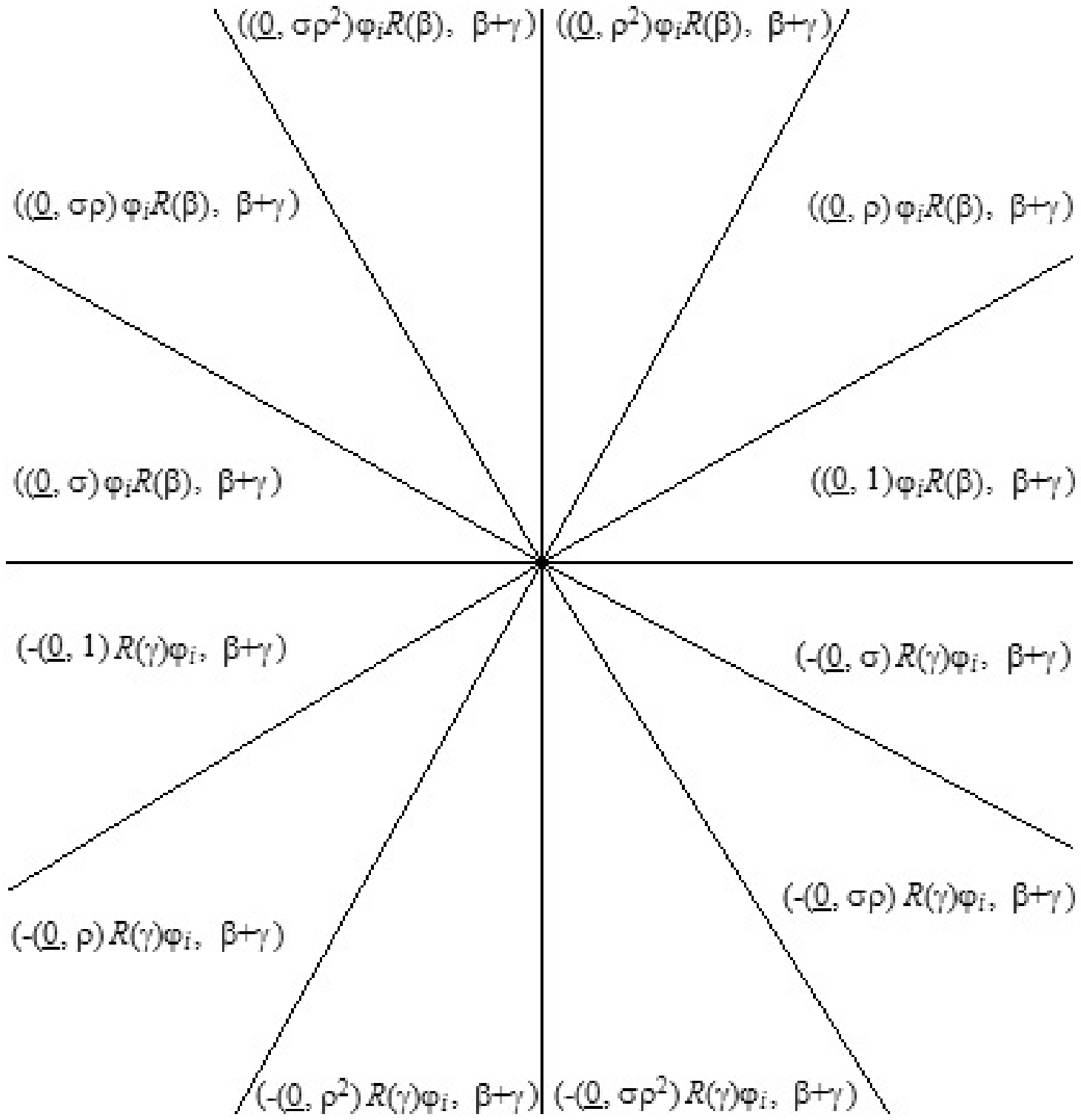,width=99mm,height=99mm}}
\caption{The upper sheet of the $t$-Riemann surface.}
\label{FCT_fig5}}
\end{center}

\begin{center}
\FIGURE[h]{
\centerline{\epsfig{file=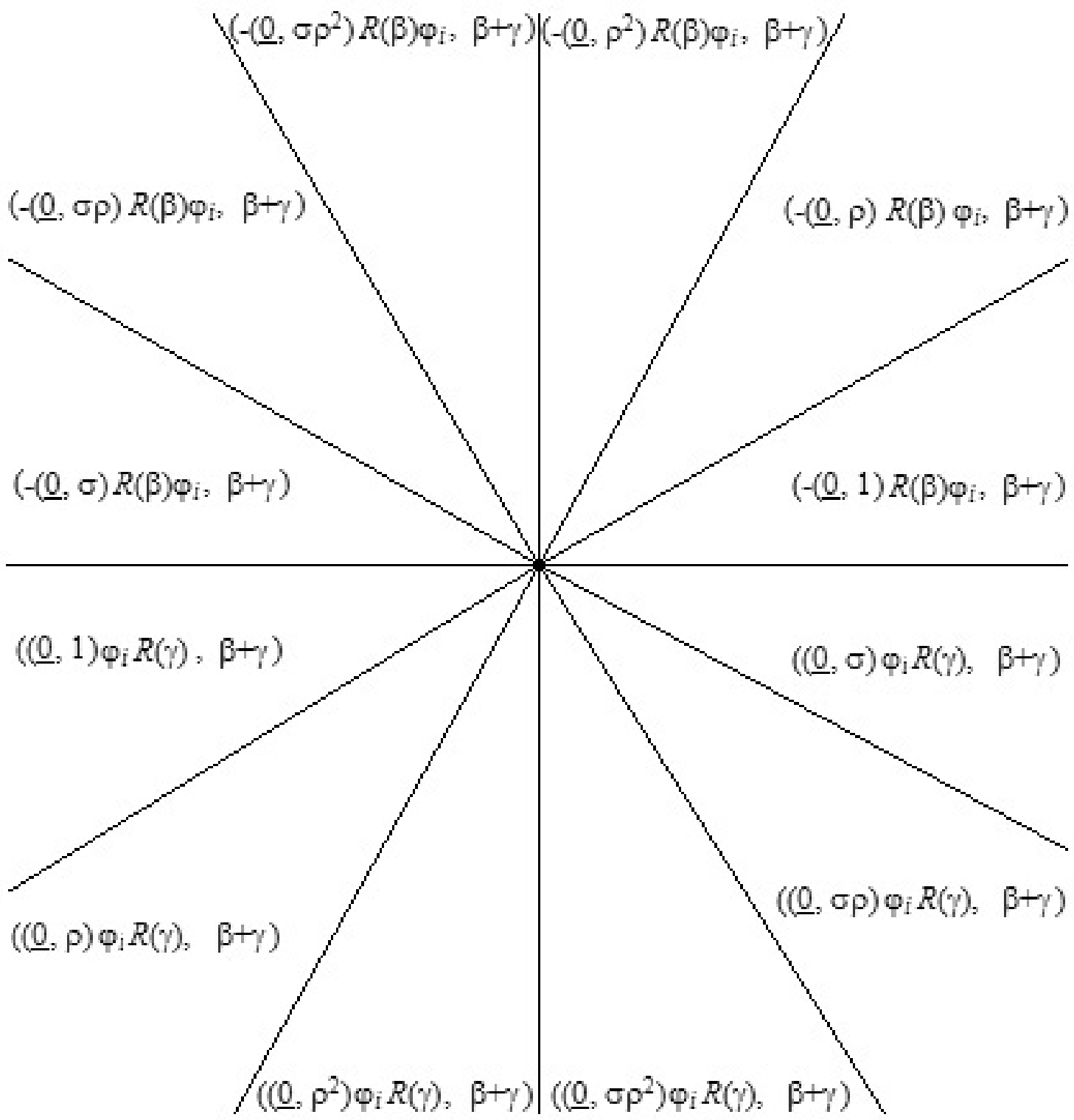,width=99mm,height=99mm}}
\caption{The lower sheet of the $t$-Riemann surface.}
\label{FCT_fig6}}
\end{center}

\vskip 2truecm %
\hskip -0.7truecm %
$\textbf{CASE 1.}$ Suppose \,$\beta \neq \gamma$.\,
\begin{itemize}
\item[]LEMMA $9$. {\textit{Referring to LEMMA $8$, fix a common representative $\varphi_{i}$ of the left and right cosets of
$S_{3}$ in \,$Sym(\textit{\textbf{Z}}_{N}]S_{3})$.\, Fix a pair \,$(\beta, \gamma) \in S_{3}\!\times\!S_{3}$\, with
\,$\beta \neq \gamma$.\, Referring to LEMMA $5$, define a subset \,$T_{(\beta,\gamma)}$\, of
\,$\textit{\textbf{Z}}(\textit{\textbf{Z}}_{N}]S_{3})]\textit{\textbf{Z}}S_{3}$\, as follows:
{\setlength\arraycolsep{2pt}
\bea \label{Theor22}
T_{(\beta,\gamma)}\,& = &\,\left\{((\underline{m},\alpha),\beta + \gamma) \mid (\underline{m},\alpha)\,\in\,
\textit{\textbf{Z}}_{N}]S_{3}\right\}
\nonumber \\
& & \,\,\,\,\,\,\,\,\,\,\,\,\,\,\,\,\,\,\,\,\,\,\,\,\,\,\,\,\,\,\,\,\,\,\,\,\,\,\,\,\,\,\,\,\bigcup
\nonumber \\
& &
\,\,\,\,\,\,\,\,\,\,\,\,\,\,\,\,\,\,\,\,\,\,\,\,\,\,\,\,\,\,\left\{(0,\,\beta + \gamma) \right\}
\nonumber \\
& & \,\,\,\,\,\,\,\,\,\,\,\,\,\,\,\,\,\,\,\,\,\,\,\,\,\,\,\,\,\,\,\,\,\,\,\,\,\,\,\,\,\,\,\,\bigcup
\nonumber \\
& &
\left\{(-(\underline{m},\alpha),\beta + \gamma) \mid (\underline{m},\alpha)\,\in\,
\textit{\textbf{Z}}_{N}]S_{3}\right\}\,.
\eea}
Referring to the preceding discussion, consider the composite function in \eq{Theor16} of the complex $z$-plane to the
$w$-Riemann surface. There is a copy of the set \,$T_{(\beta,\gamma)}$\,  on the upper sheet and a copy of the set
\,$T_{(\beta,\gamma)}$\,  on the lower sheet of the $t$-Riemann surface according to the labels of the sectors in \fig{FCT_fig5}
and \fig{FCT_fig6} with the branch point labeled by the element $(0,\,\beta + \gamma)$ of both copies. The rotation of the
$z$-plane by $\pi$ radians induces a permutation
\beq \label{Theor23}
p\!:\,\,T_{(\beta,\gamma)}\,\rightarrow\,T_{(\beta,\gamma)}
\eeq
given by
{\setlength\arraycolsep{2pt}
\bea \label{Theor24}
\Lb -(\underline{m},\alpha)R(\gamma)\varphi_{i},\,\beta + \gamma \Rb p \,& = &\,
\Lb (\underline{m},\alpha)\varphi_{i}R(\gamma),\,\beta + \gamma \Rb
\nonumber \\
(0,\,\beta + \gamma)p\,& = &\,(0,\,\beta + \gamma)
\nonumber \\
\Lb (\underline{m},\alpha)\varphi_{i}R(\beta),\,\beta + \gamma
\Rb p \,& = &\, \Lb
-(\underline{m},\alpha)R(\beta)\varphi_{i},\,\beta + \gamma \Rb\,,
\eea}
for all \,$(\underline{m},\alpha) \in \textit{\textbf{Z}}_{N}]S_{3}$,\, such that each point of the copy of
\,$T_{(\beta,\gamma)}$\, on the upper sheet moves continuously along a circular curve that winds exactly once around the
branch point, to the point superposed directly below it on the copy of \,$T_{(\beta,\gamma)}$\, on the lower sheet of the
$t$-Riemann surface.
}}
\end{itemize}

\begin{itemize}
\item[]LEMMA $10$. {\textit{Referring to LEMMA $9$, let \,$Sym(T_{(\beta,\gamma)})$\, denote the full permutation group of
the set \,$T_{(\beta,\gamma)}$.\, Let \,$<\!p\!>$\, denote the cyclic subgroup of \,$Sym(T_{(\beta,\gamma)})$\, generated by
$p$. Then \,$<\!p\!>$\, is nontrivial and acts faithfully on the set \,$T_{(\beta,\gamma)}$.
}}
\end{itemize}

\begin{itemize}
\item[]LEMMA $11$. {\textit{Referring to LEMMA $9$ and LEMMA $10$, let \,\,\,\,$1\!\!: \textit{\textbf{C}} \rightarrow
\textit{\textbf{C}}\,;\,\,z \rightarrow z$\, denote the identity and \,\,$\pi\!\!: \textit{\textbf{C}} \rightarrow
\textit{\textbf{C}}\,;\,\,z \rightarrow -z$\, denote the rotation through an angle of $\pi$ radians of the $z$-plane.
Then the two-element cyclic group $\left\{1, \pi\right\}$ acts faithfully on the set \,$<\!p\!>$\, as follows:\,
\,$p^{n}1 = p^{n}$\, and \,$p^{n}\pi = p^{1 - n}$,\, for all $n$ in $\textbf{\textbf{Z}}$.
}}
\end{itemize}

\begin{itemize}
\item[]LEMMA $12$. {\textit{Putting together LEMMA $9$, LEMMA $10$ and LEMMA $11$, there is a well-defined action of the
two-element cyclic group \,$\left\{1,\pi\right\}$\, on the set \,$T_{(\beta, \gamma)}$\, given by
{\setlength\arraycolsep{2pt}
\bea \label{Theor25}
\Lb (\underline{m},\alpha)\varphi_{i}R(\gamma),\,\beta + \gamma \Rb 1 \,& = &\,
\Lb (\underline{m},\alpha)\varphi_{i}R(\gamma),\,\beta + \gamma \Rb
\nonumber \\
(0,\,\beta + \gamma)1\,& = &\,(0,\,\beta + \gamma)
\nonumber \\
\Lb -(\underline{m},\alpha)R(\beta)\varphi_{i},\,\beta + \gamma
\Rb 1 \,& = &\, \Lb
-(\underline{m},\alpha)R(\beta)\varphi_{i},\,\beta + \gamma \Rb\,,
\eea}
and
{\setlength\arraycolsep{2pt}
\bea \label{Theor26}
\Lb (\underline{m},\alpha)\varphi_{i}R(\gamma),\,\beta + \gamma \Rb \pi \,& = &\,
\Lb -(\underline{m},\alpha)R(\gamma)\varphi_{i},\,\beta + \gamma \Rb
\nonumber \\
(0,\,\beta + \gamma) \pi\,& = &\,(0,\,\beta + \gamma)
\nonumber \\
\Lb -(\underline{m},\alpha)R(\beta)\varphi_{i},\,\beta + \gamma
\Rb \pi \,& = &\, \Lb
(\underline{m},\alpha)\varphi_{i}R(\beta),\,\beta + \gamma \Rb\,,
\eea}
for all \,$(\underline{m},\alpha)$\, in $\textit{\textbf{Z}}_{N}]S_{3}$. This action is faithful.
}}
\end{itemize}

\vskip 1truecm %
\hskip -0.7truecm %
$\textbf{CASE 2.}$ Suppose \,$\beta = \gamma$. Referring to the labeling of the sectors of the sheets of the
$t$-Riemann surface in \fig{FCT_fig5} and \fig{FCT_fig6}, note that in this case \,$R(\beta) = R(\gamma)$\, and \,$\beta + \gamma = 2\beta$\, in the group algebra $\textit{\textbf{Z}}S_{3}$.

\begin{itemize}
\item[]LEMMA $13$. {\textit{Referring to LEMMA $8$, fix a common representative $\varphi_{i}$ of the left and right cosets
of $S_{3}$ in \,$Sym(\textit{\textbf{Z}}_{N}]S_{3})$.\, Fix a pair \,$(\beta, \beta) \in S_{3}\!\times\!S_{3}$\, with
\,$\beta = \gamma$.\, Referring to LEMMA $5$, define a subset \,$T_{(\beta,\beta)}$\, of
\,$\textit{\textbf{Z}}(\textit{\textbf{Z}}_{N}]S_{3})]\textit{\textbf{Z}}S_{3}$\, as follows:
{\setlength\arraycolsep{2pt}
\bea \label{Theor27}
T_{(\beta,\beta)}\,& = &\,\left\{((\underline{m},\alpha),2\beta) \mid (\underline{m},\alpha)\,\in\,
\textit{\textbf{Z}}_{N}]S_{3}\right\}
\nonumber \\
& & \,\,\,\,\,\,\,\,\,\,\,\,\,\,\,\,\,\,\,\,\,\,\,\,\,\,\,\,\,\,\,\,\,\,\,\,\,\,\,\,\,\,\,\,\bigcup
\nonumber \\
& &
\,\,\,\,\,\,\,\,\,\,\,\,\,\,\,\,\,\,\,\,\,\,\,\,\,\,\,\,\,\,\left\{(0,\,2\beta) \right\}
\nonumber \\
& & \,\,\,\,\,\,\,\,\,\,\,\,\,\,\,\,\,\,\,\,\,\,\,\,\,\,\,\,\,\,\,\,\,\,\,\,\,\,\,\,\,\,\,\,\bigcup
\nonumber \\
& &
\left\{(-(\underline{m},\alpha),2\beta) \mid (\underline{m},\alpha)\,\in\,
\textit{\textbf{Z}}_{N}]S_{3}\right\}\,.
\eea}
Referring to the preceding discussion, consider the composite function in \eq{Theor16} of the complex $z$-plane to the
$w$-Riemann surface. There is a copy of the set \,$T_{(\beta,\beta)}$\,  on the upper sheet and a copy of the set
\,$T_{(\beta,\beta)}$\,  on the lower sheet of the $t$-Riemann surface according to the labels of the sectors in \fig{FCT_fig5}
and \fig{FCT_fig6} with the branch point labeled by the element $(0,\,2\beta)$ of both copies. The rotation of the
$z$-plane by $\pi$ radians induces a permutation
\beq \label{Theor28}
p\!:\,\,T_{(\beta,\beta)}\,\rightarrow\,T_{(\beta,\beta)}
\eeq
given by
{\setlength\arraycolsep{2pt}
\bea \label{Theor29}
\Lb -(\underline{m},\alpha)R(\beta)\varphi_{i},\,2\beta \Rb p \,& = &\,
\Lb (\underline{m},\alpha)\varphi_{i}R(\beta),\,2\beta \Rb
\nonumber \\
(0,\,2\beta)p\,& = &\,(0,\,2\beta)
\nonumber \\
\Lb (\underline{m},\alpha)\varphi_{i}R(\beta),\,2\beta
\Rb p \,& = &\, \Lb
-(\underline{m},\alpha)R(\beta)\varphi_{i},\,2\beta \Rb\,,
\eea}
for all \,$(\underline{m},\alpha) \in \textit{\textbf{Z}}_{N}]S_{3}$,\, such that each point of the copy of
\,$T_{(\beta,\beta)}$\, on the upper sheet moves continuously along a circular curve that winds exactly once around the
branch point, to the point superposed directly below it on the copy of \,$T_{(\beta,\beta)}$\, on the lower sheet of the
$t$-Riemann surface. Then \,$p = p^{-1}$\, so that \,$< \!p\!> = \left\{1, p\right\}$\, is a two-element cyclic subgroup
of the full permutation group \,$Sym(T_{(\beta, \beta)})$,\, and \,$<\!p\!>$\, acts faithfully on the set \,$T_{(\beta,\beta)}$.\,
}}
\end{itemize}

\begin{itemize}
\item[]LEMMA $14$. {\textit{Referring to LEMMA $13$, let \,\,\,\,$1\!\!: \textit{\textbf{C}} \rightarrow \textit{\textbf{C}}
\,;\,\,z \rightarrow z$\, denote the identity and \,\,$\pi\!\!: \textit{\textbf{C}} \rightarrow \textit{\textbf{C}}\,;\,\, z
\rightarrow -z$\, denote the rotation through an angle of $\pi$ radians of the $z$-plane. Then there is a well-defined action
of the two-element cyclic group \,$\left\{1, \pi\right\}$\, on the set \,$T_{(\beta,\beta)}$\, given by
{\setlength\arraycolsep{2pt}
\bea \label{Theor30}
\Lb (\underline{m},\alpha)\varphi_{i}R(\beta),\,2\beta \Rb 1 \,& = &\,
\Lb (\underline{m},\alpha)\varphi_{i}R(\beta),\,2\beta \Rb
\nonumber \\
(0,\,2\beta )1\,& = &\,(0,\,2\beta)
\nonumber \\
\Lb -(\underline{m},\alpha)R(\beta)\varphi_{i},\,2\beta
\Rb 1 \,& = &\, \Lb
-(\underline{m},\alpha)R(\beta)\varphi_{i},\,2\beta \Rb\,,
\eea}
and
{\setlength\arraycolsep{2pt}
\bea \label{Theor31}
\Lb (\underline{m},\alpha)\varphi_{i}R(\beta),\,2\beta \Rb \pi \,& = &\,
\Lb -(\underline{m},\alpha)R(\beta)\varphi_{i},\,2\beta \Rb
\nonumber \\
(0,\,2\beta) \pi\,& = &\,(0,\,2\beta)
\nonumber \\
\Lb -(\underline{m},\alpha)R(\beta)\varphi_{i},\,2\beta
\Rb \pi \,& = &\, \Lb
(\underline{m},\alpha)\varphi_{i}R(\beta),\,2\beta \Rb\,,
\eea}
for all \,$(\underline{m},\alpha)$\, in $\textit{\textbf{Z}}_{N}]S_{3}$. This action is faithful.
}}
\end{itemize}

\subsubsection{Main Construction}
Let us review the final goal of Section $2$\, and recall the main definitions we have made so far. We have defined $N$ to
be the minimal number of colors required to properly color any map from the class of all maps on the sphere. We know that
$4 \leq N \leq 6$ from Section $2.1.1$. In Section 2.1.2., we have chosen a specific map $\textit{\textbf{m}(N)}$ on the sphere which requires all of $N$ colors \,$0, 1, ..., N - 1$\, to properly color it. The map $\textit{\textbf{m}(N)}$ has been properly colored and the
regions of $\textit{\textbf{m}(N)}$ partitioned into disjoint, nonempty equivalence classes
\,$\underline{0}, \underline{1}, ..., \underline{N - 1}$\, according to the color they receive. The set
\,$\left\{\underline{0}, \underline{1}, ..., \underline{N - 1}\right\}$\, is endowed with the structure of the cyclic
group $\textit{\textbf{Z}}_{N}$ under addition modulo $N$. In Section $2.1.2$ we have also built the split extension
\,$\textit{\textbf{Z}}_{N}]S_{3}$.\, The underlying set \,$\textit{\textbf{Z}}_{N}]S_{3}$\, of cardinality $6N$ is defined to
be the point set of the Steiner system \,$S(N + 1, 2N, 6N)$\, which will be constructed in this section. We are required to
define the blocks of size $2N$ and show that every set of \,$N + 1$\, points is contained in a unique block. Once this goal
is achieved, LEMMA $3$ shows that \,$N = 4$.

\begin{itemize}
\item[]LEMMA $15$. {\textit{Let \,$\textit{\textbf{Z}}_{N}]S_{3}$\, denote the split extension defined in LEMMA $4$ and
\,$Sym(\textit{\textbf{Z}}_{N}]S_{3})$\, denote the full permutation group on the set \,$\textit{\textbf{Z}}_{N}]S_{3}$.\,
Define
\beq \label{Theor32}
\mu\!: \,Sym(\textit{\textbf{Z}}_{N}]S_{3})\,\rightarrow\,Sym(\textit{\textbf{Z}}_{N}]S_{3}),
\eeq
by
\beq \label{Theor33}
\psi\,=\,R(\gamma)\varphi_{i}\,\rightarrow\,\varphi_{i} R(\gamma)\,=\,\psi^{\mu}\,.
\eeq
Then $\mu$ is a bijection (one-to-one correspondence) of the set $Sym(\textit{\textbf{Z}}_{N}]S_{3})$ with itself.
}}
\end{itemize}

\begin{itemize}
\item[]LEMMA $16$. {\textit{Define the set $G$ as follows:
\beq \label{Theor34}
G\,=\,\left\{\begin{array}{ll}
\left(\begin{array}{ccc}
\!\psi           \!\!& \\
\,\,\psi^{\mu} \!\!& \\
\end{array} \right)
\left. \begin{array}{ccc}
\mid  \hskip 0.024truecm & \\
\mid  \hskip 0.024truecm & \\
\end{array} \right.
& \textrm{\!\!\!\!\!\!\!\!$\mid$ \,\,$\psi \in Sym(\textit{\textbf{Z}}_{N}]S_{3})$}\\
\end{array} \right\}\,=\,
\left\{\begin{array}{ll}
\left(\begin{array}{ccc}
R(\gamma)\varphi_{i}  \!\!& \\
\varphi_{i}R(\gamma)  \!\!& \\
\end{array} \right)
\left. \begin{array}{ccc}
\mid  \hskip 0.024truecm & \\
\mid  \hskip 0.024truecm & \\
\end{array} \right.
& \textrm{\!\!\!\!\!\!\!\!$\mid$}
\left. \begin{array}{ccc}
\!\!\!\!\!\!\!\!\gamma\,\in\,S_{3} & \\
\,\,i\,=\,1, ..., k      & \\
\end{array} \right.
\end{array} \right\}\,.
\eeq
Define multiplication in $G$ as follows:
\beq \label{Theor35}
\left(\begin{array}{ccc}
\psi_{1}           \!\!& \\
\,\psi_{1}^{\mu} \!\!& \\
\end{array} \right)
\left(\begin{array}{ccc}
\psi_{2}           \!\!& \\
\,\psi_{2}^{\mu} \!\!& \\
\end{array} \right)\,=\,
\left(\begin{array}{ccc}
(\psi_{1}\psi_{2})           \!\!& \\
\,\,(\psi_{1}\psi_{2})^{\mu} \!\!& \\
\end{array} \right)\,
\eeq \
i.e.
\beq \label{Theor36}
\left(\begin{array}{ccc}
R(\gamma_{1})\varphi_{i_{1}}      \!\!& \\
\,\varphi_{i_{1}}R(\gamma_{1})  \!\!& \\
\end{array} \right)
\left(\begin{array}{ccc}
R(\gamma_{2})\varphi_{i_{2}}     \!\!& \\
\,\varphi_{i_{2}}R(\gamma_{2}) \!\!& \\
\end{array} \right)\,=\,
\left(\begin{array}{ccc}
R(\gamma_{3})\varphi_{i_{3}}     \!\!& \\
\,\varphi_{i_{3}}R(\gamma_{3}) \!\!& \\
\end{array} \right)\,,
\eeq
where \,$R(\gamma_{3})\varphi_{i_{3}} $\, is the unique expression for \,$R(\gamma_{1})\varphi_{i_{1}}
R(\gamma_{2})\varphi_{i_{2}}$\, according to the right coset decomposition of $S_{3}$ in
\,$Sym(\textit{\textbf{Z}}_{N}]S_{3})$\,. Then $G$ is a group.
}}
\end{itemize}

\begin{itemize}
\item[]LEMMA $17$. {\textit{Consider the set \,$\textit{\textbf{Z}}_{N}]S_{3}$\, and let
\beq \label{Theor37}
\left(\begin{array}{ccc}
\psi            \!\!& \\
\,\,\psi^{\mu}  \!\!& \\
\end{array} \right)\,=\,
\left(\begin{array}{ccc}
R(\gamma)\varphi_{i}     \!\!& \\
\,\varphi_{i}R(\gamma) \!\!& \\
\end{array} \right)\,\in\,G\,.
\eeq
Define
\beq \label{Theor38}
\uparrow\!\!\left(\begin{array}{ccc}
\psi            \!\!& \\
\,\,\psi^{\mu}  \!\!& \\
\end{array} \right)\!:
\,\,\,\textit{\textbf{Z}}_{N}]S_{3}\,\rightarrow\,\textit{\textbf{Z}}_{N}]S_{3}\,,
\eeq
by
\beq \label{Theor39}
(\underline{m},\alpha)\,\rightarrow\,(\underline{m},\alpha)\uparrow\!\!
\left(\begin{array}{ccc}
\psi            \!\!& \\
\,\,\psi^{\mu}  \!\!& \\
\end{array} \right)\,=\,
(\underline{m},\alpha)\uparrow\!\!
\left(\begin{array}{ccc}
R(\gamma)\varphi_{i}     \!\!& \\
\,\varphi_{i}R(\gamma) \!\!& \\
\end{array} \right)\,=\,
(\underline{m},\alpha)R(\gamma)\varphi_{i}\,.
\eeq
Define
\beq \label{Theor40}
\downarrow\!\!\left(\begin{array}{ccc}
\psi            \!\!& \\
\,\,\psi^{\mu}  \!\!& \\
\end{array} \right)\!:
\,\,\,\textit{\textbf{Z}}_{N}]S_{3}\,\rightarrow\,\textit{\textbf{Z}}_{N}]S_{3}\,,
\eeq
by
\beq \label{Theor41}
(\underline{m},\alpha)\,\rightarrow\,(\underline{m},\alpha)\downarrow\!\!
\left(\begin{array}{ccc}
\psi            \!\!& \\
\,\,\psi^{\mu}  \!\!& \\
\end{array} \right)\,=\,
(\underline{m},\alpha)\downarrow\!\!
\left(\begin{array}{ccc}
R(\gamma)\varphi_{i}     \!\!& \\
\,\varphi_{i}R(\gamma) \!\!& \\
\end{array} \right)\,=\,
(\underline{m},\alpha)\varphi_{i}R(\gamma)\,.
\eeq
Then
\beq \label{Theor42}
(\underline{m},\alpha)\uparrow\!\!\left(\begin{array}{ccc}
\psi            \!\!& \\
\,\,\psi^{\mu}  \!\!& \\
\end{array} \right)\,=\,
(\underline{m},\alpha)\downarrow\!\!\left(\begin{array}{ccc}
\psi            \!\!& \\
\,\,\psi^{\mu}  \!\!& \\
\end{array} \right)
\,\,\,\, for \,\,all \,\,\,\,(\underline{m},\alpha)\,\in\,\textit{\textbf{Z}}_{N}]S_{3}\,.
\eeq
Both $\uparrow$ and $\downarrow$ are well-defined, faithful and \,$|\textit{\textbf{Z}}_{N}]S_{3}|$-transitive \footnote{The action of a group $G$ on a set $S$ is said to be \emph{$n$-transitive} if for any pair of $n$-element subsets $\{x_1, ..., x_n\}$ and $\{y_1, ..., y_n\}$ of $S$, there exists $g$ in $G$ such that $x_ig = y_ig$ for $i = 1, ...,n$.} right actions of the group $G$ on the set $\textit{\textbf{Z}}_{N}]S_{3}$.
}}
\end{itemize}

\begin{itemize}
\item[]LEMMA $18$. {\textit{Let \,$(\underline{m}_{1},\alpha_{1}), ..., (\underline{m}_{r},\alpha_{r})$\, be any \,$r$\,
distinct elements of \,$\textit{\textbf{Z}}_{N}]S_{3}$\, and let \,$(\underline{n}_{1},\beta_{1}), ...,
(\underline{n}_{s},\beta_{s})$\, be any
\,$s$\, distinct elements of \,$\textit{\textbf{Z}}_{N}]S_{3}$.\, Let
\beq \label{Theor43}
H_{r,s}\,=\,\left\{\begin{array}{ll}
\left(\begin{array}{ccc}
\psi  \!\!& \\
\,\psi^{\mu}  \!\!& \\
\end{array} \right) \in G
\left. \begin{array}{ccc}
\mid  \hskip 0.024truecm & \\
\mid  \hskip 0.024truecm & \\
\end{array} \right.
& \textrm{\!\!\!\!\!\!\!\!$\mid$}
\left. \begin{array}{ccc}
\,\,\,(\underline{m}_{i},\alpha_{i})\uparrow\!\!\left(\begin{array}{ccc}
\psi            \!\!& \\
\,\,\psi^{\mu}  \!\!& \\
\end{array} \right)\,=\,(\underline{m}_{i},\alpha_{i})\,\,\,\, for
\,\,\,\, i\,=\,1,...,r
& \\
\left. \begin{array}{ccc}
\hskip -3.9truecm and
\end{array} \right.
& \\
\,\,\,(\underline{n}_{j},\beta_{j})\downarrow\!\!\left(\begin{array}{ccc}
\psi            \!\!& \\
\,\,\psi^{\mu}  \!\!& \\
\end{array} \right)\,=\,(\underline{n}_{j},\beta_{j})\,\,\,\, for
\,\,\,\, j\,=\,1,...,s
& \\
\end{array} \right.
\end{array} \right\}\,,
\eeq
then $H_{r,s}$ is a subgroup of $G$.
}}
\end{itemize}

\begin{itemize}
\item[]LEMMA $19$. {\textit{Define
\beq \label{Theor44}
H\,=\,\left\{\begin{array}{ll}
\left(\begin{array}{ccc}
\psi  \!\!& \\
\,\psi^{\mu}  \!\!& \\
\end{array} \right) \in G
\left. \begin{array}{ccc}
\mid  \hskip 0.024truecm & \\
\mid  \hskip 0.024truecm & \\
\end{array} \right.
& \textrm{\!\!\!\!\!\!\!\!$\mid$}
\left. \begin{array}{ccc}
\,\,\,(\underline{m},1)\uparrow\!\!\left(\begin{array}{ccc}
\psi            \!\!& \\
\,\,\psi^{\mu}  \!\!& \\
\end{array} \right)\,=\,(\underline{m},1)\,\,\,\, for \,\,all
\,\,\,\, \underline{m} \in \textit{\textbf{Z}}_{N}
& \\
\left. \begin{array}{ccc}
\hskip -3.7truecm and
\end{array} \right.
& \\
\hskip -3.1truecm (\underline{0},\sigma)\downarrow\!\!\left(\begin{array}{ccc}
\psi            \!\!& \\
\,\,\psi^{\mu}  \!\!& \\
\end{array} \right)\,=\,(\underline{0},\sigma)
& \\
\end{array} \right.
\end{array} \right\}\,.
\eeq
Then given
\,$
\left(\begin{array}{ccc}
\psi            \!\!& \\
\,\,\psi^{\mu}  \!\!& \\
\end{array} \right)\,\in\,H
$\, \\
either
\beq \label{Theor45}
(\underline{m},\alpha)\downarrow\!\!\left(\begin{array}{ccc}
\psi            \!\!& \\
\,\,\psi^{\mu}  \!\!& \\
\end{array} \right)\,=\,(\underline{m},\alpha)\,\,\,\, for \,\,all
\,\,\,\, (\underline{m},\alpha) \in \textit{\textbf{Z}}_{N}]S_{3}
\eeq
or
\beq \label{Theor46}
(\underline{m},\alpha)\downarrow\!\!\left(\begin{array}{ccc}
\psi            \!\!& \\
\,\,\psi^{\mu}  \!\!& \\
\end{array} \right)\,=\,(\underline{m},\alpha^{\sigma})\,\,\,\, for \,\,all
\,\,\,\, (\underline{m},\alpha) \in \textit{\textbf{Z}}_{N}]S_{3}\,.
\eeq
}}
\end{itemize}

\begin{itemize}
\item[]LEMMA $20$. {\textit{
Let $H$ be the subgroup of $G$ defined in LEMMA $19$. Then $H$ is a nontrivial group of involutions \footnote{The order of an element $h$ of a group $H$ is the least positive integer $k$ such that $h^k = e$. An \emph{involution} is an element of order 2 in the group $H$.} of the set \,$\textit{\textbf{Z}}_{N}]S_{3}$\,. In particular, every nontrivial element of $H$ is of order 2.
}}
\end{itemize}

\begin{itemize}
\item[]LEMMA $21$. {\textit{Denote the right cosets of \,$\textit{\textbf{Z}}_{N}$\, in \,$\textit{\textbf{Z}}_{N}]S_{3}$\,
by \,$\textit{\textbf{Z}}_{N}, \textit{\textbf{Z}}_{N}\rho, \textit{\textbf{Z}}_{N}\rho^{2}, \textit{\textbf{Z}}_{N}\sigma,
\textit{\textbf{Z}}_{N}\sigma\rho, \textit{\textbf{Z}}_{N}\sigma\rho^{2}$. \\
Define
\beq \label{Theor47}
Fix\!\downarrow\!\!(H)\,=\,\left\{\begin{array}{ll}
(\underline{m},\alpha)
\in
\textit{\textbf{Z}}_{N}]S_{3}
\left. \begin{array}{ccc}
\mid  \hskip 0.024truecm & \\
\mid  \hskip 0.024truecm & \\
\end{array} \right.
& \textrm{\!\!\!\!\!\!\!\!$\mid$}
\left. \begin{array}{ccc}
\,\,\,(\underline{m},\alpha)\downarrow\!\!\left(\begin{array}{ccc}
\psi            \!\!& \\
\,\,\psi^{\mu}  \!\!& \\
\end{array} \right)\,=\,(\underline{m},\alpha)\,\,\,\, for \,\,all
\,\,\,\,
\left(\begin{array}{ccc}
\psi            \!\!& \\
\,\,\psi^{\mu}  \!\!& \\
\end{array} \right)
\in
H
\end{array} \right.
\end{array} \right\}\,.
\eeq
Then \\
\beq \label{Theor48}
Fix\!\downarrow\!\!(H)\,=\,\left\{(\underline{m},\alpha) \in \textit{\textbf{Z}}_{N}]S_{3} \,\mid\, \alpha\,=\,1\,\,\,\, or\,\,\,\,
\alpha\,=\,\sigma\right\}\,.
\eeq
$Fix\!\downarrow\!\!(H)$ is the set of elements of \,$Z_N]S_3$\, that are fixed by the action $\downarrow$ of all elements
of the group $H$. The action $\downarrow$ of a nontrivial element of $H$ transposes the coset
\,$\textit{\textbf{Z}}_{N}\rho$\, with the coset \,$\textit{\textbf{Z}}_{N}\rho^{2}$\, and also transposes the coset
\,$\textit{\textbf{Z}}_{N}\sigma\rho$\, with the coset \,$\textit{\textbf{Z}}_{N}\sigma\rho^{2}$.
}}
\end{itemize}

\begin{itemize}
\item[]LEMMA $22$. {\textit{Let \,$Norm_{{}_G}(H)$\, denote the normalizer \footnote{The \emph{normalizer} of a
subgroup $H$ of a group $G$ is the largest subgroup of $G$ in which $H$ is normal. It is defined as \,$$Norm_{{}_G}(H) = \left\{g \in G\!: \,gH = Hg\right\}.$$} of $H$
in $G$. The action $\downarrow$ of $G$ on \,$\textit{\textbf{Z}}_{N}]S_{3}$\, is restricted to an action $\downarrow$ of
\,$Norm_{{}_G}(H)$\, on \,$Fix\!\downarrow\!(H)$\, which is \,$(|\textit{\textbf{Z}}_{N}| + 1)$-transitive.
}}
\end{itemize}

\begin{itemize}
\item[]LEMMA $23$. {\textit{There exists a Steiner system \,$S(N + 1, 2N, 6N )$,\, where the points are the elements of the
set \,$\textit{\textbf{Z}}_{N}]S_{3}$,\, and the set of the blocks is
\beq \label{Theor49}
\left\{\begin{array}{ll}
Fix\!\downarrow\!\!(H) \downarrow\!\!
\left(\begin{array}{ccc}
\psi            \!\! & \\
\,\,\psi^{\mu}  \!\! & \\
\end{array} \right)
\left. \begin{array}{ccc}
\mid  \hskip 0.024truecm & \\
\mid  \hskip 0.024truecm & \\
\end{array} \right.
& \textrm{\!\!\!\!\!\!\!\!$\mid$}
\,\,\left(\begin{array}{ccc}
\psi            \!\!& \\
\,\,\psi^{\mu}  \!\!& \\
\end{array} \right)
\in
G
\end{array} \right\}\,.
\eeq
}}
\end{itemize}

\vskip 1truecm
\begin{itemize}
\item[]THEOREM. {\textit{Any map on the sphere can be properly colored by using at most four
colors.}} \\
\vskip -0.5truecm
PROOF: Referring to Section $2.1.1$, we have defined $N$ to be the minimal number of colors required to properly color any map
from the class of all maps on the sphere. Based on the definition of $N$, we have selected a specific map
$\textit{\textbf{m}(N)}$ on the sphere which requires no fewer than $N$ colors to be properly colored. Based on the
definition of the map $\textit{\textbf{m}(N)}$, we have selected the proper coloring of its regions using $N$ colors
\,$0, 1, ..., N-1$.\, Working with the fixed number $N$, the fixed map $\textit{\textbf{m}(N)}$, and the fixed proper
coloring of the regions of the map $\textit{\textbf{m}(N)}$, LEMMA $23$ has explicitly constructed the Steiner system
\,$S(N + 1, 2N, 6N)$.\, Now LEMMA $3$ implies that $N = 4$. $\Box$

\end{itemize}

\subsection{Schr\"{o}dinger disks and $t$-Riemann Surfaces}
It is well known that the quantum-mechanical behavior of a particle is completely described by the \emph{wave function}
\,$\Psi(\texttt{x},\texttt{y},\texttt{z},\texttt{t})\,=\,e^{i(k_{\texttt{x}}\texttt{x} + k_{\texttt{y}}\texttt{y} +
k_{\texttt{z}}\texttt{z} - \omega \texttt{t})}$\, which is a solution of the
Schr\"{o}dinger wave equation
\beq \label{Schr}
-\,\hbar^{2}\FFr{\partial^{2}\!\Psi}{\partial \textsf{t}^{2}}\,=\,-\,\hbar^{2}c^{2}\,\nabla^{2}\!\Psi\,+\,m^{2}c^{4}\Psi\,.
\eeq
At each point \,$(\texttt{x},\texttt{y},\texttt{z},\texttt{t})$\, of space-time, the value of the wave function
corresponds to a point on the boundary of a disc $\textit{\textbf{D}}$ centered at the origin of the complex plane \,$\textit{\textbf{C}}$\, as shown below in \fig{FCT_fig7}. We call $\textit{\textbf{D}}$ a \emph{Schr\"{o}dinger disc}.

\begin{center}
\FIGURE[h]{
\centerline{\epsfig{file=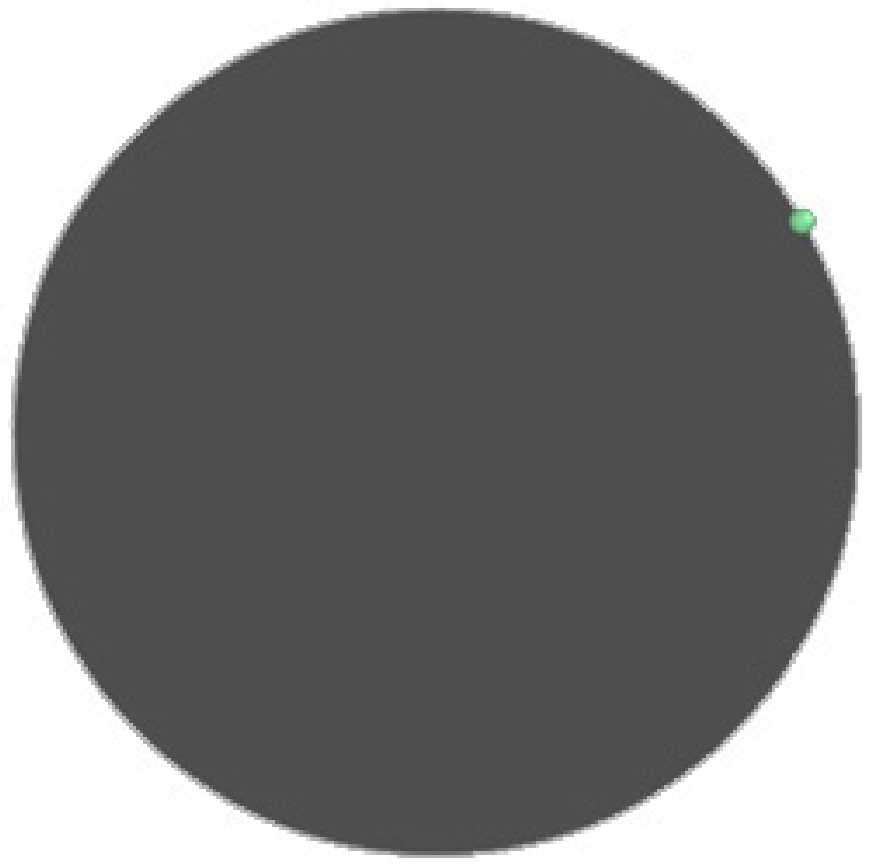,width=55mm,height=55mm}}
\caption{A particle is represented by a Schr\"{o}dinger disc.}
\label{FCT_fig7}}
\end{center}

By the uncertainty principle, it is impossible to specify precisely and simultaneously both the position and the momentum of the particle. We have defined the Schr\"{o}dinger disc $\textit{\textbf{D}}$ representing the particle at a
precisely specified position \,$(\texttt{x},\texttt{y},\texttt{z},\texttt{t})$\, in space-time, with an uncertain
momentum vector $\textit{\textbf{p}}$. Instead, we may define $\textit{\textbf{D}}$ representing a particle with a precisely
specified momentum vector $\textit{\textbf{p}}$ and an uncertain position \,$(\texttt{x},\texttt{y},\texttt{z},\texttt{t})$\,
in space-time. Then $\textit{\textbf{D}}$ may be oriented in one of two possible ways: clockwise or counter-clockwise,
depending on whether we choose the normal vector to the complex plane according to the left-hand or the right-hand rule. We align this normal vector with the momentum vector $\textit{\textbf{p}}$ of the particle in the case that $\textit{\textbf{p}}$
is precisely specified. The left-handed orientation of $\textit{\textbf{D}}$ will represent a particle of left-handed
helicity and the right-handed orientation will represent a particle of right-handed helicity. Note that helicity is conserved for massless particles, which always travel with the velocity of light, but not for massive particles. According to special relativity, the direction of the momentum vector $\textit{\textbf{p}}$ is reversed relative to any reference frame that moves faster than the particle. In the case of massive particles which always travel with a velocity less than light, it is possible for an observer to change to a reference frame that moves faster than the spinning particle, in which case the particle will appear to move backwards and its helicity will be reversed. Thus, for a particle with positive mass, the helicity cannot be conserved with respect to all reference frames. Since we now know that the neutrinos have positive mass \cite{Kami}, all the fermions are known to be massive. Hence helicity is not a relativistic invariant for the fermions.

We now build the SM of particles from copies of oriented Schr\"{o}dinger discs, arranged in a certain way as dictated by the mathematical proof of the four color theorem, following \cite{Dhar2}. After the proof of the four color theorem (Sections $2.1.1$,\, $2.1.2$,\, $2.1.3$,\, $2.1.4$) is complete, a posteriori we know that \,$N = 4$.\, From now on, our description is more concrete and pictorial, as opposed to \fig{FCT_fig3}, \fig{FCT_fig4}, \fig{FCT_fig5} and \fig{FCT_fig6}. The four colors \,$0, 1, 2, 3$\, are represented by the palette shown in \fig{FCT_fig8}.
\begin{center}
\FIGURE[h]{
\centerline{\epsfig{file=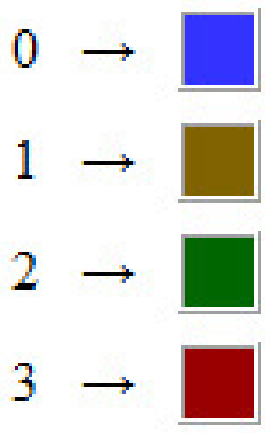,width=20mm,height=30mm}}
\caption{The palette of the four colors \,$0, 1, 2, 3$.}
\label{FCT_fig8}}
\end{center}
We select the map \,$\textit{\textbf{m}}(4)$\, on the surface of the sphere, with its proper coloring as shown in \fig{FCT_fig9}.
\begin{center}
\FIGURE[h]{
\centerline{\epsfig{file=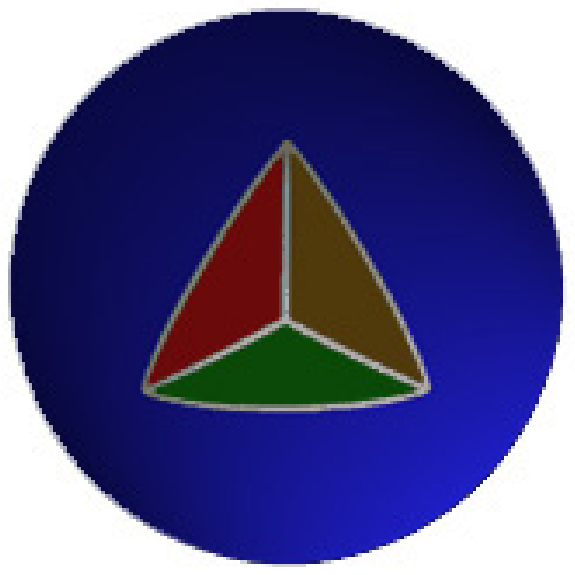,width=47mm,height=47mm}}
\caption{The map \,$\textit{\textbf{m}}(4)$\, on the sphere.}
\label{FCT_fig9}}
\end{center}
According to the color each region receives, the regions are partitioned into four equivalence classes that
form a cyclic group \,$\textit{\textbf{Z}}_{4} = \left\{\underline{0}, \underline{1}, \underline{2}, \underline{3}\right\}$\,
under addition modulo $4$. By boring a small hole in the blue region $\underline{0}$, we may deform the surface of the sphere until it is flat, in order to obtain a copy of the map \,$\textit{\textbf{m}}(4)$\, on the complex plane $\textit{\textbf{C}}$ (stereographic
projection). We may perform the deformation of the map in such a way that both the origin and the boundary of the disc
$\textit{\textbf{D}}$ are contained entirely inside the blue region of the map. Thus we obtain the map \,$\textit{\textbf{m}}(4)$\, inside $\textit{\textbf{D}}$, with the origin inside the blue region, as shown in \fig{FCT_fig10}.
\begin{center}
\FIGURE[h]{
\centerline{\epsfig{file=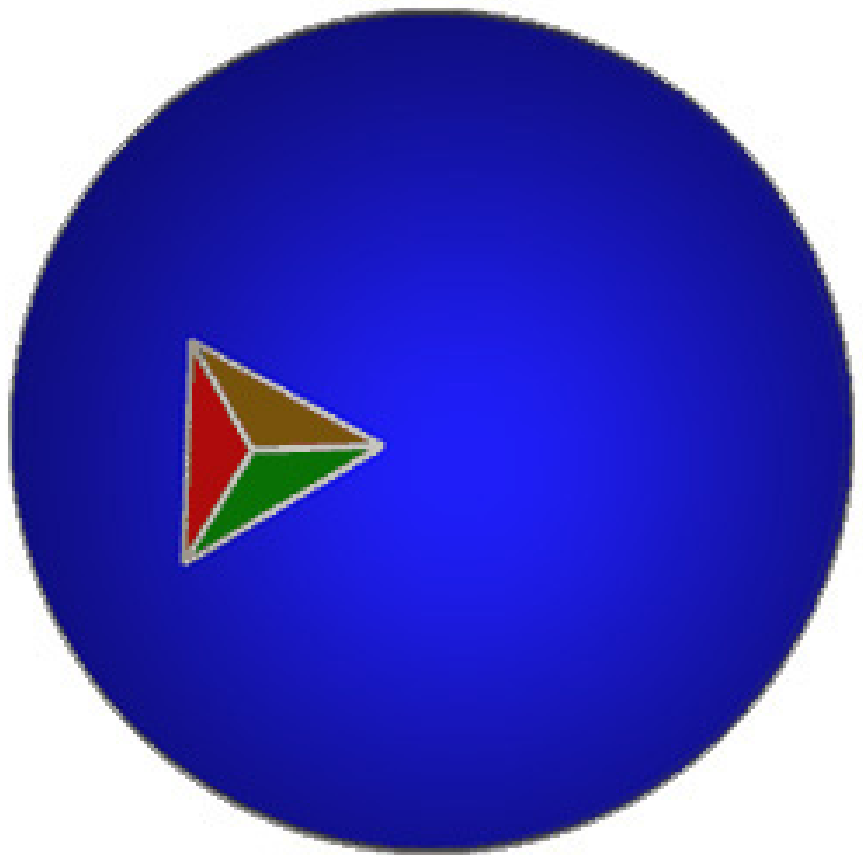,width=47mm,height=47mm}}
\caption{The map \,$\textit{\textbf{m}}(4)$\, inside the disc $\textit{\textbf{D}}$.}
\label{FCT_fig10}}
\end{center}
Next, we cut $\textit{\textbf{D}}$ along the positive real axis. This cut has an upper and a lower edge, as shown in \fig{FCT_fig11}.
\begin{center}
\FIGURE[h]{
\centerline{\epsfig{file=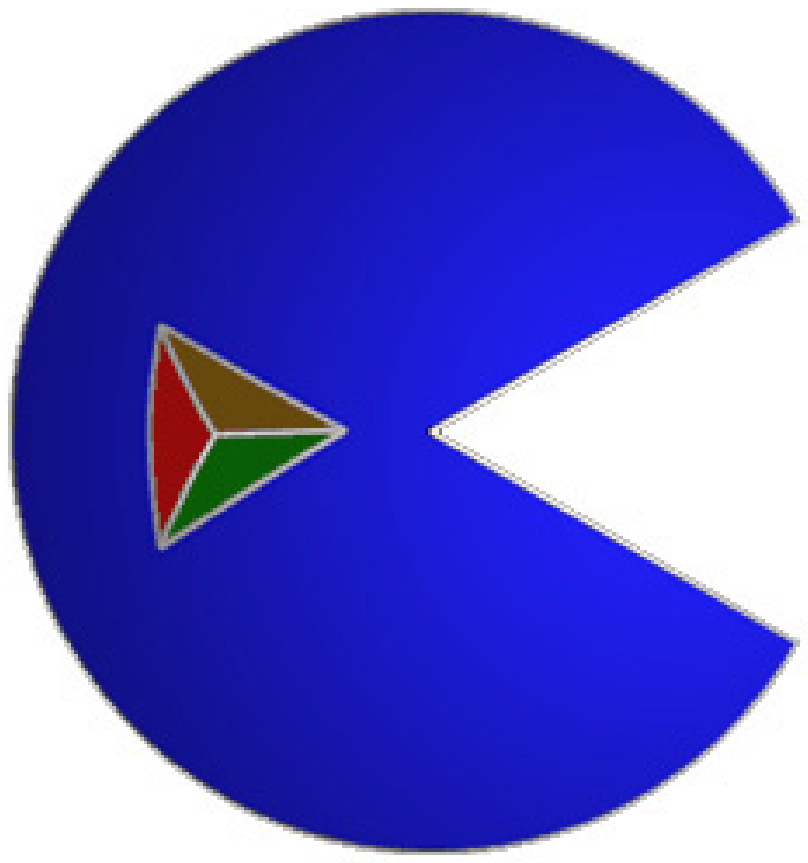,width=46mm,height=46mm}}
\caption{The map \,$\textit{\textbf{m}}(4)$\, inside the disc $\textit{\textbf{D}}$ with the cut.}
\label{FCT_fig11}}
\end{center}
For \,$N = 4$\, the construction of the $\textit{t}$-Riemann surface is the same as in Section $2.1.3$. Consider the composition of the functions as in \eq{Theor15}. Then the composite is given by the assignment as in \eq{Theor16}. Take twenty-four identical copies of the map $\textit{\textbf{m}}(4)$ on the disc with the cut, labeled \,$k = 1, ..., 24$ as shown in \fig{FCT_fig12}. For \,$k = 1, ..., 23$\, attach the lower edge of the cut of the disc $k$ with the upper edge of the cut of the disc \,$k+ 1$.\, To complete the cycle, attach the lower edge of the cut of the disc $24$ with the upper edge of the cut of the disc $1$. Recall that this forms the $w$-Riemann surface, and the branch point $w = 0$ connects all discs.
\begin{center}
\FIGURE[h]{
\centerline{\epsfig{file=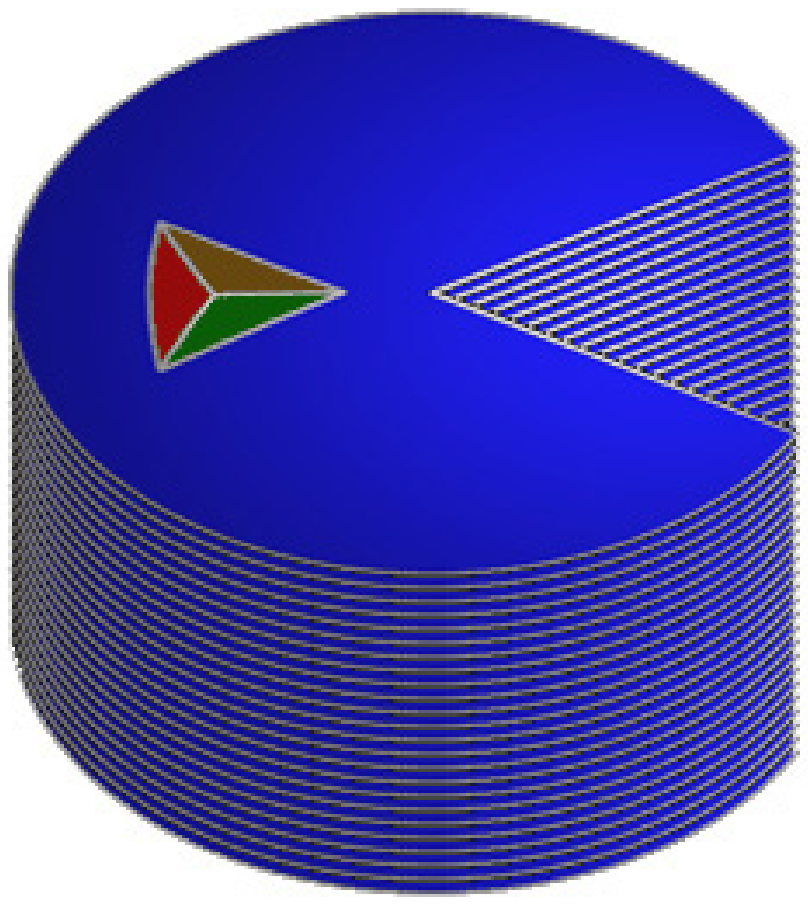,width=46mm,height=46mm}}
\caption{Twenty-four identical copies of the map \,$\textit{\textbf{m}}(4)$.}
\label{FCT_fig12}}
\end{center}
There are twenty-four superposed copies of the map $\textit{\textbf{m}}(4)$ on the $w$-Riemann surface corresponding to the twenty-four sectors in \eq{Theor17} on the $z$-plane. These are divided into two sets. The first set consists of twelve superposed copies of the map $\textit{\textbf{m}}(4)$ corresponding to the sectors in \eq{Theor18} of the upper half of the $z$-plane which comprise the upper sheet of the $t$-Riemann surface. The second set consists of twelve superposed copies of the map $\textit{\textbf{m}}(4)$ corresponding to the sectors in \eq{Theor19} of the lower half of the $z$-plane which
comprise the lower sheet of the $t$-Riemann surface. The $t$-Riemann surface is orientable since every orientation of a disc is carried over to the disc next to it (see \fig{FCT_fig13}).

\begin{center}
\FIGURE[h]{
\centerline{\epsfig{file=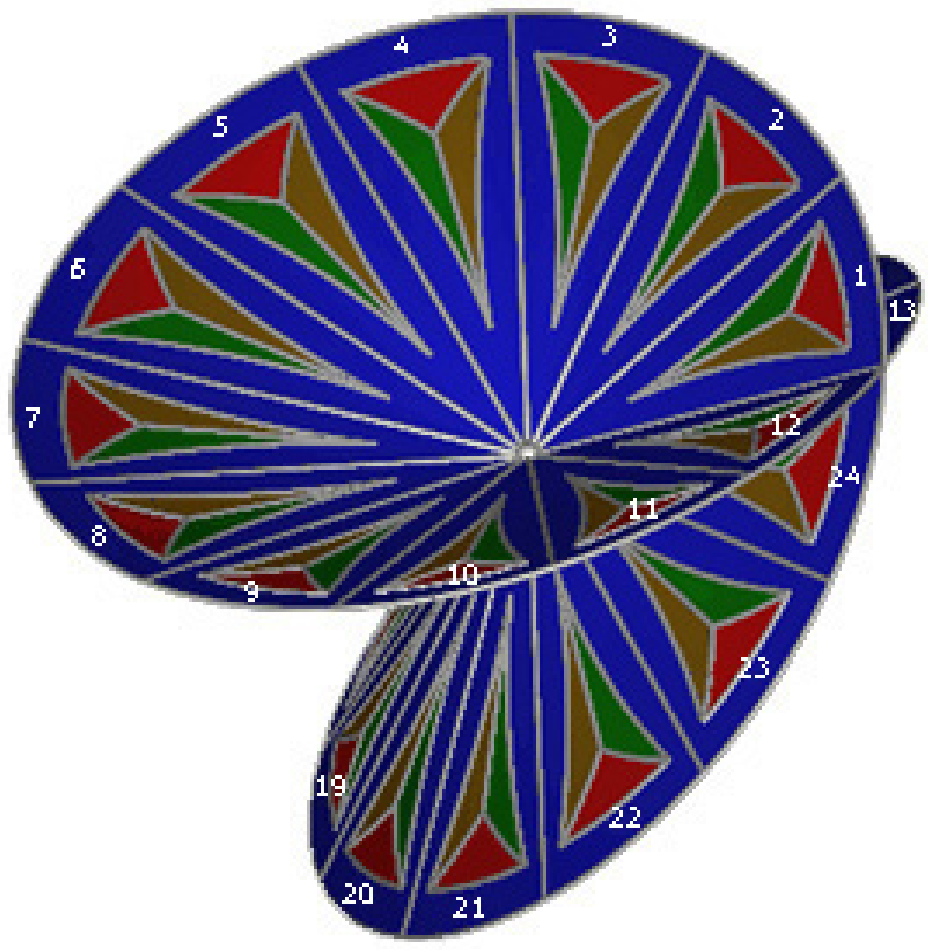,width=65mm,height=65mm}}
\caption{The $\textit{t}$-Riemann surface.}
\label{FCT_fig13}}
\end{center}

Recall from Section 2.1 that corresponding to the trivial representation of $S_{3}$, we regard $\textit{\textbf{Z}}_{4}$
as its Eilenberg module and form the split extension \,$\textit{\textbf{Z}}_{4}]S_{3}$\, that is abstractly isomorphic to
the direct product \,$\textit{\textbf{Z}}_{4}\!\times\!S_{3}$.\, Next, we form the integral group algebras
\,$\textit{\textbf{Z}}(\textit{\textbf{Z}}_{4}]S_{3})$\, and \,$\textit{\textbf{Z}}S_{3}$.\, Again, corresponding to the
trivial representation of \,$\textit{\textbf{Z}}S_{3}$,\, we regard \,$\textit{\textbf{Z}}(\textit{\textbf{Z}}_{4}]S_{3})$\,
as its Eilenberg module and form the spilt extension
\,$\textit{\textbf{Z}}(\textit{\textbf{Z}}_{4}]S_{3})]\textit{\textbf{Z}}S_{3}$\, that is abstractly isomorphic to the
direct product \,$\textit{\textbf{Z}}(\textit{\textbf{Z}}_{4}]S_{3})\!\times\!\textit{\textbf{Z}}S_{3}$.\, Let
\,$Sym(\textit{\textbf{Z}}_{4}]S_{3})$\, denote the symmetric group of order $24!$ on
\,$|\textit{\textbf{Z}}_{4}]S_{3}| = 24$\, letters. Then $S_{3}$ is embedded in \,$Sym(\textit{\textbf{Z}}_{4}]S_{3})$\,
via the Cayley right regular representation $R$. Select a common system of representatives
\,$\left\{\varphi_{i}\!\mid i = 1,2,3,...,24!/6\right\}$\, for the left and right cosets of the embedded subgroup $S_{3}$
in the group \,$Sym(\textit{\textbf{Z}}_{4}]S_{3})$.\, Fix a common coset representative $\varphi_{i}$ of $S_{3}$ in
\,$Sym(\textit{\textbf{Z}}_{4}]S_{3})$\, and fix a pair \,$(\beta, \gamma) \in S_{3}\!\times\!S_{3}$.\, The regions of the maps on the $t$-Riemann surface are labeled by the elements of the split extension
\,$\textit{\textbf{Z}}(\textit{\textbf{Z}}_{4}]S_{3})]\textit{\textbf{Z}}S_{3}$,\, with the branch point of the surface which connects all discs, labeled \,$(0, \beta + \gamma)$\, (see LEMMA $9$). The labeling
scheme of the regions of the maps on the $t$-Riemann surface is shown in \fig{FCT_fig14}, \fig{FCT_fig15}, \fig{FCT_fig16} and \fig{FCT_fig17}.

\begin{center}
\FIGURE[h]{
\centerline{\epsfig{file=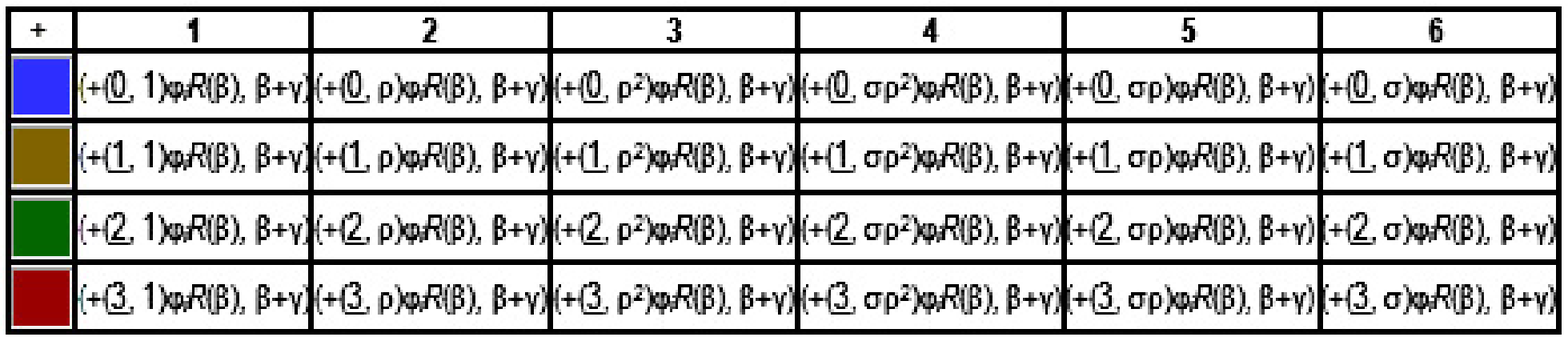,width=160mm,height=35mm}}
\caption{The labeling scheme of the regions of the maps on the upper half of the upper sheet of the $t$-Riemann surface
(all labels have a positive sign).}
\label{FCT_fig14}}
\end{center}

\begin{center}
\FIGURE[h]{
\centerline{\epsfig{file=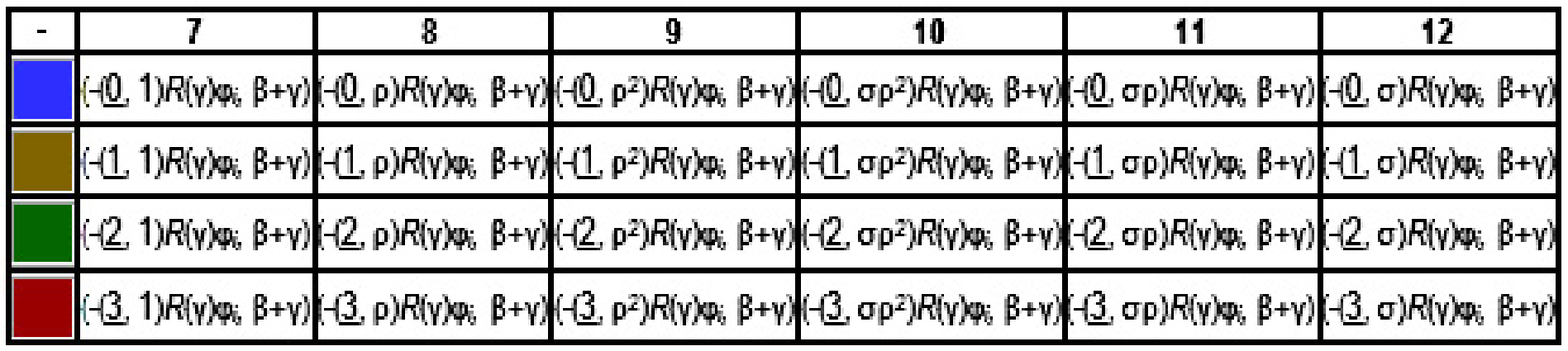,width=160mm,height=36mm}}
\caption{The labeling scheme of the regions of the maps on the lower half of the upper sheet of the $t$-Riemann surface
(all labels have a negative sign).}
\label{FCT_fig15}}
\end{center}

\begin{center}
\FIGURE[h]{
\centerline{\epsfig{file=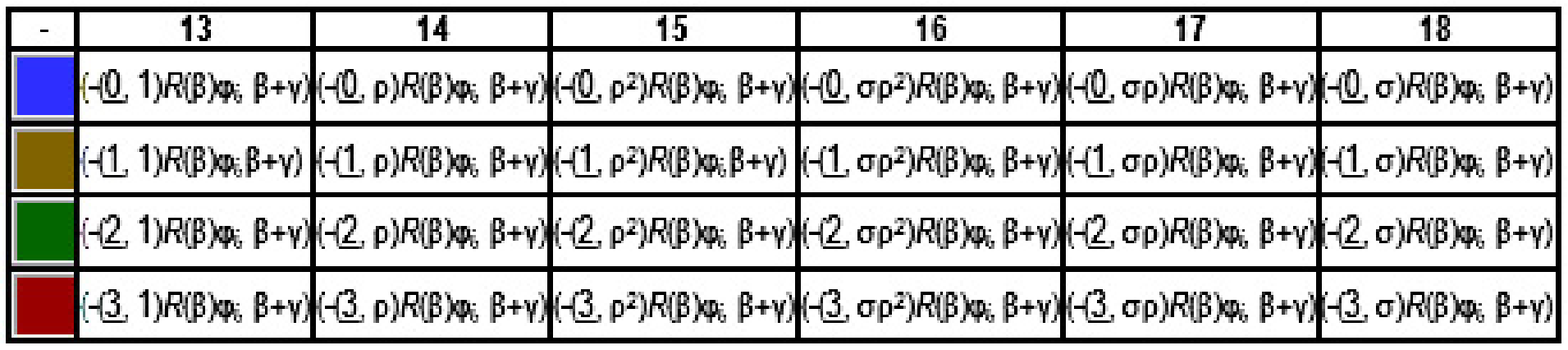,width=160mm,height=36mm}}
\caption{The labeling scheme of the regions of the maps on the upper half of the lower sheet of the $t$-Riemann surface
(all labels have a negative sign).}
\label{FCT_fig16}}
\end{center}

\begin{center}
\FIGURE[h]{
\centerline{\epsfig{file=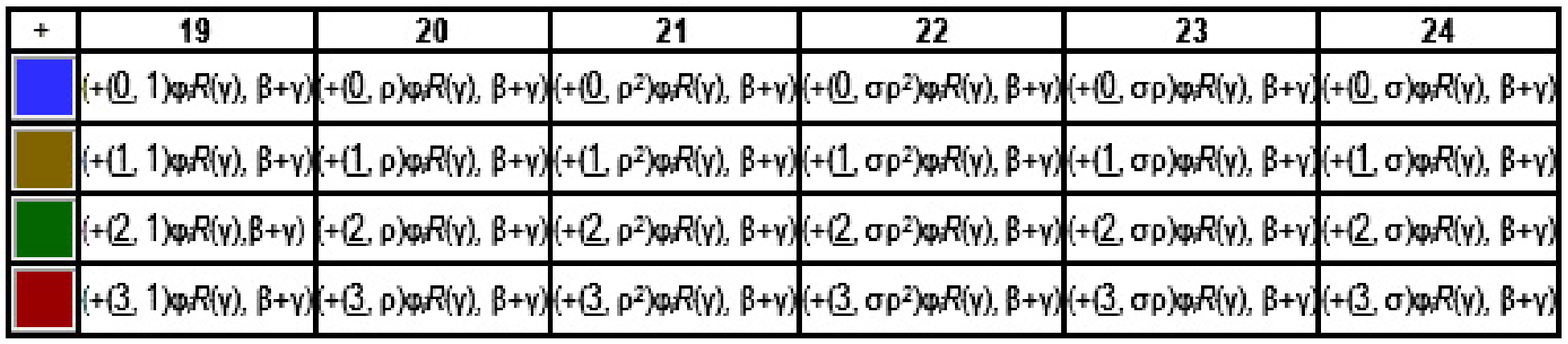,width=160mm,height=35mm}}
\caption{The labeling scheme of the regions of the maps on the lower half of the lower sheet of the $t$-Riemann surface
(all labels have a positive sign).}
\label{FCT_fig17}}
\end{center}

The main construction in the proof of the four color theorem (Section $2.1.4$) now defines the Steiner system \,$S(5, 8, 24)$.\, The $24$ points of this Steiner system are the elements of the underlying set of \,$\textit{\textbf{Z}}_{4}]S_{3}$\, (see Table $1$), and each block of
this Steiner system consists of $8$ points such that any set of $5$ points is contained in a unique block.

\begin{center}
\begin{normalsize}
\TABLE[h]{
\begin{tabular}{l l l l l l l l}
%\hline
\
\, & \hskip 0.1truecm (\underline{0},\,$1$) & \hskip 0.1truecm (\underline{0},\,$\rho$)
& \hskip 0.1truecm (\underline{0},\,$\rho^{2}$) & \hskip 0.1truecm (\underline{0},\,$\sigma\rho^{2}$)
& \hskip 0.1truecm (\underline{0},\,$\sigma\rho$) & \hskip 0.1truecm (\underline{0},\,$\sigma$)  \,  \\
%\hline %\hline\
\, & \hskip 0.1truecm (\underline{1},\,$1$) & \hskip 0.1truecm (\underline{1},\,$\rho$)
& \hskip 0.1truecm (\underline{1},\,$\rho^{2}$) & \hskip 0.1truecm (\underline{1},\,$\sigma\rho^{2}$)
& \hskip 0.1truecm (\underline{1},\,$\sigma\rho$) & \hskip 0.1truecm (\underline{1},\,$\sigma$)   \,  \\
%\hline %\hline\
\, & \hskip 0.1truecm (\underline{2},\,$1$) & \hskip 0.1truecm (\underline{2},\,$\rho$)
& \hskip 0.1truecm (\underline{2},\,$\rho^{2}$) & \hskip 0.1truecm (\underline{2},\,$\sigma\rho^{2}$)
& \hskip 0.1truecm (\underline{2},\,$\sigma\rho$) & \hskip 0.1truecm (\underline{2},\,$\sigma$)  \,   \\
%\hline %\hline\
\, & \hskip 0.1truecm (\underline{3},\,$1$) &  \hskip 0.1truecm (\underline{3},\,$\rho$)
& \hskip 0.1truecm (\underline{3},\,$\rho^{2}$) & \hskip 0.1truecm (\underline{3},\,$\sigma\rho^{2}$)
& \hskip 0.1truecm (\underline{3},\,$\sigma\rho$) & \hskip 0.1truecm (\underline{3},\,$\sigma$) & \, \\
%\hline
\end{tabular}
\caption{The $24$ points of the Steiner system \,$S(5,\,8,\,24)$.}}
\end{normalsize}
\end{center}

In the APPENDIX we show how to explicitly calculate all the blocks of the Steiner system \,$S(5, 8, 24)$\, according to the lemmas of the proof of the four color theorem.

\section{The Particle Frame}
\label{sec:Frame}
We specify a general mathematical framework from which all particles of the SM will be defined, together with their basic
physical properties: \emph{spin}, \emph{charge} and \emph{mass}. We call the labeled \textit{t}-Riemann surface, constructed
in Section $2.2$, a \emph{particle frame}. Each kind of particle in the SM will be defined by selecting a particular disc
or the intersection of a particular set of discs from the particle frame. At a time, there can be only one particle on the
particle frame, and only the selected discs will be active. The selected discs are the Schr\"{o}dinger discs that
determine the quantum-mechanical behavior of the particle at a space-time point \,$(\texttt{x}, \texttt{y}, \texttt{z},
\texttt{t})$.\, In mathematical terminology, particle frames associated with space-time points constitute a \emph{vector bundle},
and a \emph{section} of this vector bundle is specified as a \emph{gauge}. Thus, physical symmetries associated with sets of particles
defined on the particle frame correspond to \emph{gauge transformations}. The algebraic labeling and topological structure of the
\textit{t}-Riemann surface according to the proof of the four color theorem provide us with a set of precise rules for
the definition of the particles, antiparticles and force fields that make up the SM. For this purpose, it is convenient to draw the
particle frame embedded in a flat Euclidean three-dimensional space \,$(\texttt{x}, \texttt{y}, \texttt{z})$,\, and associate
with the drawing an independent time dimension $\texttt{t}$. This makes it easy to see all parts of the particle frame and
visualize how the spin, charge, mass and other rules work. The previous figures of the discs (including the maps) in the
complex plane have also been drawn as discs embedded in the flat Euclidean three-dimensional space. Such an embedding in
\fig{FCT_fig13} has self-intersections.

A blank particle frame (without selecting any particular disc) which corresponds to a space-time point in vacuum, with
its position at various angles for visualization, is shown in \fig{GU_fig1}, \fig{GU_fig2}, \fig{GU_fig3}, \fig{GU_fig4}, \fig{GU_fig5} and \fig{GU_fig6}. Remember that the maps must always be drawn on the top side of the upper and lower sheets of the particle frame. One should always view the particle frame from the top, such as in \fig{GU_fig1}.

\DOUBLEFIGURE[h]{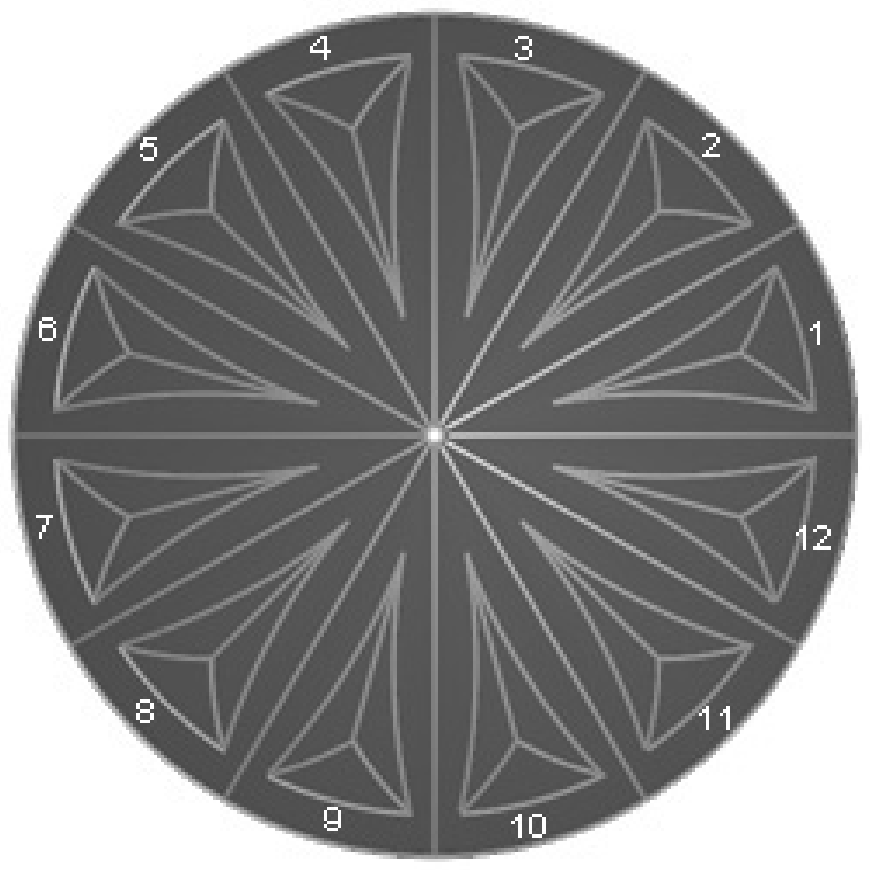,width=48mm,height=48mm}{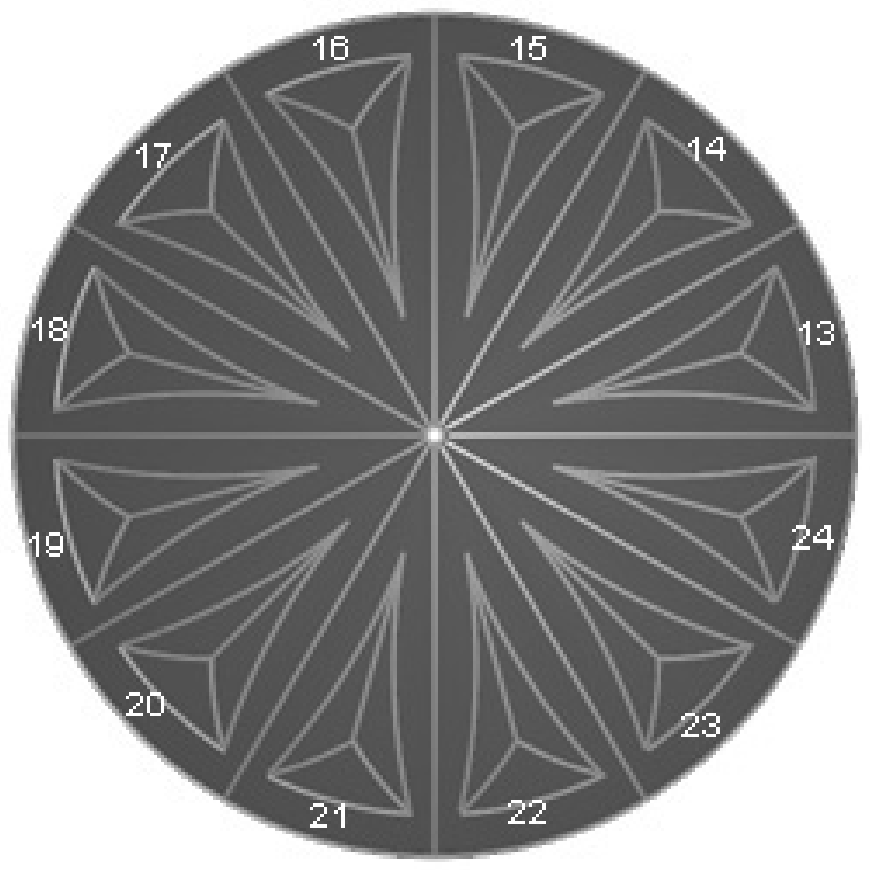,width=48mm,height=48mm}
{The view of the numbered upper sheet when its image lies on the plane perpendicular to the ray of sight.
\label{GU_fig1}}
{The view of the numbered lower sheet when its image lies on the plane perpendicular to the ray of sight (assuming that the upper sheet is invisible).
\label{GU_fig2}}

\DOUBLEFIGURE[h]{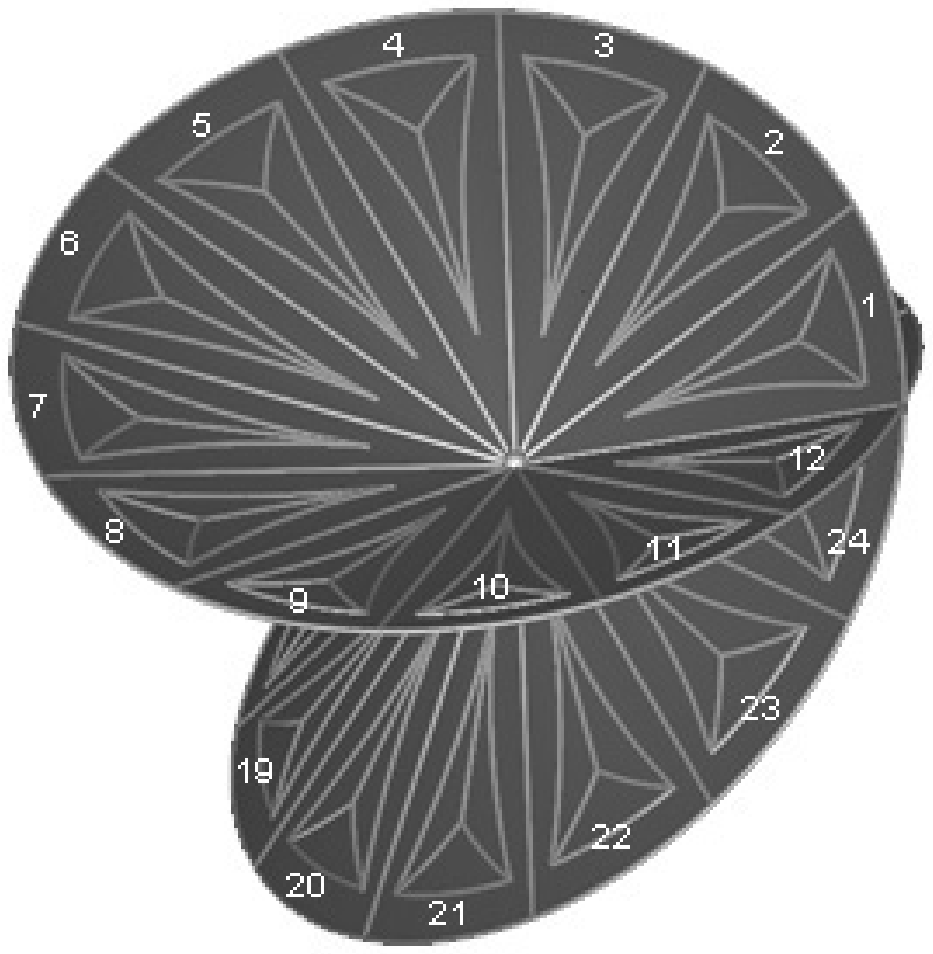,width=48mm,height=48mm}{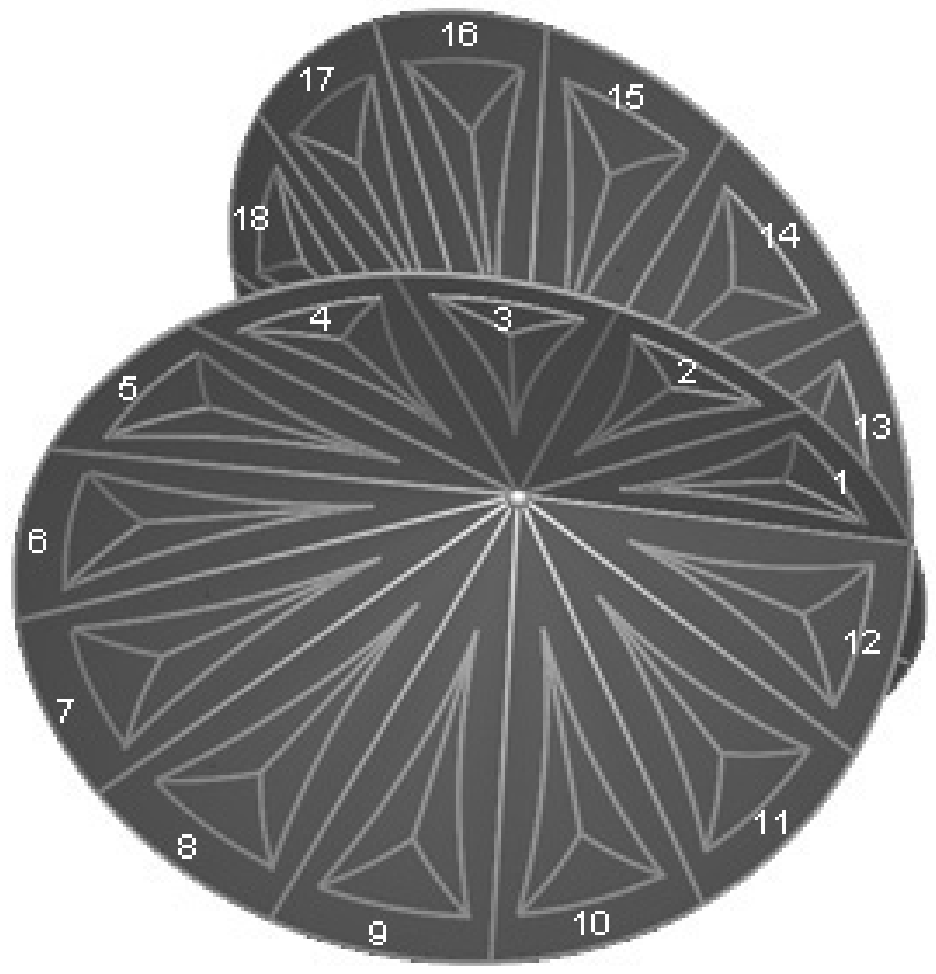,width=48mm,height=48mm}
{The numbered particle frame in \protect\fig{GU_fig1} rotated around the $x$ axis by $+\,45^{\circ}$.
\label{GU_fig3}}
{The numbered particle frame in \protect\fig{GU_fig1} rotated around the $x$ axis by $-\,45^{\circ}$.
\label{GU_fig4}}

\DOUBLEFIGURE[h]{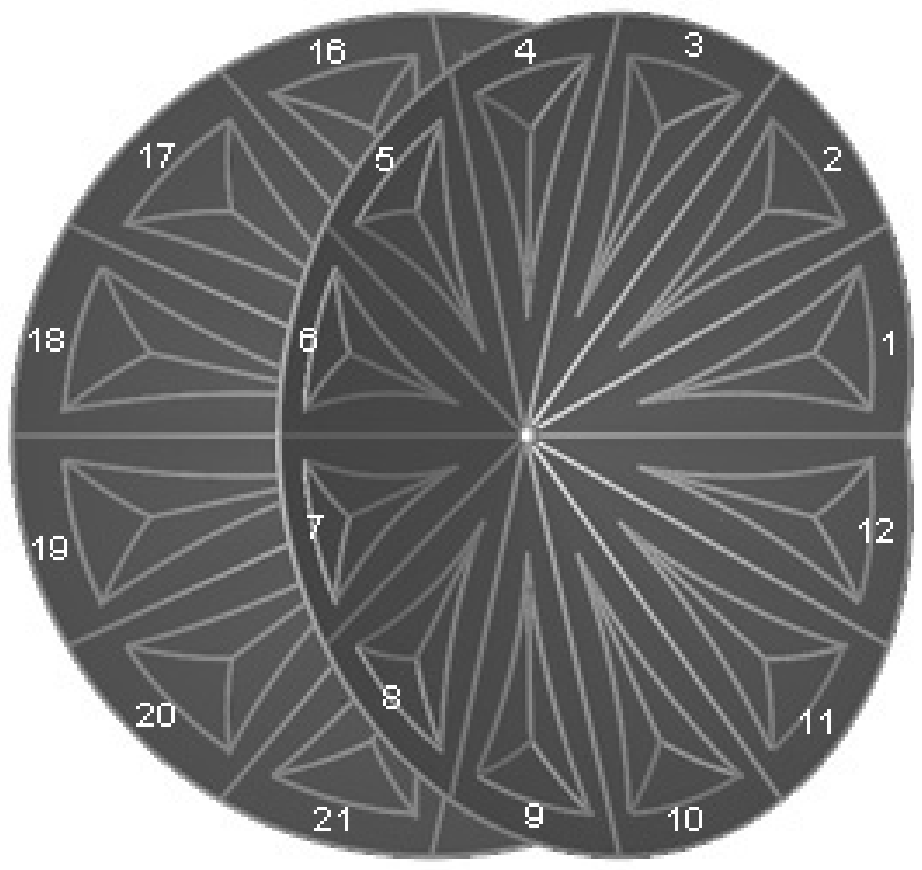,width=48mm,height=48mm}{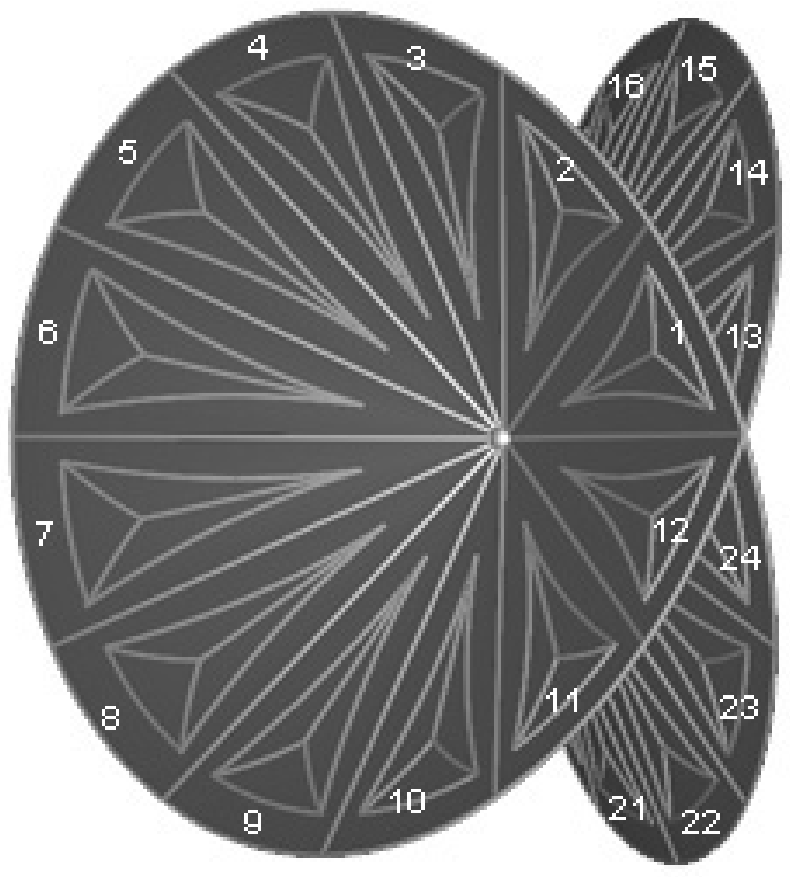,width=48mm,height=48mm}
{The numbered particle frame in \protect\fig{GU_fig1} rotated around the $y$ axis by $+\,45^{\circ}$.
\label{GU_fig5}}
{The numbered particle frame in \protect\fig{GU_fig1} rotated around the $y$ axis by $-\,45^{\circ}$.
\label{GU_fig6}}

\subsection{The Fermion, Boson and Higgs Selection Rules}
{\textbf{$\mathcal{A}$) The Fermion Selection Rule.}} Distinct particle frames with fermions defined on them cannot be
superposed at a point in space-time because of the Pauli exclusion principle. A fermion-type particle will be selected
from the particle frame as follows. First select a disc out of the $24$ discs and then select a region of the map on that
selected disc, as shown in \fig{GU_fig7}.
\begin{center}
\FIGURE[h]{
\centerline{\epsfig{file=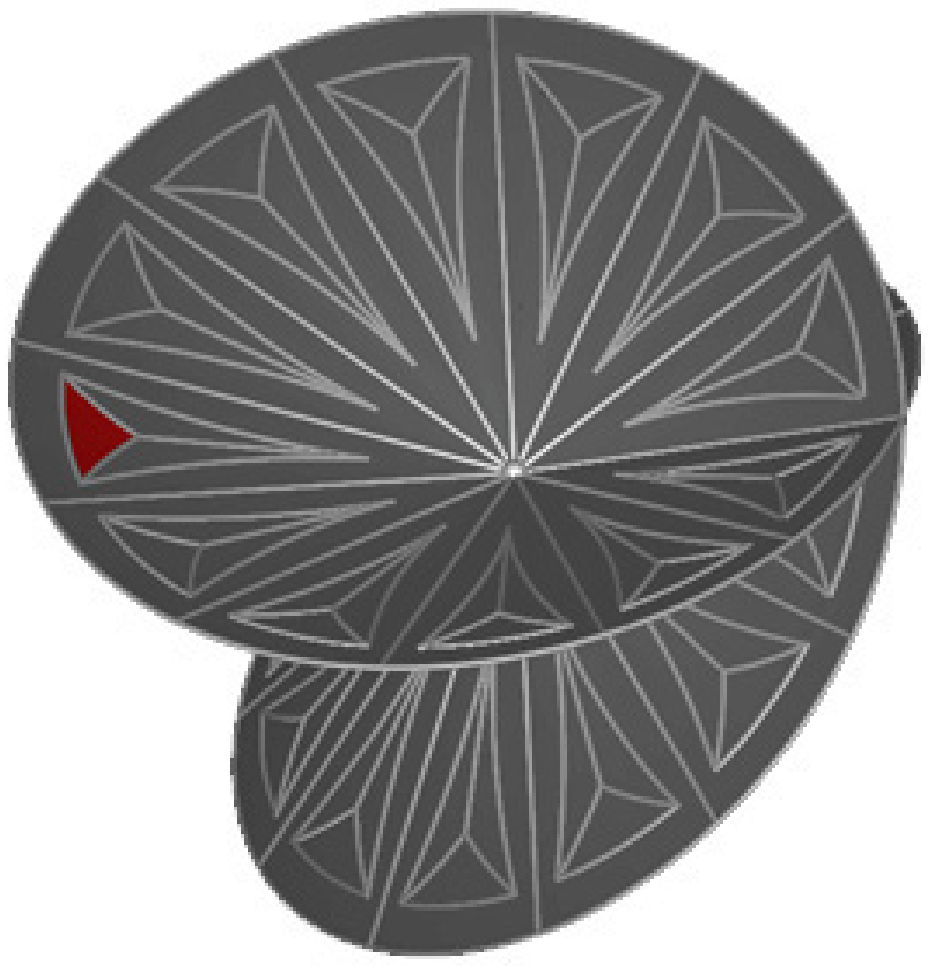,width=50mm,height=50mm}}
\caption{A fermion-type particle.}
\label{GU_fig7}}
\end{center}
Referring to the labeling in \fig{FCT_fig14}, \fig{FCT_fig15}, \fig{FCT_fig16} and \fig{FCT_fig17}, there are two types of fermions and each type comes in three generations. The type $1$ fermions (leptons) consist of a disc corresponding to the
label \,$1$ (generation I), \,$\rho$ (generation II) \,and \,$\rho^{2}$ (generation III) of the \textit{t}-Riemann surface.
The type $2$ fermions (quarks) consist of a disc corresponding to the label \,$\sigma$ (generation I), \,$\sigma\rho$
(generation II) and \,$\sigma\rho^{2}$ (generation III) of the \textit{t}-Riemann surface. Each generation consists of one
lepton doublet and one quark doublet, as in the SM. In Section $4$, we will see how to use this rule to define the quarks
and antiquarks of the SM (for the definition of the leptons see \cite{Dhar2}).

\vskip 1truecm
\hskip -0.725truecm
{\textbf{$\mathcal{B}$) The Boson Selection Rule.}} Many distinct particle frames with bosons defined on them can be
superposed at a point in space-time, since the Pauli exclusion principle does not apply to bosons.
\DOUBLEFIGURE[h]{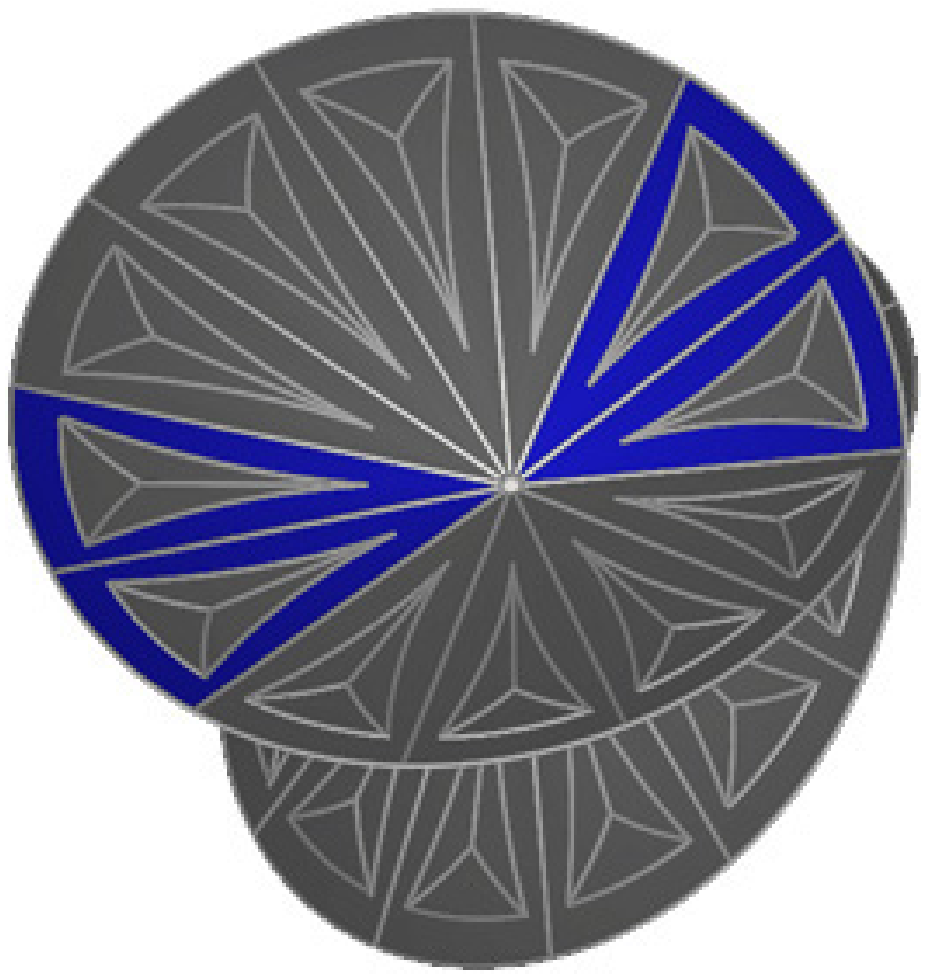,width=50mm,height=50mm}{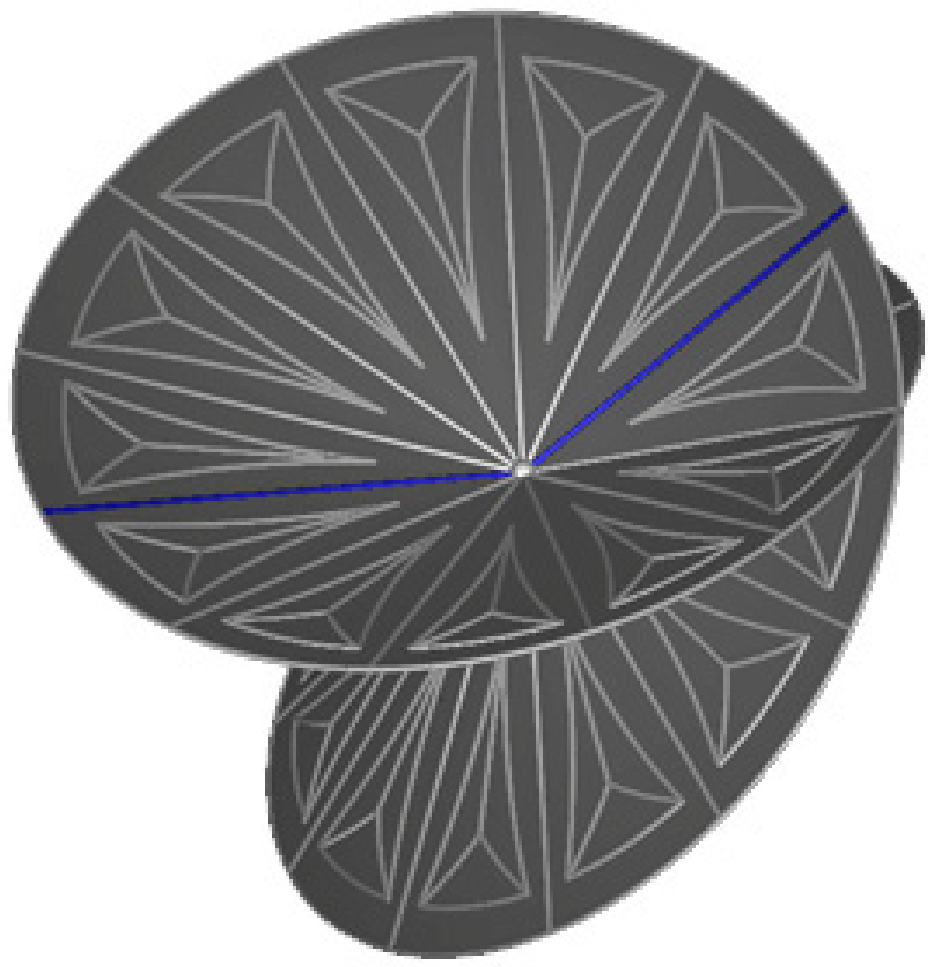,width=50mm,height=50mm}
{A boson-type particle as two pairs of discs.
\label{GU_fig8}}
{A boson-type particle as a pair of rays.
\label{GU_fig9}}
A boson-type particle will be selected from the particle frame as follows. First select a pair of fermion-type particles from the $24$ discs (with selected regions of the same color respectively) such that the two discs have an intersecting boundary (a ray on the particle frame). Then select another pair of fermion-type particles with selected regions of the same color as before, however, in such a way that the corresponding ray on the particle frame is distinct, as shown in \fig{GU_fig8}. Thus, we may select the boson-type particle by choosing a pair of rays on the particle frame with a particular color, as shown in \fig{GU_fig9}. In Section $4$, we will see how to use this rule to define the gauge vector bosons of the SM. In particular, two pairs of fermion-type particles that define a boson are interpreted as \emph{creation} and \emph{annihilation} operators during interactions in which the boson is exchanged \cite{Dhar2}.

\FIGURE[h]{
\begin{minipage}{60mm}{
\centerline{\epsfig{file=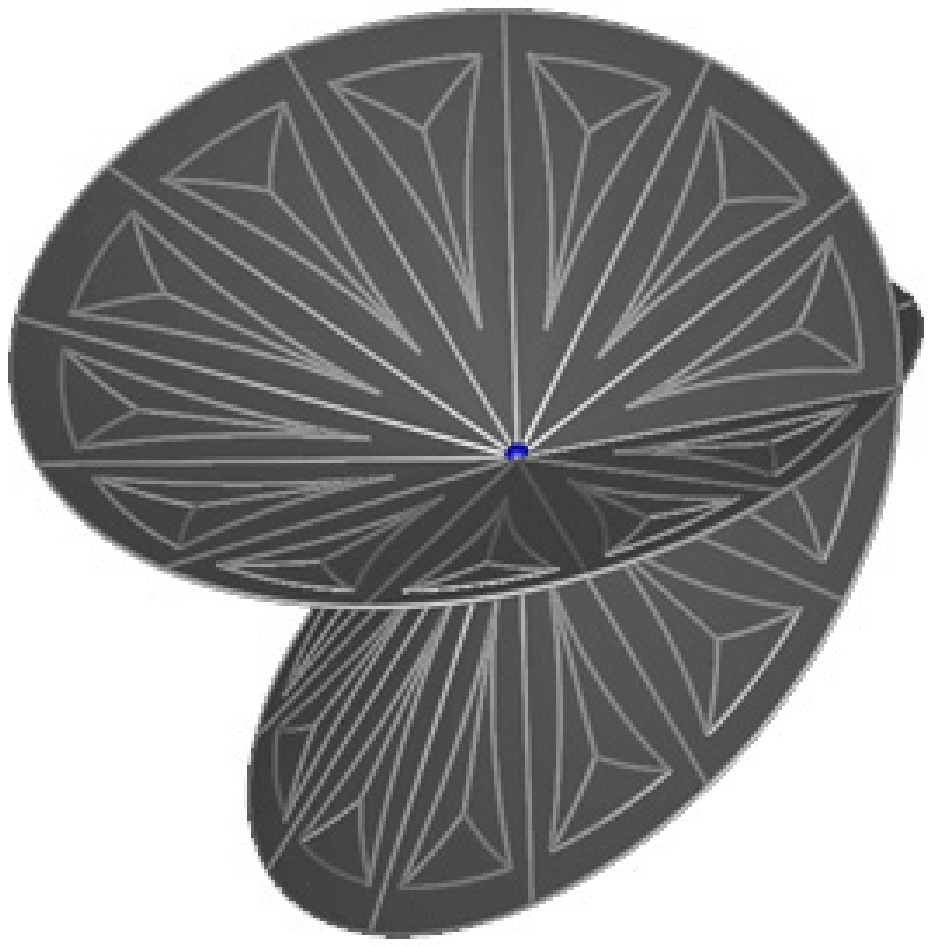,width=50mm}}
}
\end{minipage}
\caption{The Higgs-type particle.}
\label{GU_fig10}
}

\hskip -0.725truecm
{\textbf{$\mathcal{C}$) The Higgs Selection Rule.}} A Higgs-type particle is a scalar boson, i.e., it does not select a
preferred direction in space like a vector boson. It is selected as the intersection of all $24$ discs of the particle frame. This is the branch point of the \textit{t}-Riemann surface, and this selection of the Higgs-type particle is unique (\fig{GU_fig10}). The origins of the upper and lower sheets of the \textit{t}-Riemann surface (the centers in \fig{GU_fig1} and \fig{GU_fig2}) are interpreted as forming a Cooper pair \cite{Coop3, Coop2, Coop1} and the Higgs particle undergoes Bose condensation, plunging into the lowest energy state possible. This idea will be fortified in Section $5$, where we calculate the mass of the Higgs boson and discuss the mass creating mechanism using the Mass Rule $\mathcal{H}$ defined below. \\

\vskip0.5truecm
\subsection{The Spin, Electric Charge and Weak Isospin Rules}
\vskip0.5truecm
{\textbf{$\mathcal{D}$) The Spin Rule.}} The particle frame consists of four half-surfaces:
\begin{quote}
$\bullet$ The upper half of the upper sheet. \\
$\bullet$ The lower half of the upper sheet. \\
$\bullet$ The upper half of the lower sheet. \\
$\bullet$ The lower half of the lower sheet.
\end{quote}
Given a particle as a selection $S$ of the intersection of a set of discs or as a pair of rays, count the number $n$ of
half-surfaces of the particle frame that intersect with a whole segment of $S$. Define \,$s = n/2$\, to be the
\emph{spin} of the particle. With this definition, the fermion-type particle shown in \fig{GU_fig7} has \,$n = 1$\, and spin
\,$s = 1/2$;\, the boson-type particle shown in \fig{GU_fig9} has \,$n = 2$\, and spin \,$s = 2/2 = 1$;\, the Higgs-type
particle shown in \fig{GU_fig10} has \,$n = 0$\, and spin \,$s = 0/2 = 0$.\, In Section $4$, this rule is used to
calculate the spin of each quark and vector boson of the SM.
\vskip0.5truecm
\hskip -0.725truecm
{\textbf{$\mathcal{E}$) The Electric Charge Rule.}} We first associate each color with a unique absolute value of the electric charge according to the scheme shown in \fig{GU_fig11}.
\begin{center}
\FIGURE[h]{
\centerline{\epsfig{file=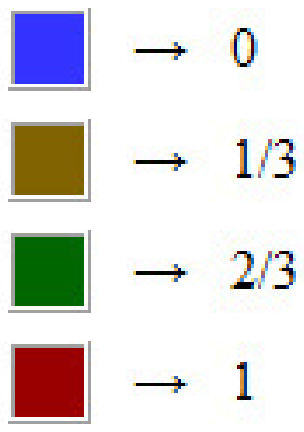,width=20mm,height=30mm}}
\caption{The scheme for the assignment of electric charge.}
\label{GU_fig11}}
\end{center}
The particle frame is labeled according to the labeling scheme of the \textit{t}-Riemann surface shown in \fig{FCT_fig14},
\fig{FCT_fig15}, \fig{FCT_fig16} and \fig{FCT_fig17}. The signs of the labels according to these figures are as follows:
\begin{quote}
$\bullet$ The upper half of the upper sheet has a \,$+$\, sign. \\
$\bullet$ The lower half of the upper sheet has a \,$-$\, sign. \\
$\bullet$ The upper half of the lower sheet has a \,$-$\, sign. \\
$\bullet$ The lower half of the lower sheet has a \,$+$\, sign.
\end{quote}
Given a particle as a selection $S$ of the intersection of a set of discs or of a pair of rays, assign a signed electric
charge to the particle according to this scheme. This is defined to be the electric charge of the particle. With this
definition, the fermion-type particle shown in \fig{GU_fig7} has the electric charge \,$-1$;\, the boson-type particle
shown in \fig{GU_fig9} has the electric charge $0$; the Higgs-type particle shown in \fig{GU_fig10} has the electric charge
$0$. In Section $4$, this rule is used to calculate explicitly the electric charge of each quark and vector boson of the SM.

\hskip -0.725truecm
{\textbf{$\mathcal{F}$) The Weak Isospin Rule.}}
Again, we use the signs of the labels of the particle frame as described in $\mathcal{E}$ above. Given a particle as a selection $S$ of the intersection of a set of discs or as a pair of rays, count $m_{1}$ as the number of half-surfaces (with sign $+$) that intersect with a whole
segment of $S$, and $m_{2}$ as the number of half-surfaces (with sign $-$) that intersect with a whole
segment of $S$. Define
\begin{displaymath}
T_{z}\,=\,\FFr{(+\,m_{1})}{2}\,+\,\FFr{(-\,m_{2})}{2}
\end{displaymath}
to be the \emph{weak isospin} of the particle. With this definition, the fermion-type particle shown in \fig{GU_fig7} has
\,$T_{z} = -1/2$;\, the boson-type particle shown in \fig{GU_fig9} has \,$T_{z} = 0$;\, the Higgs type particle shown in
\fig{GU_fig10} has \,$T_{z} = 0$.\, In Section $4$, this rule is used to calculate the weak isospin of the quarks and vector bosons in the SM.

We observed at the beginning of Section $2.2$ that helicity is not a relativistic invariant for the fermions. However, one may still speak of left-handed and right-handed fermions in terms of \emph{chirality} in a particular frame of reference. In the SM, left-handed quarks and leptons have \,$T = 1/2$,\, and are grouped into weak-isospin doublets with \,$T_{z} = \pm\,1/2$\, that behave the same under the weak interaction. For example, the up-type quarks \,$(u,\,c,\,t)$\,
have \,$T_{z} = +\,1/2$,\, and always transform into down-type quarks \,$(d,\,s,\,b)$\, that have \,$T_{z} = -\,1/2$\, (and vice versa). Right-handed fermions have \,$T = 0$\, and form singlets that do not undergo the weak interaction. It has been observed that only the left-handed fermions interact with the weak interaction. In most circumstances, two left-handed fermions interact more strongly than right-handed fermions. Experiments sensitive to this effect imply that the universe has a preference for left-handedness. In the SM matched onto the particle frame, the left-handed quarks and leptons have the same values of the weak isospin as
above. On the particle frame, the right-handed fermions (according to the definition of helicity) compose doublets because of the symmetry of the number of regions on the upper and lower sheets, and have values of  isospin opposite to the left-handed case. To account for this difference between helicity and chirality, we can define $m_1$ and $m_2$ to be zero for the right-handed fermions (according to the definition of chirality) on the particle frame. Then all fermions in the SM have the correct value of weak isospin when it is matched with the particle frame.

For the vector bosons in the SM, it is known that the nonbreakable local gauge symmetry \,$SU(2)\!\times\!U(1)$\,
requires the existence of four massless vector bosons, $W^{+}$, $W^{-}$, $W^{0}$ and $B^{0}$: the $W$-bosons being the gauge fields
of the weak isospin group $SU(2)$, and the $B^{0}$-boson being the gauge field of the weak hypercharge group $U(1)$.
Nonetheless, the gauge \,$SU(2)\!\times\!U(1)$-symmetry is spontaneously broken, in consequence of which the weak gauge bosons
acquire masses. From the well-known expression \,$N = 2\,T + 1$,\, the bosons $W$ have \,$T = 1$\, with three different
values of $T_{z}$. They are emitted in following transitions:
\begin{itemize}
\item[a)] $W^{+}$ boson $(T_{z} = +1)$\, is emitted in transitions \,$(T_{z} = +1/2) \rightarrow (T_{z} = -1/2)$;
\item[b)] $W^{-}$ boson $(T_{z} = -1)$\, is emitted in transitions \,$(T_{z} = -1/2) \rightarrow (T_{z} = +1/2)$;
\item[c)] Theoretically, the boson $W^{0}$ $(T_{z} = 0)$ would be emitted in such reactions where $T_{z}$ could not change.
However, under electroweak unification, the boson $W^{0}$ mixes with the weak hypercharge gauge boson $B^{0}$, resulting
in the observed massive boson $Z^{0}$ and massless photon by the following mutually-orthogonal linear superposition
\begin{displaymath}
A\,=\,B^{0}\cos{\!\Theta_{W}}\,+W^{0}\sin{\!\Theta_{W}}\,,
\end{displaymath}
\begin{displaymath}
Z\,=\,- B^{0}\sin{\!\Theta_{W}}\,+W^{0}\cos{\!\Theta_{W}}
\end{displaymath}
of the fields $W^{0}$ and $B^{0}$ of the nonbreakable group structure of $SU(2)\!\times\!U(1)$. Here $A$ and $Z$ are the
fields of the photon and $Z^{0}$ boson, respectively, and $\Theta_{W}-$the Weinberg angle (see Section $4$).
\end{itemize}
The weak interactions involving the bosons $W^{\pm}$ occur exclusively on left-handed fermions and right-handed antiparticles
of the fermions. The boson $Z^{0}$ interacts with both left-handed fermions and their antiparticles. For the SM matched onto the particle frame, the left-handed and right-handed vector bosons have the same values of the weak isospin as above:
\begin{center}
\,$W^{+}$ $(T_{z} = +1)$,\, \,$W^{-}$ $(T_{z} = -1)$,\, \,$Z^{0}$ $(T_{z} = 0)$.
\end{center}

\hskip -0.725truecm
{\textbf{$\mathcal{G}$) The Strong Charge Rule.}}
Here we follow 't Hooft's \cite{Ho} description of the SM, interpreted in our context \footnote{Note that to avoid obvious
confusion with our use of the word \emph{color}, we use the word ``strong" charge to denote what is usually called the
``color" charge in QCD. This terminology also seems to be more appropriate for a grand unified theory since the types of
the charges are now denoted by the names of the associated force fields: electromagnetic, weak, strong and gravitational.}.
When we speak of charge without any further specification, we always mean the electric charge. In addition to the electric
charge, a particle must also have electromagnetic, weak, strong and gravitational charges that are used to describe the
corresponding field theories.

The electromagnetic and weak charges are neutral as far as the strong force is concerned, hence they are regarded as neutral
strong charges \footnote{Note that a neutral strong charge corresponds to a ``colorless'' combination of ``color'' charges
in QCD.}. Then, according to QCD, $N_{c}$ is defined to be the number of unsigned strong charges for a particular type of
force. The electromagnetic, weak, strong and gravitational charges are also given a sign (charge or anticharge), exactly
as for the electric charge in the Electric Charge Rule $\mathcal{E}$. Since the gravitational charge will not play a role
in the context of our paper, we refer the reader to \cite{Dhar2} for a detailed description of the Gravitational Charge
Rule.

\begin{itemize}
\item[a)] There is only one neutral strong charge possible for each lepton-type fermion. Thus, each lepton-type fermion has
exactly one neutral unsigned strong charge, so \,$N_{c} = 1$.

\item[b)] Each quark-type fermion may have one of three kinds of unsigned strong charges \,$\sigma$,\, \,$\sigma\rho$\,
and \,$\sigma\rho^{2}$,\, so \,$N_{c} = 3$\, (the same picture as in QCD).\, After assigning a sign according to this notation,
the strong charge for a quark-type fermion is one of \,$\pm\sigma$,\, \,$\pm\sigma\rho$\, and \,$\pm\sigma\rho^{2}$, exactly as in QCD
\footnote{Thus the ``strong" charge-anticharge pairs \,$\pm\sigma$,\, \,$\pm\sigma\rho$\, and
\,$\pm\sigma\rho^{2}$\, correspond precisely to the ``color" charge-anticharge pairs in QCD: red/anti-red, green/anti-green
and blue/anti-blue, respectively.}.

\item[c)] The photon carries a neutral strong charge-anticharge pair with \,$+$\, and \,$-$\, signs, respectively. Thus, there is only one unsigned strong charge for the photon, and \,$N_{c}=1$. The parameter of the group $U(1)$ corresponds to the neutral strong charge $1$, and its generator corresponds to the photon.

\item[d)] Each of the three vector bosons $Z^{0}$, $W^{+}$ and $W^{-}$ carries a neutral strong charge-anticharge (with \,$+$\,
and \,$-$\, signs), charge-charge (with \,$+$\, and \,$+$\, signs) or anticharge-anticharge (with \,$-$\, and \,$-$\, signs)
pair, respectively. Thus for each of these bosons there is only one unsigned strong charge, and \,$N_{c} = 1$ for each.\, The
parameters of the group $SU(2)$ correspond to the two weak charges $\rho$ and $\rho^{2}$, and its three generators correspond
to three vector bosons under consideration.
\end{itemize}

Referring to the Mass Rule $\mathcal{H}$ next and \fig{FCT_fig14}, \fig{FCT_fig15}, \fig{FCT_fig16} and \fig{FCT_fig17}, we will use one of $\beta$ or $\gamma$ to specify the unsigned strong charge of a particle on the particle frame.

\subsection{The Mass Rule}
{\textbf{$\mathcal{H}$) The Mass Rule.}} We define the assignment of the rest mass of a particle, of the SM matched onto the particle frame, in the following way.
Recall that each element $\psi$ of \,$Sym(\textit{\textbf{Z}}_{4}]S_{3})$\, is a permutation of the underlying set
\,$\textit{\textbf{Z}}_{4}]S_{3}$.\, Each permutation $\psi$ may be thought of as representing the entropy or disorder of
the set \,$\textit{\textbf{Z}}_{4}]S_{3}$\, \footnote{Recall that the Boltzmann entropy hypothesis relates the entropy $E$
of a system in a particular state to the probability $p$ of finding it in that state: \,$E = k \log{p} + c$,\, where $k$ is
the Boltzmann constant and $c$ is another constant \cite{Schro}. For our purposes, the particle frame is the system under
consideration, and the permutation of the set \,$\textit{\textbf{Z}}_{4}]S_{3}$\, that labels the regions is the state of the
system.}.

Let us suppose that each kind of particle $S$ is associated with a unique permutation $\psi_{S}$ which, in turn, is associated
with a unique value of energy. We interpret the rest mass of the particle $S$ as being created by the energy of the uniquely
associated permutation $\psi_{S}$, as follows. Write \,$\psi_{S} = R(\delta)\varphi_{j}$\, as the unique expression in terms
of a common coset representative $\varphi_{j}$, and \,$\delta \in S_{3}$.\, By definition
\,$\psi_S^\mu = \varphi_{j} R(\delta)$,\, as described in LEMMA $15$. Then $\psi_{S}$ and $\psi_{S}^\mu$ act on any given
region \,$(\underline{m}, \alpha)$\, by means of the $\uparrow$ and $\downarrow$ group actions, respectively. Note that by
LEMMA $17$, the two group actions $\uparrow$ and $\downarrow$ are equal, so
\,$(\underline{m}, \alpha)\psi_S = (\underline{m}, \alpha)\psi_S^\mu$\, for any selected region. The given particle is
represented on the particle frame corresponding to the $t$-Riemann surface with $\varphi_{i}$, $\beta$, $\gamma$ chosen
accordingly. This means that for any region of the type \,$(\underline{m}, \alpha)$\, in the selection $S$ that represents
the particle on the frame, we have the uniquely associated particle rest mass
\,$(\underline{m}, \alpha)\psi_S = (\underline{m}, \alpha)\psi_S^\mu$.\, By the Antiparticle Rule {\textbf{$\mathcal{J}$}}
below, a particle and antiparticle will always have the same rest mass. We have special cases for bosons:
\begin{quote}
$\bullet$ For the massless bosons $S$, we assume that the uniquely associated permutation \,$\psi_{S} = R(\delta)\varphi_{j}$\,
has $\varphi_{j}$ equal to the identity permutation. This assumption is forced if the selection of the photon or gluon (or
graviton) \cite{Dhar2} is to have required properties according to all other rules, e.g. the correct values of $N_{c}$.
\end{quote}
\begin{quote}
$\bullet$ For the massive bosons $S$, we assume that the uniquely associated permutation \,$\psi_{S} = R(\delta)\varphi_{j}$\,
has \,$\delta = 1$.\, This assumption is forced if the vector bosons are to have required properties according to all other
rules, e.g. the correct values of $N_{c}$.
\end{quote}
The rest mass of a particle is usually determined from experimental observations. However, it is shown in \cite{Dhar2} that
the rest masses of all particles on the particle frame cannot be independent and most of their mass ratios must be fixed
quite precisely due to the structure of the particle frame \footnote{This is because the experimentally
observed rest mass of a particle is only obtained after renormalization in quantum field theory (the mass of a ``bare''
particle diverges to infinity). Thus we must calculate the cut-offs corresponding to the energy scales for the
renormalization. In \cite{Dhar2} we show how the particle frame evolves along the cosmological time-line or equivalent
energy scales. This necessarily implies that the permutations representing the rest masses of the particles will have certain
fixed ratios of fixed points at different energy scales. Thus, the renormalization procedure gives rise to fixed mass ratios
of the particles on the particle frame.}.

\subsection{The other Selection Rules}
{\textbf{$\mathcal{I}$) The Equivalence Rule.}} Given two different selections $S_{1}$ and $S_{2}$ of particles: if the
resulting particles have the same spin, electric charge, weak isospin and mass according to the above rules
$\mathcal{D}$, $\mathcal{E}$, $\mathcal{F}$ and $\mathcal{H}$ then we regard $S_{1}$ and $S_{2}$ as representing the same
kind of particles in the SM on the particle frame.

\hskip -0.725truecm
{\textbf{$\mathcal{J}$) The Antiparticle Rule.}} Let the function $\pi$ denote a rotation of the $z$-plane by $\pi$ radians
(the $z$-plane is defined in Section $2.1.3$). Then $\pi$ induces a rotation of the $\textit{t}$-Riemann surface by $2\pi$
radians. Any point on the particle frame is transformed by $\pi$ into the point superposed directly above or below it by a
continuous rotation that winds exactly once around the branch point. Given a particle as a selection $S$ of the intersection
of a set of discs or as a pair of rays, the image of the particle under the function $\pi$ is called its \emph{antiparticle}.
Note that by the above rules, an antiparticle is of the same type as the original particle, with identical spin and mass but
the opposite charge. If a particle has no charge then we cannot distinguish between the particle and its antiparticle. By
the equivalence rule $\mathcal{I}$, a particle with no charge is equivalent to its antiparticle in the SM on the particle
frame. For example, as we will see later, the $Z^{0}$ vector boson is its own antiparticle.

\hskip -0.725truecm
{\textbf{$\mathcal{K}$) The Helicity Rule.}} The helicity of a particle is defined by selecting one of two possible
orientations (left-handed or right-handed) for all active Schr\"{o}dinger discs on the particle frame. This orientation
carries over to neighboring discs and defines the orientation of the \textit{t}-Riemann surface.

\hskip -0.725truecm
{\textbf{$\mathcal{L}$) CP Transformation Rule.}} Given a particle defined on the particle frame at a space-time point
\,$(\texttt{x}, \texttt{y}, \texttt{z}, \texttt{t})$\, with momentum vector $\textit{\textbf{p}}$, we define the
transformations $\textbf{C}$ and $\textbf{P}$  as follows:
\begin{quote}
$\bullet$ $\textbf{P}$ reverses the spatial coordinates to \,$(-\texttt{x}, -\texttt{y}, -\texttt{z})$.\, Note that this
means that the direction of the momentum vector is also reversed to \,$-\textit{\textbf{p}}$.\, Hence, the orientation of
the particle frame is reversed and by the helicity rule $\mathcal{K}$, the helicity of the particle is reversed. Left-handed
particles are transformed into right-handed particles of the same kind and vice versa. \\
$\bullet$ $\textbf{C}$ transforms the particle into its antiparticle. In particular, the charge of the particle is reversed.
\end{quote}
When both transformations are performed together, the particle is said to undergo $\textbf{CP}$ transformation.
For example, $\textbf{CP}$ transformation of a left-handed electron gives a right-handed positron. Since particle
interactions are mediated by bosons and their associated force fields, one can characterize the symmetry of the particle
interactions by applying $\textbf{CP}$ transformations to the gauge mediating bosons:
\begin{quote}
$\bullet$ If \,$\textbf{CP}$ transformation applied to a boson in the SM on the particle frame yields an equivalent boson
then one says that $\textbf{CP}$ symmetry is preserved in interactions involving that boson. For example, $\textbf{CP}$
symmetry is preserved by interactions involving the photon, i.e. electromagnetic interactions. The strong interaction also
seems to be invariant under the combined $\textbf{CP}$ transformation operation.  \\
$\bullet$ If $\textbf{CP}$ transformation applied to a boson in the SM on the particle frame does not yield an equivalent
boson then one says that $\textbf{CP}$ symmetry is violated in interactions involving that boson. For example, $\textbf{CP}$
symmetry is violated (verified experimentally) by interactions involving $W^{+}$ and $W^{-}$ vector bosons, i.e. weak
interactions.
\end{quote}

\hskip -0.725truecm
{\textbf{$\mathcal{M}$) The Standard Model Completion Rule.}}
If all particle frames corresponding to all particles in the universe were to be superimposed (hypothetically, of course)
then the fermions and bosons should fit together perfectly according to the above rules, forming the complete SM on the
particle frame.
\begin{center}
\FIGURE[h]{
\centerline{\epsfig{file=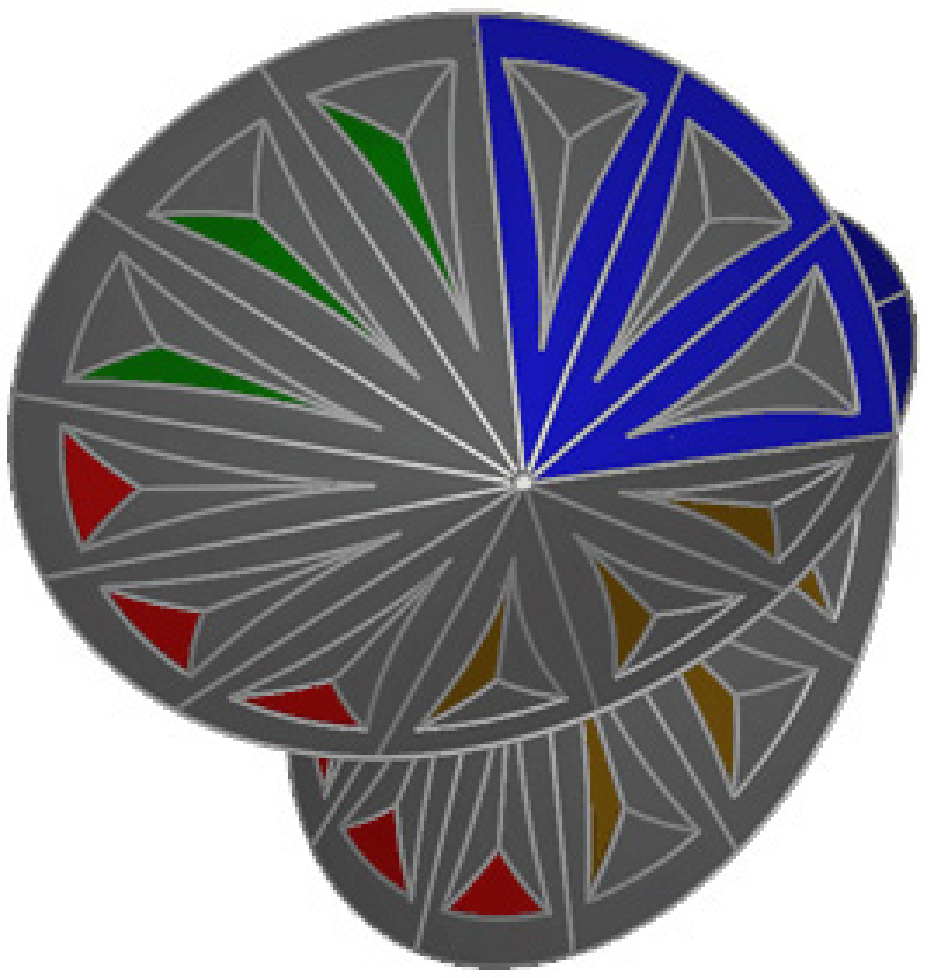,width=50mm,height=50mm}}
\caption{The perfect fitting of the fermions in the SM.}
\label{GU_fig12}}
\end{center}
Each of the $24$ discs of the particle frame represents the Schr\"{o}dinger disc of a unique fermion in the SM, respecting
all the above rules. There cannot be any other fermions in the SM of our mathematical construction (see \fig{GU_fig12}).
\begin{center}
\FIGURE[h]{
\centerline{\epsfig{file=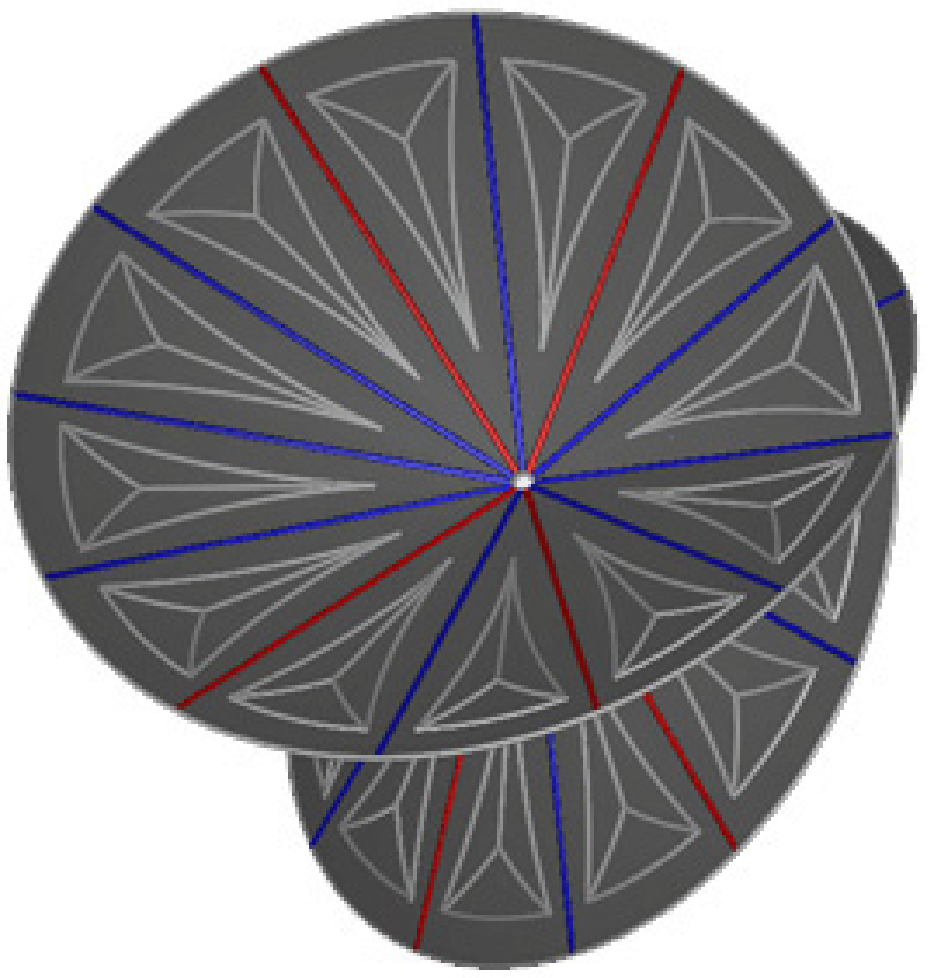,width=50mm,height=50mm}}
\caption{The perfect fitting of the bosons in the SM.}
\label{GU_fig13}}
\end{center}
Each of the $24$ pairs of rays of the particle frame represent four Schr\"{o}dinger discs of a unique boson in the SM,
respecting all the above rules. There cannot be any other bosons in the SM of our mathematical construction (see \fig{GU_fig13}), except for the unique Higgs boson.

\section{Quarks and Gauge Vector Bosons on the Particle Frame}
\label{sec:QuarkBoson}
In this section we define the quarks and gauge vector bosons on the particle frame, making use of the specified selection
rules of the previous section.
\subsection{Quarks}
\begin{enumerate}
\item Let \,$\psi_{u}$,\, \,$\psi_{d}$,\, \,$\psi_{c}$,\, \,$\psi_{s}$,\, \,$\psi_{t}$,\, \,$\psi_{b}$ be the unique permutations associated with the quark particles, according to the Mass
Rule $\mathcal{H}$.

\item
\begin{itemize}
\item[a)] Select $\varphi_{i}$ and $\beta$ according to the unique expression \,$\psi_{u} = \varphi_{i}R(\beta)$\, and
\,$\gamma = \sigma$,\, \,$\sigma\rho$\, and \,$\sigma\rho^{2}$.
\item[b)] Select $\varphi_{i}$ and $\gamma$ according to the unique expression \,$\psi_{d} = R(\gamma)\varphi_{i}$\, and
\,$\beta = \sigma$,\, \,$\sigma\rho$\, and \,$\sigma\rho^{2}$.
\item[c)] Select $\varphi_{i}$ and $\beta$ according to the unique expression \,$\psi_{c} = \varphi_{i}R(\beta)$\, and
\,$\gamma = \sigma$,\, \,$\sigma\rho$\, and \,$\sigma\rho^{2}$.
\item[d)] Select $\varphi_{i}$ and $\gamma$ according to the unique expression \,$\psi_{s} = R(\gamma)\varphi_{i}$\, and
\,$\beta = \sigma$,\, \,$\sigma\rho$\, and \,$\sigma\rho^{2}$.
\item[e)] Select $\varphi_{i}$ and $\beta$ according to the unique expression \,$\psi_{t} = \varphi_{i}R(\beta)$\, and
\,$\gamma = \sigma$,\, \,$\sigma\rho$\, and \,$\sigma\rho^{2}$.
\item[f)] Select $\varphi_{i}$ and $\gamma$ according to the unique expression \,$\psi_{b} = R(\gamma)\varphi_{i}$\, and
\,$\beta = \sigma$,\, \,$\sigma\rho$\, and \,$\sigma\rho^{2}$.
\end{itemize}

\item Then the particle frame corresponds to the \textit{t}-Riemann surface with these choices of $\varphi_{i}$, $\beta$ and
$\gamma$.

\item
\begin{itemize}
\item[a)] We select the disc $6$ on the upper sheet of the particle frame and its green region $\underline{2}$ that has the
label \,$\Lb +(\underline{2},\,\sigma)\varphi_{i}R(\beta),\,\beta + \gamma \Rb$\, according to \fig{FCT_fig13} and
\fig{FCT_fig14}. This represents the Schr\"{o}dinger disc of the $u$ quark on the particle frame, according to \fig{FCT_fig7}
and \fig{GU_fig14}.

\item[b)] We select the disc $12$ on the upper sheet of the particle frame and its yellow region $\underline{1}$ that has the
label \,$\Lb -(\underline{1},\,\sigma)R(\gamma)\varphi_{i},\,\beta + \gamma \Rb$\, according to \fig{FCT_fig13} and
\fig{FCT_fig15}. This represents the Schr\"{o}dinger disc of the $d$ quark on the particle frame, according to \fig{FCT_fig7}
and \fig{GU_fig16}.

\item[c)] We select the disc $5$ on the upper sheet of the particle frame and its green region $\underline{2}$ that has the
label \,$\Lb +(\underline{2},\,\sigma\rho)\varphi_{i}R(\beta),\,\beta + \gamma \Rb$\, according to \fig{FCT_fig13} and
\fig{FCT_fig14}. This represents the Schr\"{o}dinger disc of the $c$ quark on the particle frame, according to \fig{FCT_fig7}
and \fig{GU_fig18}.

\item[d)] We select the disc $11$ on the upper sheet of the particle frame and its yellow region $\underline{1}$ that has the
label \,$\Lb -(\underline{1},\,\sigma\rho)R(\gamma)\varphi_{i},\,\beta + \gamma \Rb$\, according to \fig{FCT_fig13} and
\fig{FCT_fig15}. This represents the Schr\"{o}dinger disc of the $s$ quark on the particle frame, according to \fig{FCT_fig7}
and \fig{GU_fig20}.

\item[e)] We select the disc $4$ on the upper sheet of the particle frame and its green region $\underline{2}$ that has the
label \,$\Lb +(\underline{2},\,\sigma\rho^{2})\varphi_{i}R(\beta),\,\beta + \gamma \Rb$\, according to \fig{FCT_fig13} and
\fig{FCT_fig14}. This represents the Schr\"{o}dinger disc of the $t$ quark on the particle frame, according to \fig{FCT_fig7}
and \fig{GU_fig22}.

\item[f)] We select the disc $10$ on the upper sheet of the particle frame and its yellow region $\underline{1}$ that has the
label \,$\Lb -(\underline{2},\,\sigma\rho^{2})R(\gamma)\varphi_{i},\,\beta + \gamma \Rb$\, according to \fig{FCT_fig13} and
\fig{FCT_fig15}. This represents the Schr\"{o}dinger disc of the $b$ quark on the particle frame, according to \fig{FCT_fig7}
and \fig{GU_fig24}.
\end{itemize}

\item
Using the Antiparticle Rule $\mathcal{J}$,
\begin{itemize}
\item[a)] We select the disc $18$ on the lower sheet of the particle frame and its green region $\underline{2}$ that has
the label \,$\Lb -(\underline{2},\,\sigma)R(\beta)\varphi_{i},\,\beta + \gamma \Rb$\, according to \fig{FCT_fig13} and
\fig{FCT_fig16}. This represents the Schr\"{o}dinger disc of the $\bar{u}$ antiquark on the particle frame, according to
\fig{FCT_fig7} and \fig{GU_fig15}.

\item[b)] We select the disc $24$ on the lower sheet of the particle frame and its yellow region $\underline{1}$ that has
the label \,$\Lb +(\underline{1},\,\sigma)\varphi_{i}R(\gamma),\,\beta + \gamma \Rb$\, according to \fig{FCT_fig13} and
\fig{FCT_fig17}. This represents the Schr\"{o}dinger disc of the $\bar{d}$ antiquark on the particle frame, according to
\fig{FCT_fig7} and \fig{GU_fig17}.

\item[c)] We select the disc $17$ on the lower sheet of the particle frame and its green region $\underline{2}$ that has
the label \,$\Lb -(\underline{2},\,\sigma\rho)R(\beta)\varphi_{i},\,\beta + \gamma \Rb$\, according to \fig{FCT_fig13} and
\fig{FCT_fig16}. This represents the Schr\"{o}dinger disc of the $\bar{c}$ antiquark on the particle frame, according to
\fig{FCT_fig7} and \fig{GU_fig19}.

\item[d)] We select the disc $23$ on the lower sheet of the particle frame and its yellow region $\underline{1}$ that has
the label \,$\Lb +(\underline{1},\,\sigma\rho)\varphi_{i}R(\gamma),\,\beta + \gamma \Rb$\, according to \fig{FCT_fig13} and
\fig{FCT_fig17}. This represents the Schr\"{o}dinger disc of the $\bar{s}$ antiquark on the particle frame, according to
\fig{FCT_fig7} and \fig{GU_fig21}.

\item[e)] We select the disc $16$ on the lower sheet of the particle frame and its green region $\underline{2}$ that has
the label \,$\Lb -(\underline{2},\,\sigma\rho^{2})R(\beta)\varphi_{i},\,\beta + \gamma \Rb$\, according to \fig{FCT_fig13}
and \fig{FCT_fig16}. This represents the Schr\"{o}dinger disc of the $\bar{t}$ antiquark on the particle frame, according to
\fig{FCT_fig7} and \fig{GU_fig23}.

\item[f)] We select the disc $22$ on the lower sheet of the particle frame and its yellow region $\underline{1}$ that has
the label \,$\Lb +(\underline{1},\,\sigma\rho^{2})\varphi_{i}R(\gamma),\,\beta + \gamma \Rb$\, according to \fig{FCT_fig13}
and \fig{FCT_fig17}. This represents the Schr\"{o}dinger disc of the $\bar{b}$ antiquark on the particle frame, according to
\fig{FCT_fig7} and \fig{GU_fig25}.
\end{itemize}

\DOUBLEFIGURE[h]{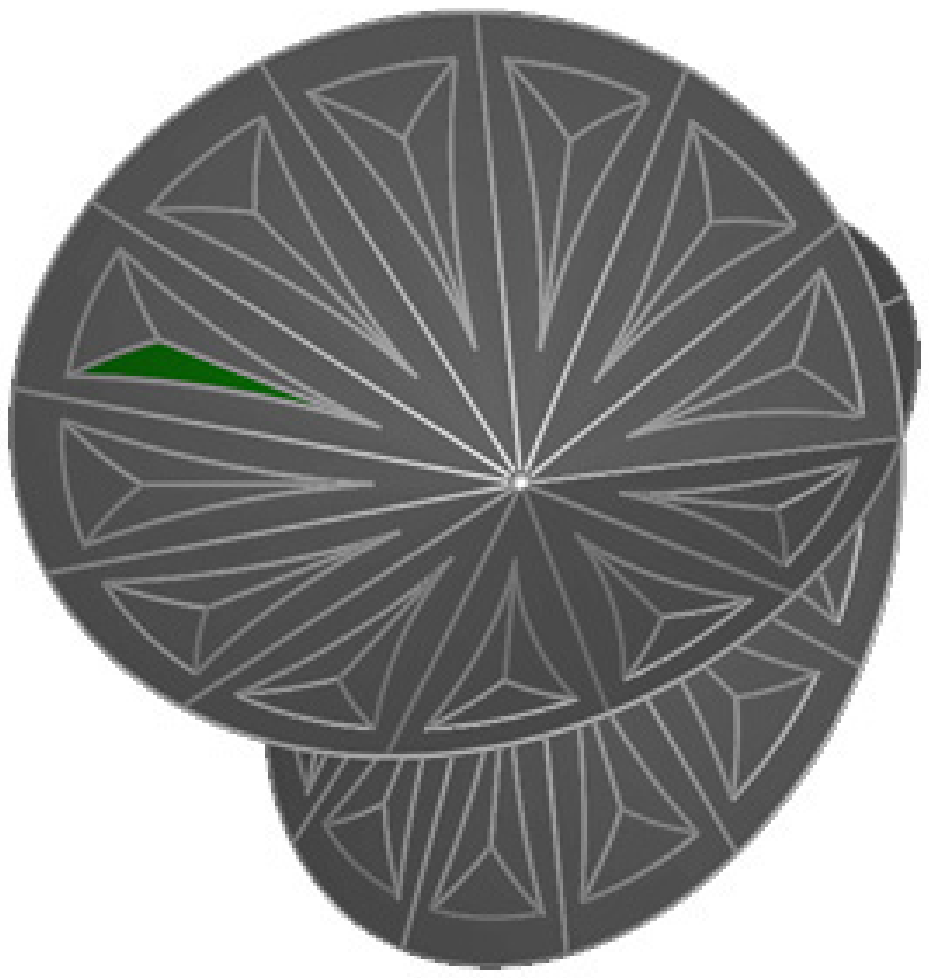,width=50mm,height=50mm}{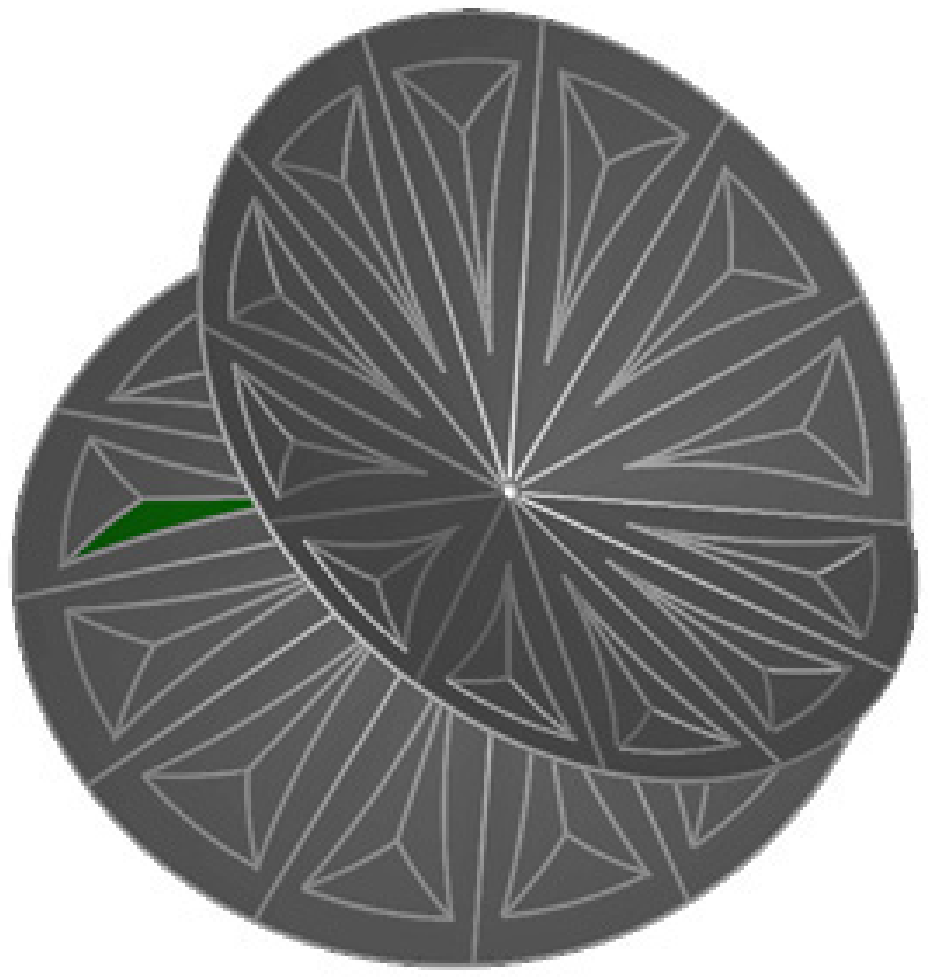,width=50mm,height=50mm}
{$u$ quark on the particle frame.
\label{GU_fig14}}
{$\bar{u}$ antiquark on the particle frame.
\label{GU_fig15}}

\DOUBLEFIGURE[h]{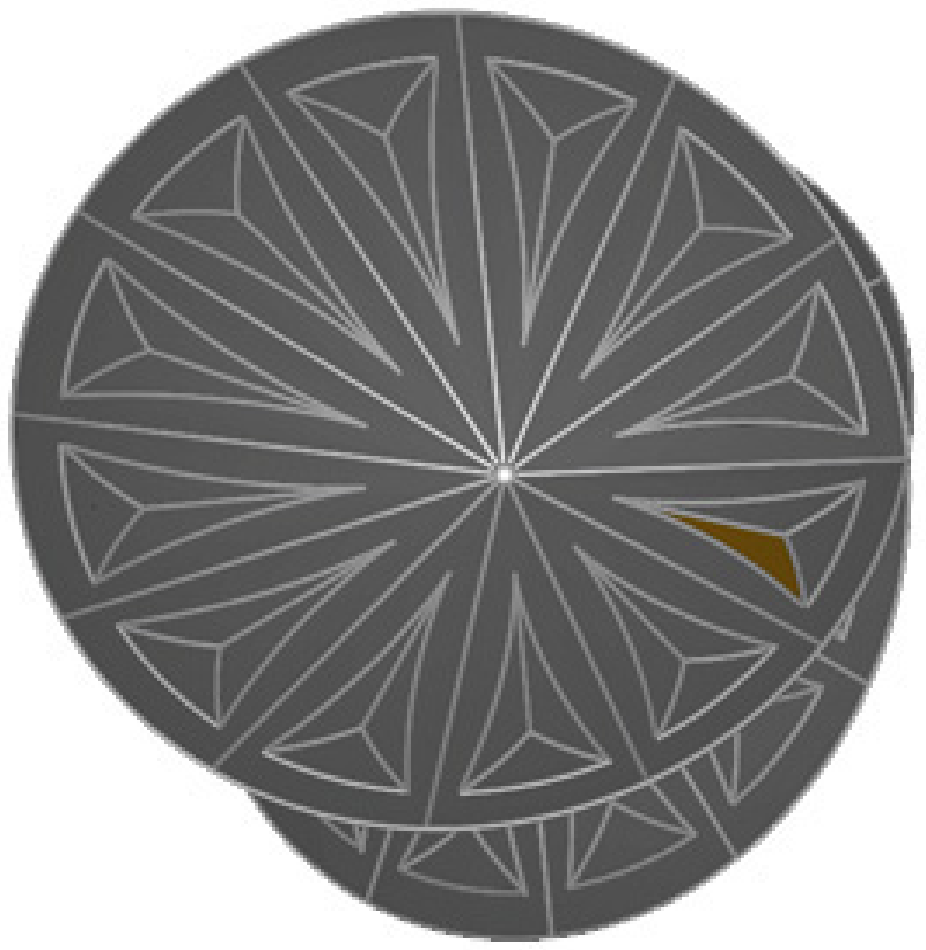,width=50mm,height=50mm}{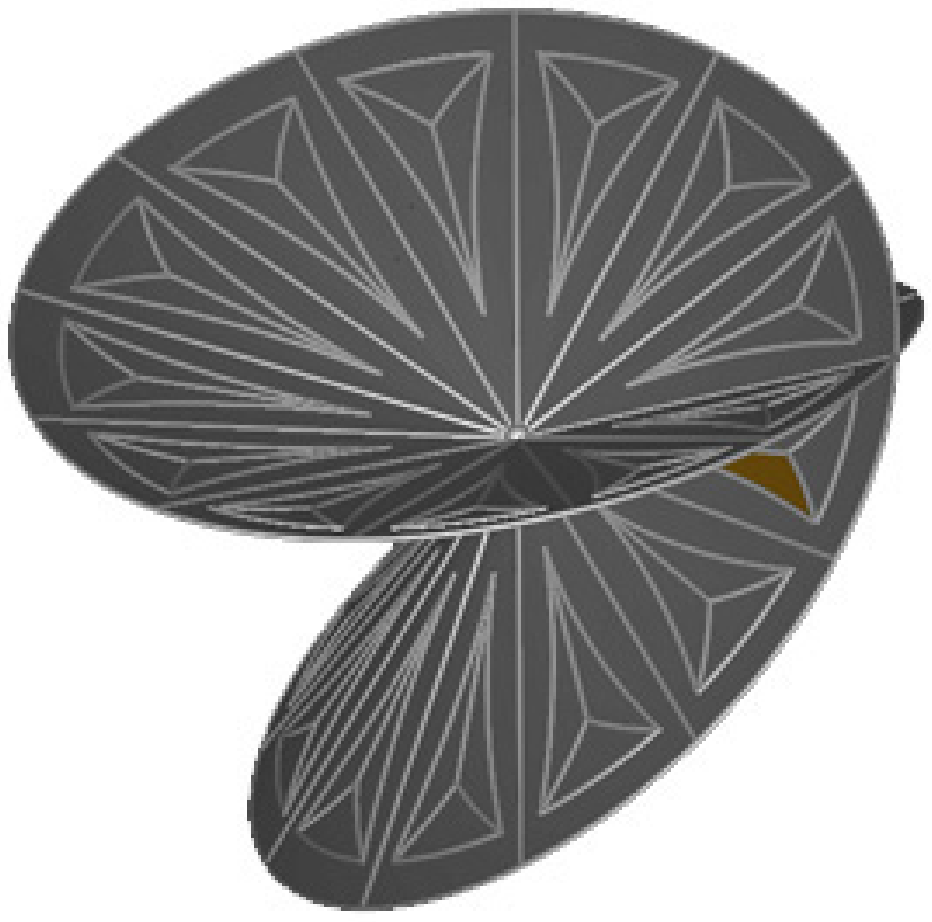,width=50mm,height=50mm}
{$d$ quark on the particle frame.
\label{GU_fig16}}
{$\bar{d}$ antiquark on the particle frame.
\label{GU_fig17}}

\DOUBLEFIGURE[h]{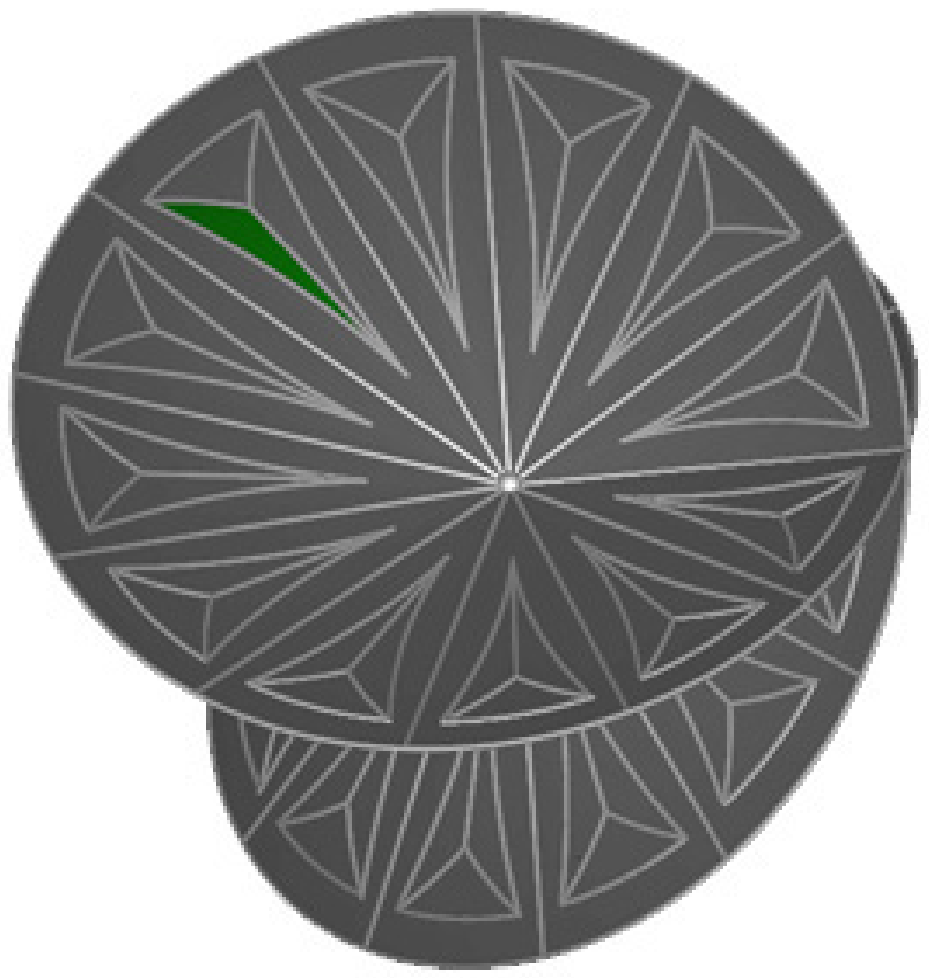,width=50mm,height=50mm}{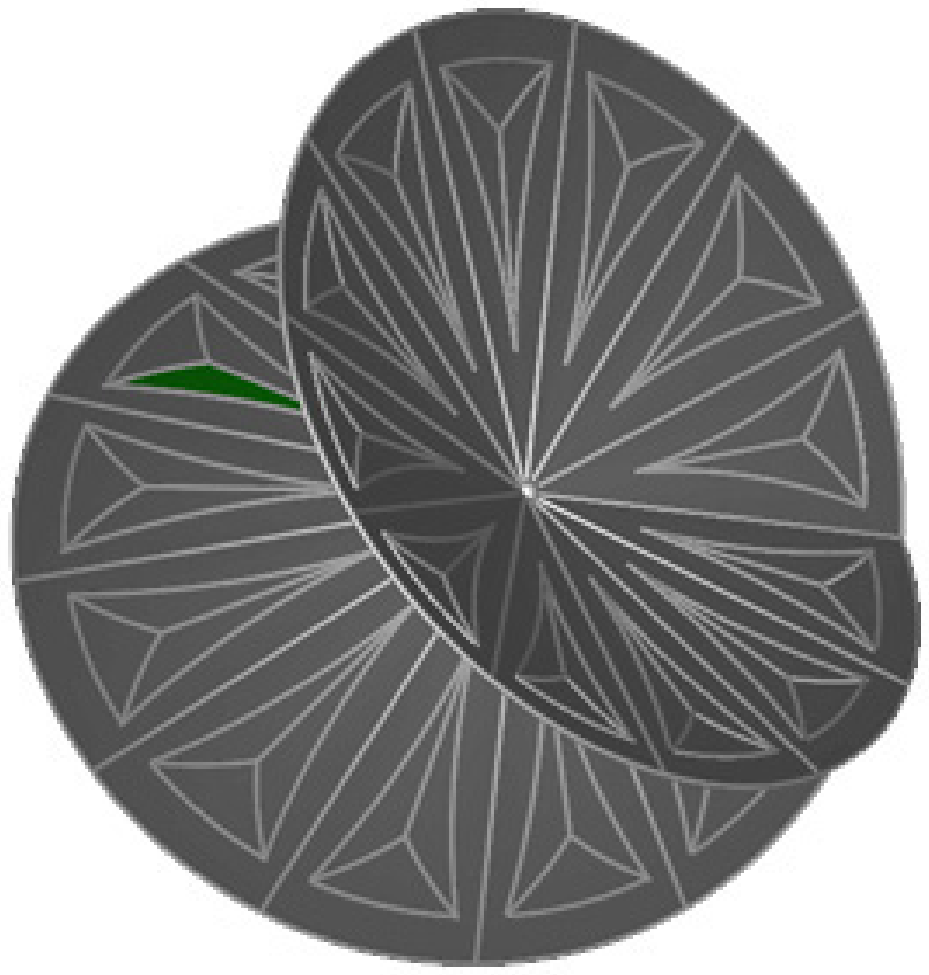,width=50mm,height=50mm}
{$c$ quark on the particle frame.
\label{GU_fig18}}
{$\bar{c}$ antiquark on the particle frame.
\label{GU_fig19}}

\DOUBLEFIGURE[h]{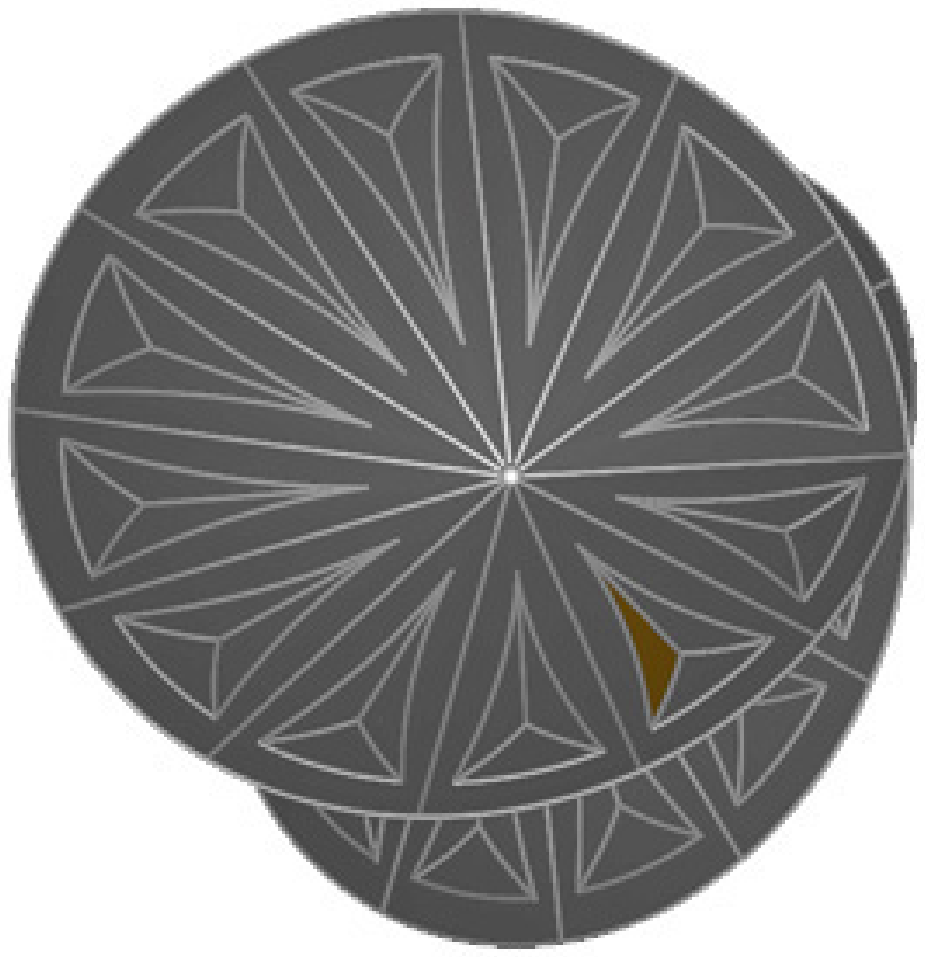,width=50mm,height=50mm}{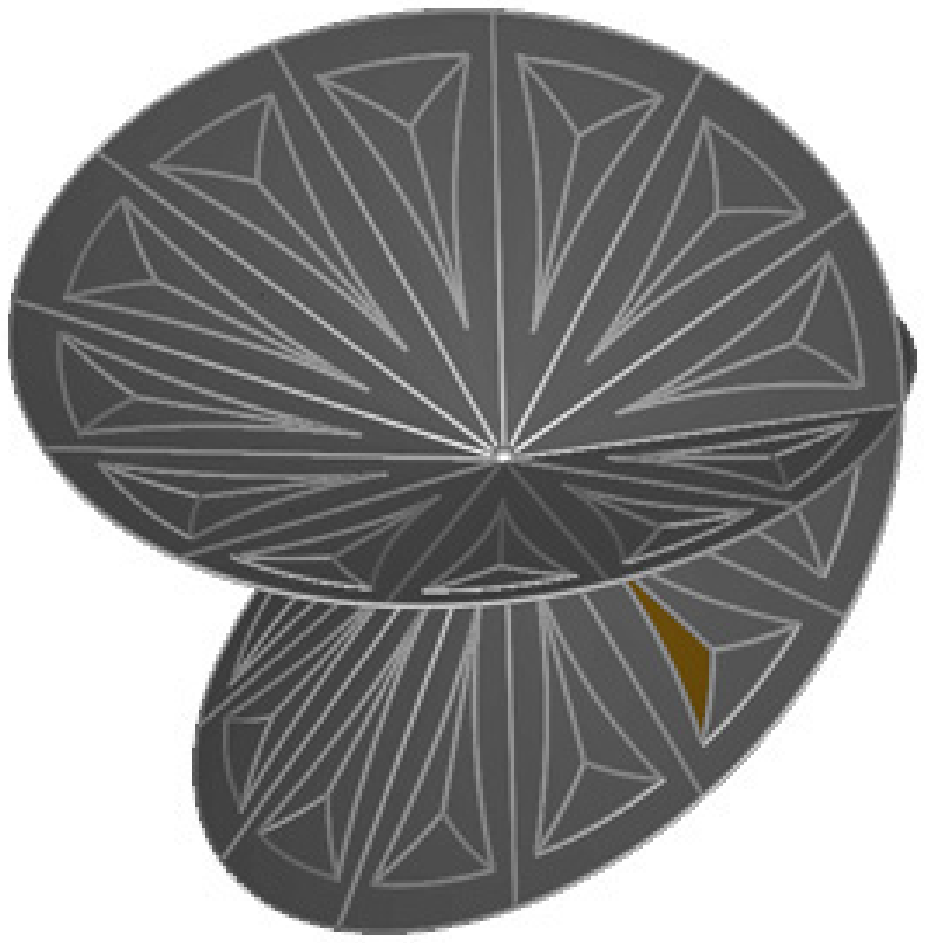,width=50mm,height=50mm}
{$s$ quark on the particle frame.
\label{GU_fig20}}
{$\bar{s}$ antiquark on the particle frame.
\label{GU_fig21}}

\DOUBLEFIGURE[h]{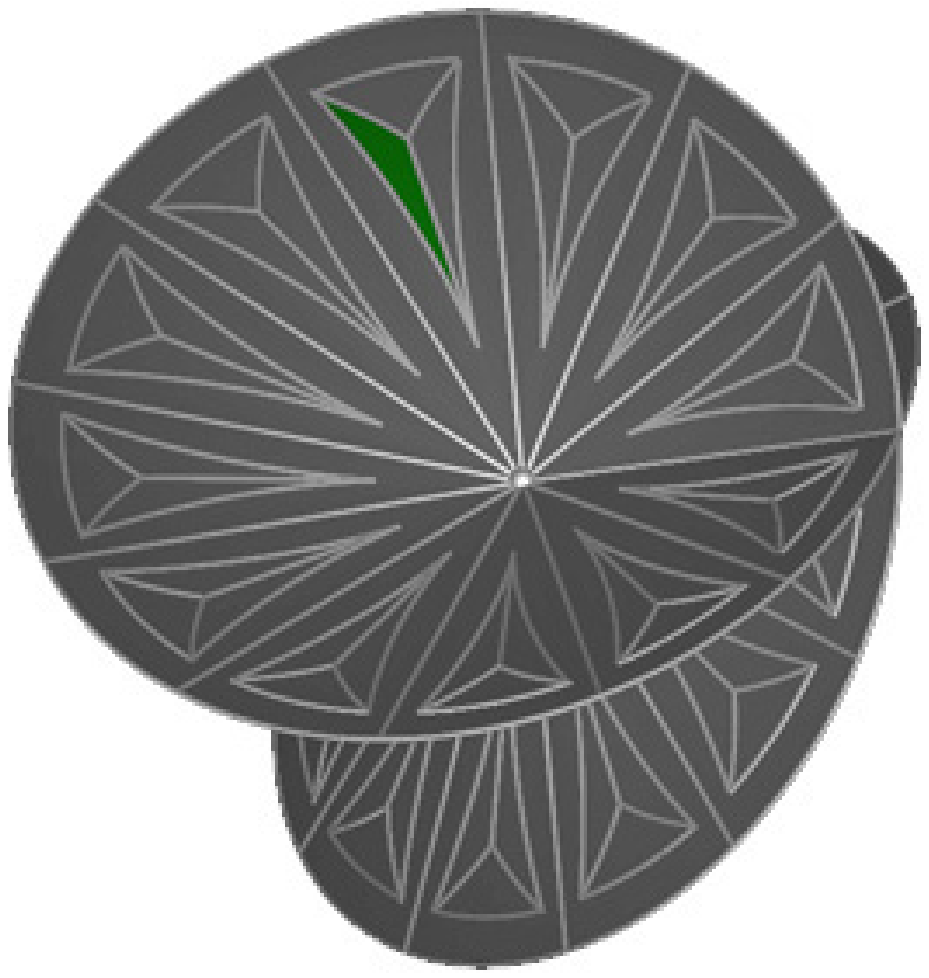,width=50mm,height=50mm}{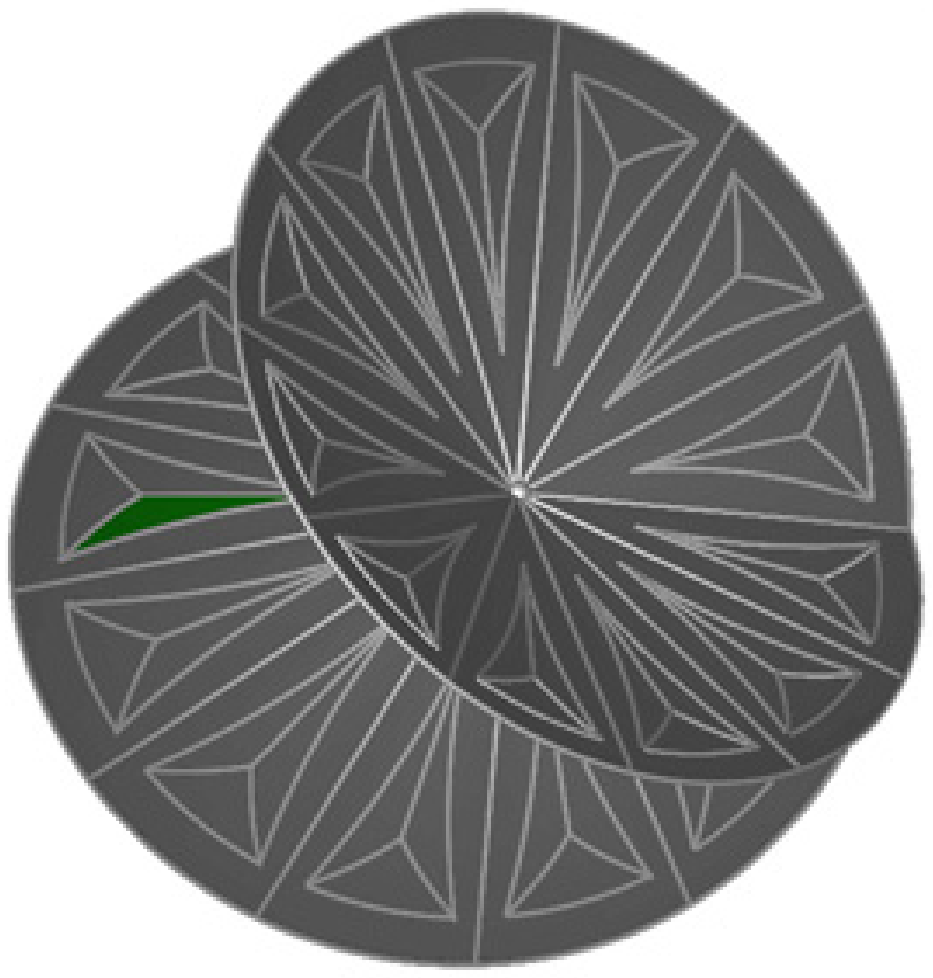,width=50mm,height=50mm}
{$t$ quark on the particle frame.
\label{GU_fig22}}
{$\bar{t}$ antiquark on the particle frame.
\label{GU_fig23}}

\DOUBLEFIGURE[h]{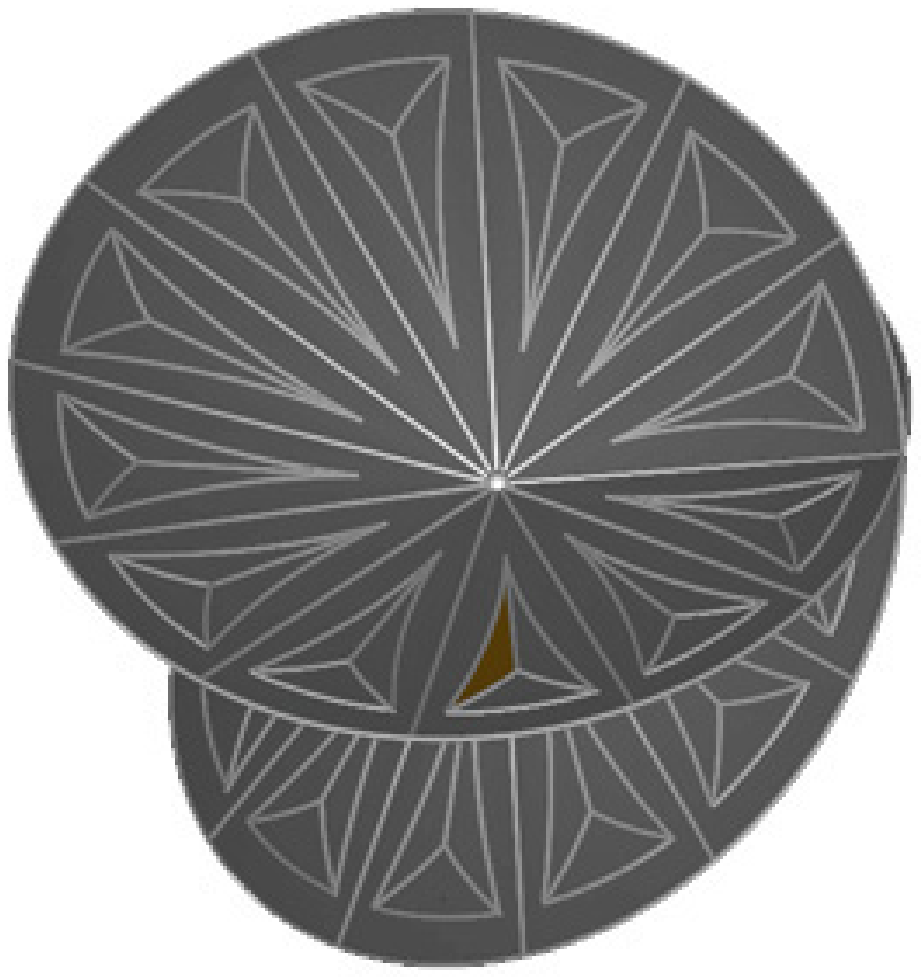,width=50mm,height=50mm}{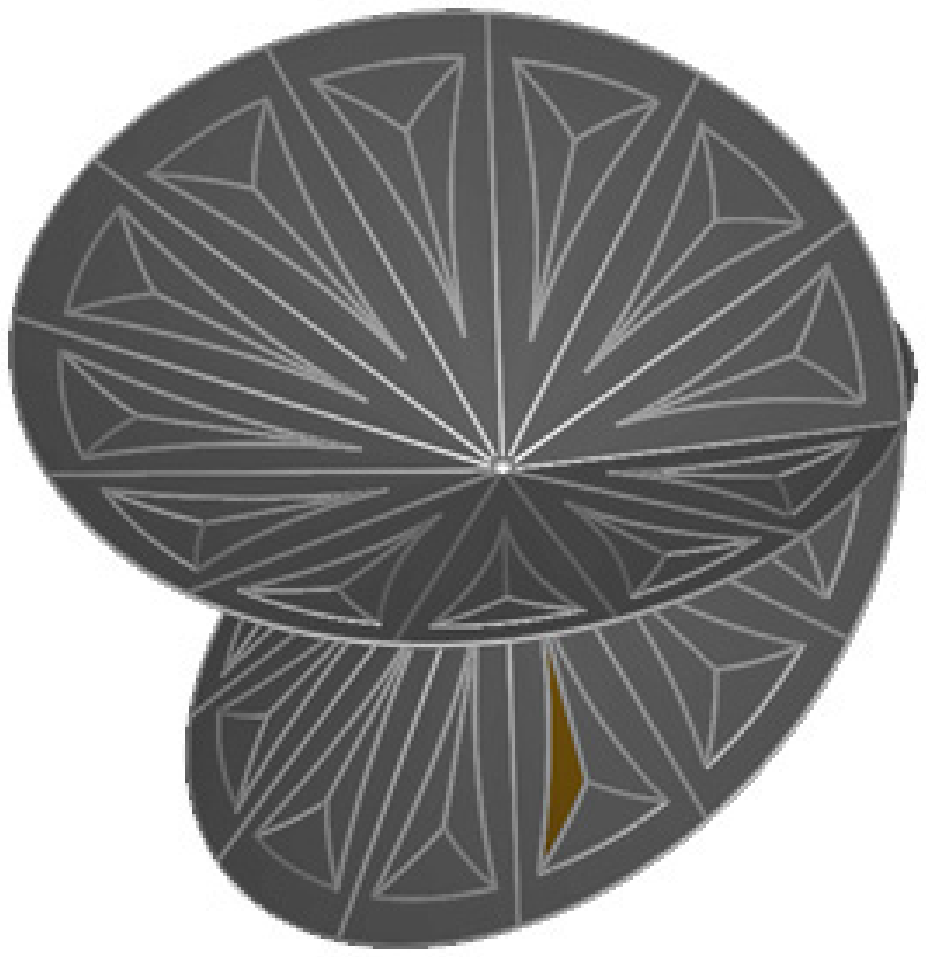,width=50mm,height=50mm}
{$b$ quark on the particle frame.
\label{GU_fig24}}
{$\bar{b}$ antiquark on the particle frame.
\label{GU_fig25}}

\item By the Fermion Selection Rule $\mathcal{A}$ - \,$u$, $\bar{u}$, $d$, $\bar{d}$\, are quarks of generation I;
\,$c$, $\bar{c}$, $s$, $\bar{s}$\, are quarks of generation II; \,$t$, $\bar{t}$, $b$, $\bar{b}$\, are quarks of
generation III.

\item By the Spin Rule $\mathcal{D}$, the spin of all \,$u$, $\bar{u}$, $d$, $\bar{d}$, $c$, $\bar{c}$, $s$, $\bar{s}$,
$t$, $\bar{t}$, $b$, $\bar{b}$\, quarks and antiquarks is \,$1/2$.

\item By the Electric Charge Rule $\mathcal{E}$, the electric charge of \,$u$, $c$, $t$\, is \,$+2/3$;\,the electric charge of  \,$d$, $s$, $b$\, is \,$-1/3$;\,the electric charge of  \,$\bar{u}$, $\bar{c}$, $\bar{t}$\, is \,$-2/3$;\,the electric charge of  \,$\bar{d}$, $\bar{s}$, $\bar{b}$\, is \,$+1/3$.

\vskip1.0truecm
\item By the Weak Isospin Rule $\mathcal{F}$,
\begin{itemize}
\item[a)] The weak isospin of $u_{_{\!L}}$, $c_{_{\!L}}$, $t_{_{\!L}}$\, is \,$+1/2$;\,the weak isospin of \,$d_{_{\!L}}$, $s_{_{\!L}}$, $b_{_{\!L}}$\, is \,$-1/2$;\,the weak isospin of
\,$\bar{u}_{_{\!L}}$, $\bar{d}_{_{\!L}}$, $\bar{c}_{_{\!L}}$, $\bar{s}_{_{\!L}}$, $\bar{t}_{_{\!L}}$, $\bar{b}_{_{\!L}}$\, is
\,$0$;
\item[b)] The weak isospin of $\bar{u}_{_{\!R}}$, $\bar{c}_{_{\!R}}$, $\bar{t}_{_{\!R}}$\, is \,$-1/2$;\,the weak isospin of \,$\bar{d}_{_{\!R}}$, $\bar{s}_{_{\!R}}$,
$\bar{b}_{_{\!R}}$\, is \,$+1/2$;\,the weak isospin of \,$u_{_{\!R}}$, $d_{_{\!R}}$, $c_{_{\!R}}$, $s_{_{\!R}}$, $t_{_{\!R}}$, $b_{_{\!R}}$\,
is \,$0$.
\end{itemize}

\item By the Strong Charge Rule $\mathcal{G}$,
\begin{itemize}
\item[a)] $u$, $c$, $t$\, have one of three possible strong charges \,$+\sigma$, $+\sigma\rho$, $+\sigma\rho^{2}$;
\item[b)] $d$, $s$, $b$\, have one of three possible strong charges \,$-\sigma$, $-\sigma\rho$, $-\sigma\rho^{2}$;
\item[c)] $\bar{u}$, $\bar{c}$, $\bar{t}$\, have one of three possible strong charges
\,$-\sigma$, $-\sigma\rho$, $-\sigma\rho^{2}$;
\item[d)] $\bar{d}$, $\bar{s}$, $\bar{b}$\, have one of three possible strong charges
\,$+\sigma$, $+\sigma\rho$, $+\sigma\rho^{2}$;
\end{itemize}
so $N_{c} = 3$.

\item From indirect experimental observations, the effective rest masses of the quarks are - \,$m_{u,\bar{u}} \sim 0.003\,GeV$,
\,$m_{d,\bar{d}} \sim 0.006\,GeV$, \,$m_{c,\bar{c}} \sim 1.337\,GeV$, \,$m_{s,\bar{s}} \sim 0.1\,GeV$,
\,$m_{t,\bar{t}} \sim 171\,GeV$ and \,$m_{b,\bar{b}} \sim 4.2\,GeV$, which can be attributed by the Mass Rule $\mathcal{H}$
to the Higgs-Kibble mechanism \cite{Hig1,Hig2,Hig3,Hig4,Kib,Kib1,Djo1,Bed}.

\item In agreement with the Helicity Rule $\mathcal{K}$, all quarks and antiquarks can be theoretically observed with both
left-handed and right-handed helicities.

\item By the \textbf{CP} Transformation Rule $\mathcal{L}$ - \,$u_{_{\!L}} \Leftrightarrow \bar{u}_{_{\!R}}$,\,
\,$\bar{u}_{_{\!L}} \Leftrightarrow u_{_{\!R}}$;\, \,$d_{_{\!L}} \Leftrightarrow \bar{d}_{_{\!R}}$,\,
\,$\bar{d}_{_{\!L}} \Leftrightarrow d_{_{\!R}}$;\, \,$c_{_{\!L}} \Leftrightarrow \bar{c}_{_{\!R}}$,\,
\,$\bar{c}_{_{\!L}} \Leftrightarrow c_{_{\!R}}$;\, \,$s_{_{\!L}} \Leftrightarrow \bar{s}_{_{\!R}}$,\,
\,$\bar{s}_{_{\!L}} \Leftrightarrow s_{_{\!R}}$;\, \,$t_{_{\!L}} \Leftrightarrow \bar{t}_{_{\!R}}$,\,
\,$\bar{t}_{_{\!L}} \Leftrightarrow t_{_{\!R}}$;\, \,$b_{_{\!L}} \Leftrightarrow \bar{b}_{_{\!R}}$,\,
\,$\bar{b}_{_{\!L}} \Leftrightarrow b_{_{\!R}}$.
\end{enumerate}

In Table $2$ we represent the values of the related physical magnitudes of the left-handed quarks.

\begin{center}
\begin{normalsize}
\TABLE[h]{
\begin{tabular}{|l|l|l|l|l|l|l|l|}
\hline\
\hskip 0truecm \textbf{Quarks (left-handed)} & \hskip 0.1truecm \textbf{Symbol} & \hskip 0.1truecm \textbf{Generation}
& \hskip 0.1truecm \textbf{Charge} & \hskip 0.1truecm \textbf{Weak isospin} & \hskip 0.1truecm \textbf{Mass (MeV)}  \\
\hline
\hline %
\hskip 1.2truecm up quark & \hskip 0.6truecm $u_{_{\!L}}$ & \hskip 1.1truecm I
& \hskip 0.3truecm $+\,2/3$ & \hskip 0.7truecm $+\,1/2$ &
\hskip 0.9truecm $\sim\,3$  \\
\hline %
\hskip 0.9truecm up antiquark & \hskip 0.6truecm $\bar{u}_{_{\!L}}$ & \hskip 1.1truecm I
& \hskip 0.3truecm $-\,2/3$ & \hskip 1.2truecm $0$ &
\hskip 0.9truecm $\sim\,3$  \\
\hline %
\hskip 1.0truecm down quark & \hskip 0.6truecm $d_{_{\!L}}$ & \hskip 1.1truecm I
& \hskip 0.3truecm $-\,1/3$ & \hskip 0.7truecm $-\,1/2$ &
\hskip 0.9truecm $\sim\,6$  \\
\hline %
\hskip 0.7truecm down antiquark & \hskip 0.6truecm $\bar{d}_{_{\!L}}$ & \hskip 1.1truecm I
& \hskip 0.3truecm $+\,1/3$ & \hskip 1.2truecm $0$ &
\hskip 0.9truecm $\sim\,6$  \\
\hline %
\hskip 0.9truecm charm quark & \hskip 0.6truecm $c_{_{\!L}}$ & \hskip 1.0truecm II
& \hskip 0.3truecm $+\,2/3$ & \hskip 0.7truecm $+\,1/2$ &
\hskip 0.6truecm $\sim\,1337$  \\
\hline %
\hskip 0.6truecm charm antiquark & \hskip 0.6truecm $\bar{c}_{_{\!L}}$ & \hskip 1.0truecm II
& \hskip 0.3truecm $-\,2/3$ & \hskip 1.2truecm $0$ &
\hskip 0.6truecm $\sim\,1337$  \\
\hline %
\hskip 0.9truecm strange quark & \hskip 0.6truecm $s_{_{\!L}}$ & \hskip 1.0truecm II
& \hskip 0.3truecm $-\,1/3$ & \hskip 0.7truecm $-\,1/2$ &
\hskip 0.7truecm $\sim\,100$  \\
\hline %
\hskip 0.6truecm strange antiquark & \hskip 0.6truecm $\bar{s}_{_{\!L}}$ & \hskip 1.0truecm II
& \hskip 0.3truecm $+\,1/3$ & \hskip 1.2truecm $0$ &
\hskip 0.7truecm $\sim\,100$  \\
\hline %
\hskip 1.2truecm top quark & \hskip 0.6truecm $t_{_{\!L}}$ & \hskip 0.9truecm III
& \hskip 0.3truecm $+\,2/3$ & \hskip 0.7truecm $+\,1/2$ &
\hskip 0.4truecm $\sim\,171000$  \\
\hline %
\hskip 0.9truecm top antiquark & \hskip 0.6truecm $\bar{t}_{_{\!L}}$ & \hskip 0.9truecm III
& \hskip 0.3truecm $-\,2/3$ & \hskip 1.2truecm $0$ &
\hskip 0.4truecm $\sim\,171000$  \\
\hline %
\hskip 0.9truecm bottom quark & \hskip 0.6truecm $b_{_{\!L}}$ & \hskip 0.9truecm III
& \hskip 0.3truecm $-\,1/3$ & \hskip 0.7truecm $-\,1/2$ &
\hskip 0.6truecm $\sim\,4200$  \\
\hline %
\hskip 0.6truecm bottom antiquark & \hskip 0.6truecm $\bar{b}_{_{\!L}}$ & \hskip 0.9truecm III
& \hskip 0.3truecm $+\,1/3$ & \hskip 1.2truecm $0$ &
\hskip 0.6truecm $\sim\,4200$  \\
\hline %
\end{tabular}
\caption{The left-handed quarks in the SM.}}
\end{normalsize}
\end{center}

\subsection{$Z$ and $W^{\pm}$ Bosons}
\begin{enumerate}
\item If we select the unique permutations \,$\psi_{Z^{0}} = R(\beta)\varphi_{i}$,\, \,$\psi_{W^{+}} = R(\beta)\varphi_{i}$,\, \,$\psi_{W^{-}} = R(\beta)\varphi_{i}$, associated with the massive vector bosons then we must have \,$\beta = 1$\, according to the Mass Rule $\mathcal{H}$ and we also select \,$\gamma = 1$.

\item Then the particle frame corresponds to the \textit{t}-Riemann surface with these choices of $\varphi_{i}$,
$\beta$ and $\gamma$.

\item
\begin{itemize}
\item[a)] For the first ray we select discs $3$, $4$ on the upper sheet of the particle frame and their blue regions
$\underline{0}$ that have the labels \,$\Lb +(\underline{0},\,\rho^{2})\varphi_{i}R(\beta),\,\beta + \gamma \Rb$,\,
\,$\Lb +(\underline{0},\,\sigma\rho^{2})\varphi_{i}R(\beta),\,\beta + \gamma \Rb$,\, respectively, according to
\fig{FCT_fig13} and \fig{FCT_fig14}.
For the second ray we select discs $9$, $10$ on the upper sheet of the particle frame and their blue regions
$\underline{0}$ that have the labels \,$\Lb -(\underline{0},\,\rho^{2})R(\gamma)\varphi_{i},\,\beta + \gamma \Rb$,\,
\,$\Lb -(\underline{0},\,\sigma\rho^{2})R(\gamma)\varphi_{i},\,\beta + \gamma \Rb$,\, respectively, according to
\fig{FCT_fig13} and \fig{FCT_fig15}.
These together represent the Schr\"{o}dinger discs corresponding to the neutral component of the weak field, and the pair
of rays represent the $Z^{0}$ boson on the particle frame, according to \fig{FCT_fig7} and \fig{GU_fig26}.

\item[b)] For the first ray we select discs $2$, $3$ on the upper sheet of the particle frame and their red regions
$\underline{3}$ that have the labels \,$\Lb +(\underline{3},\,\rho)\varphi_{i}R(\beta),\,\beta + \gamma \Rb$,\,
\,$\Lb +(\underline{3},\,\rho^{2})\varphi_{i}R(\beta),\,\beta + \gamma \Rb$,\, respectively, according to
\fig{FCT_fig13} and \fig{FCT_fig14}.
For the second ray we select discs $20$, $21$ on the lower sheet of the particle frame and their red regions
$\underline{3}$ that have the labels \,$\Lb +(\underline{3},\,\rho)\varphi_{i}R(\gamma),\,\beta + \gamma \Rb$,\,
\,$\Lb +(\underline{3},\,\rho^{2})\varphi_{i}R(\gamma),\,\beta + \gamma \Rb$,\, respectively, according to
\fig{FCT_fig13} and \fig{FCT_fig17}.
These together represent the Schr\"{o}dinger discs corresponding to the positive component of the weak field, and the pair
of rays represent the $W^{+}$ boson on the particle frame, according to \fig{FCT_fig7} and \fig{GU_fig28}.

\item[c)] For the first ray we select discs $16$, $17$ on the lower sheet of the particle frame and their red regions
$\underline{3}$ that have the labels \,$\Lb -(\underline{3},\,\sigma\rho^{2})R(\beta)\varphi_{i},\,\beta + \gamma \Rb$,\,
\,$\Lb -(\underline{3},\,\sigma\rho)R(\beta)\varphi_{i},\,\beta + \gamma \Rb$,\, respectively, according to
\fig{FCT_fig13} and \fig{FCT_fig16}.
For the second ray we select discs $10$, $11$ on the upper sheet of the particle frame and their red regions
$\underline{3}$ that have the labels \,$\Lb -(\underline{3},\,\sigma\rho^{2})R(\gamma)\varphi_{i},\,\beta + \gamma \Rb$,\,
\,$\Lb -(\underline{3},\,\sigma\rho)R(\gamma)\varphi_{i},\,\beta + \gamma \Rb$,\, respectively, according to
\fig{FCT_fig13} and \fig{FCT_fig15}.
These together represent the Schr\"{o}dinger discs corresponding to the negative component of the weak field, and the pair
of rays represent the $W^{-}$ boson on the particle frame, according to \fig{FCT_fig7} and \fig{GU_fig30}.
\end{itemize}

\item
By the Antiparticle Rule $\mathcal{J}$,
\begin{itemize}
\item[a)] for the first ray we select discs $15$, $16$ on the lower sheet of the particle frame and their blue regions
$\underline{0}$ that have the labels \,$\Lb -(\underline{0},\,\rho^{2})R(\beta)\varphi_{i},\,\beta + \gamma \Rb$,\,
\,$\Lb -(\underline{0},\,\sigma\rho^{2})R(\beta)\varphi_{i},\,\beta + \gamma \Rb$,\, respectively, according to
\fig{FCT_fig13} and \fig{FCT_fig16}.
For the second ray we select discs $21$, $22$ on the lower sheet of the particle frame and their blue regions
$\underline{0}$ that have the labels \,$\Lb +(\underline{0},\,\rho^{2})\varphi_{i}R(\gamma),\,\beta + \gamma \Rb$,\,
\,$\Lb +(\underline{0},\,\sigma\rho^{2})\varphi_{i}R(\gamma),\,\beta + \gamma \Rb$,\,
respectively, according to \fig{FCT_fig13} and \fig{FCT_fig17}.
These together represent the Schr\"{o}dinger discs corresponding to the neutral component of the weak field, and the pair
of rays represent the antiparticle $\bar{Z}^{0}$ on the particle frame, according to \fig{FCT_fig7} and \fig{GU_fig27}.

\item[b)] for the first ray we select discs $14$, $15$ on the lower sheet of the particle frame and their red regions
$\underline{3}$ that have the labels \,$\Lb -(\underline{3},\,\rho)R(\beta)\varphi_{i},\,\beta + \gamma \Rb$,\,
\,$\Lb -(\underline{3},\,\rho^{2})R(\beta)\varphi_{i},\,\beta + \gamma \Rb$,\, respectively, according to
\fig{FCT_fig13} and \fig{FCT_fig16}.
For the second ray we select discs $8$, $9$ on the upper sheet of the particle frame and their red regions
$\underline{3}$ that have the labels \,$\Lb -(\underline{3},\,\rho)R(\gamma)\varphi_{i},\,\beta + \gamma \Rb$,\,
\,$\Lb -(\underline{3},\,\rho^{2})R(\gamma)\varphi_{i},\,\beta + \gamma \Rb$,\,
respectively, according to \fig{FCT_fig13} and \fig{FCT_fig15}.
These together represent the Schr\"{o}dinger discs corresponding to the negative component of the weak field, and the pair
of rays represent the antiparticle $\bar{W}^{+}$ on the particle frame, according to \fig{FCT_fig7} and \fig{GU_fig29}.

\item[c)] for the first ray we select discs $4$, $5$ on the upper sheet of the particle frame and their red regions
$\underline{3}$ that have the labels \,$\Lb +(\underline{3},\,\sigma\rho^{2})\varphi_{i}R(\beta),\,\beta + \gamma \Rb$,\,
\,$\Lb +(\underline{3},\,\sigma\rho)\varphi_{i}R(\beta),\,\beta + \gamma \Rb$,\, respectively, according to
\fig{FCT_fig13} and \fig{FCT_fig14}.
For the second ray we select discs $22$, $23$ on the lower sheet of the particle frame and their red regions
$\underline{3}$ that have the labels \,$\Lb +(\underline{3},\,\sigma\rho^{2})\varphi_{i}R(\gamma),\,\beta + \gamma \Rb$,\,
\,$\Lb +(\underline{3},\,\sigma\rho)\varphi_{i}R(\gamma),\,\beta + \gamma \Rb$,\, respectively, according to
\fig{FCT_fig13} and \fig{FCT_fig17}.
These together represent the Schr\"{o}dinger discs corresponding to the positive component of the weak field, and the pair
of rays represent the antiparticle $\bar{W}^{-}$ on the particle frame, according to \fig{FCT_fig7} and \fig{GU_fig31}.
\end{itemize}

\DOUBLEFIGURE[h]{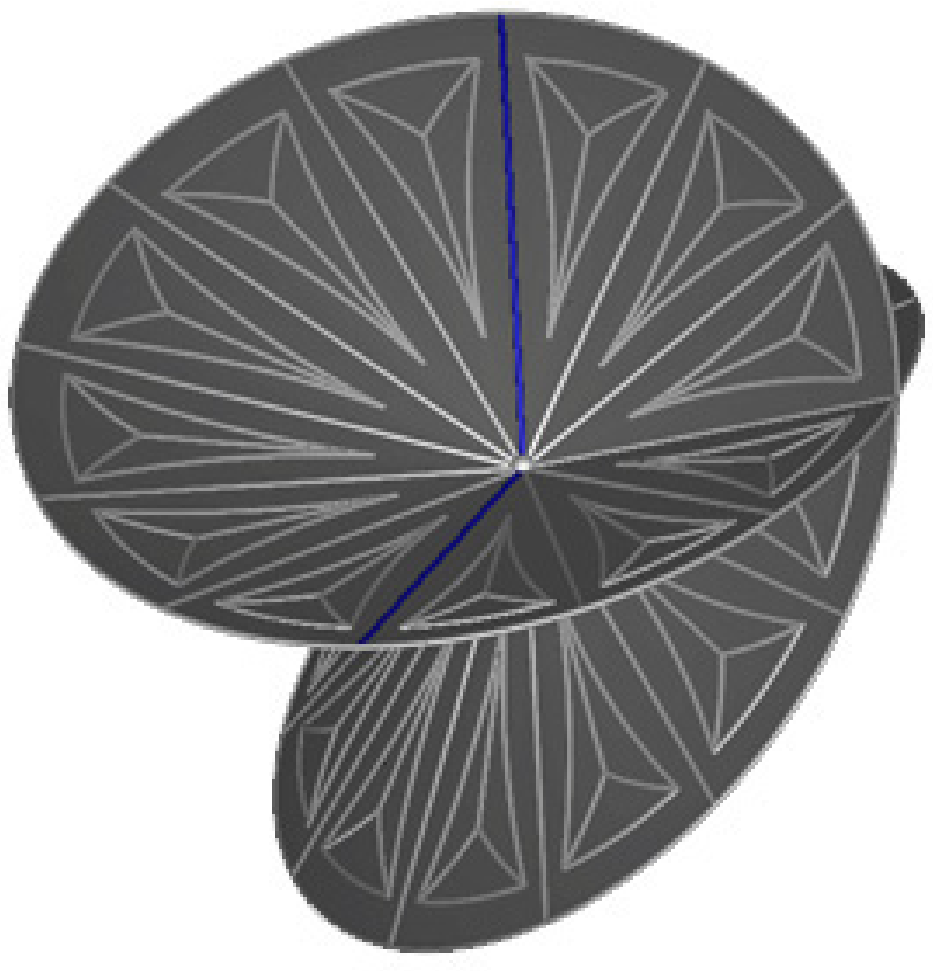,width=50mm,height=50mm}{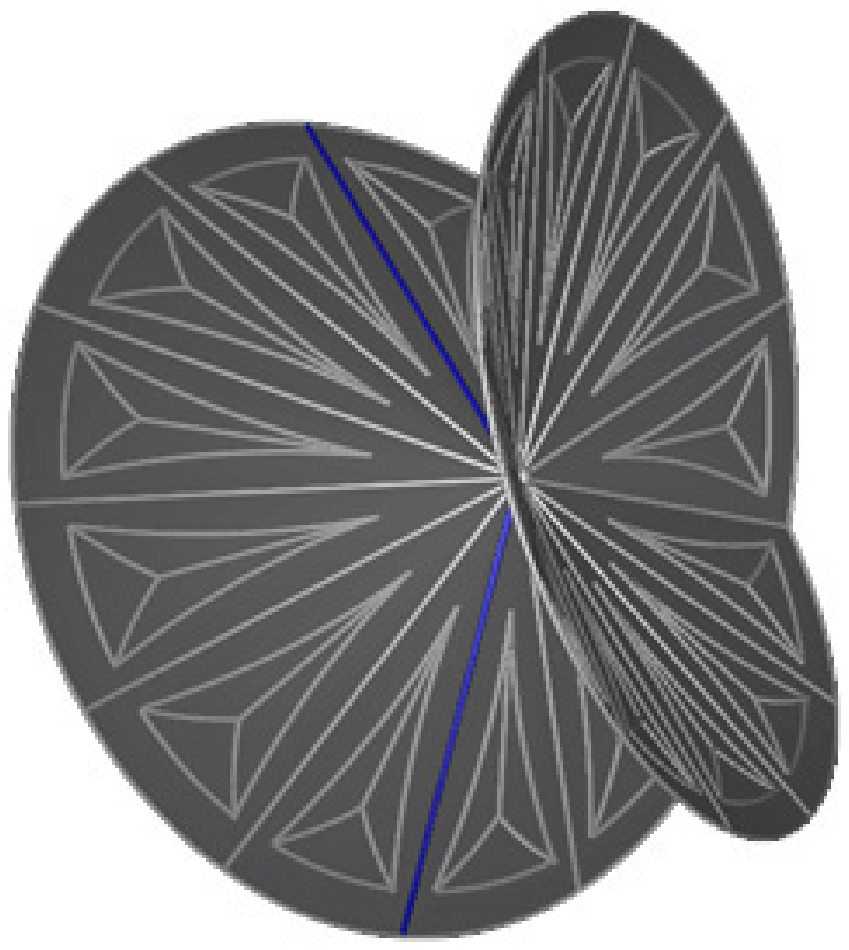,width=50mm,height=50mm}
{$Z^{0}$ boson on the particle frame.
\label{GU_fig26}}
{$Z^{0}$ boson's antiparticle on the particle frame (which is equivalent to the $Z^{0}$ boson).
\label{GU_fig27}}

\DOUBLEFIGURE[h]{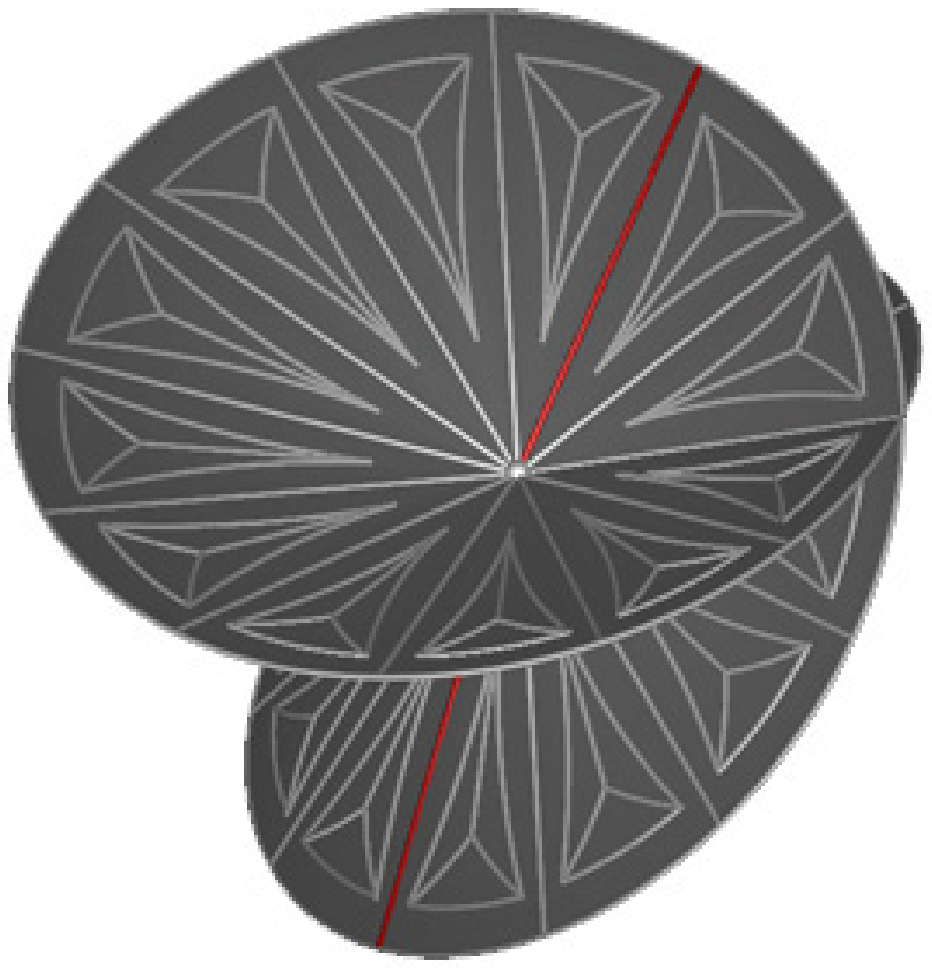,width=50mm,height=50mm}{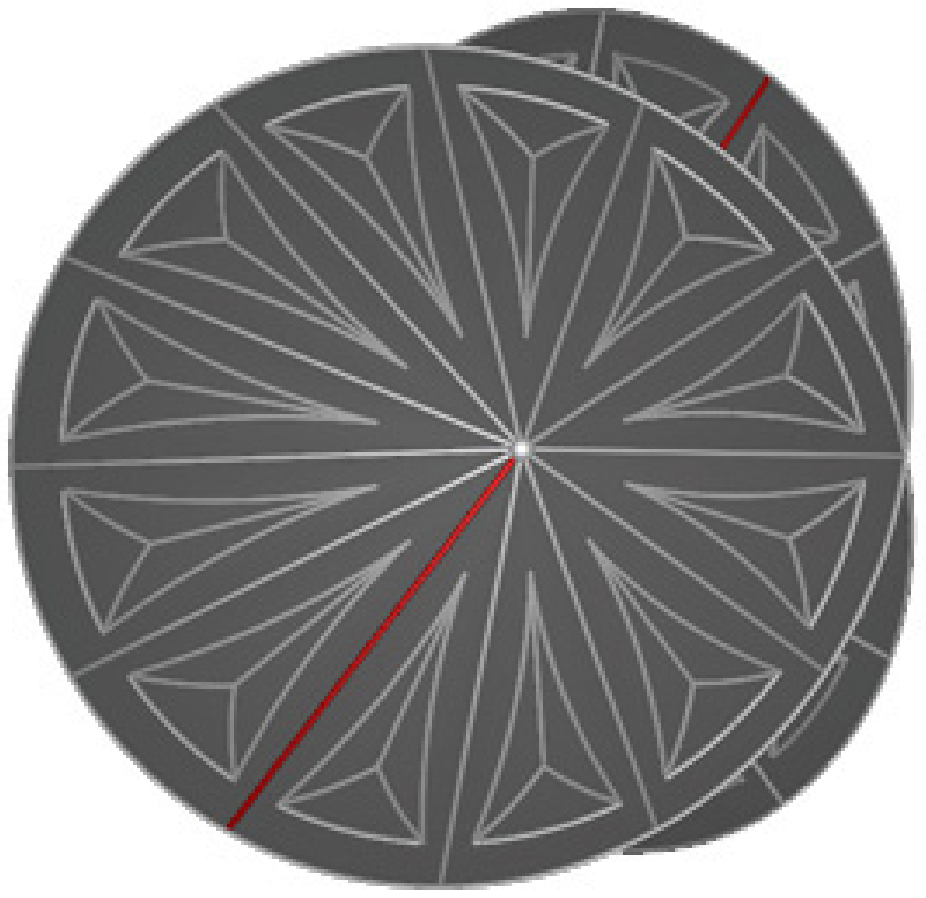,width=50mm,height=50mm}
{$W^{+}$ boson on the particle frame.
\label{GU_fig28}}
{$W^{+}$ boson's antiparticle on the particle frame (which is equivalent to the $W^{-}$ boson).
\label{GU_fig29}}

\DOUBLEFIGURE[h]{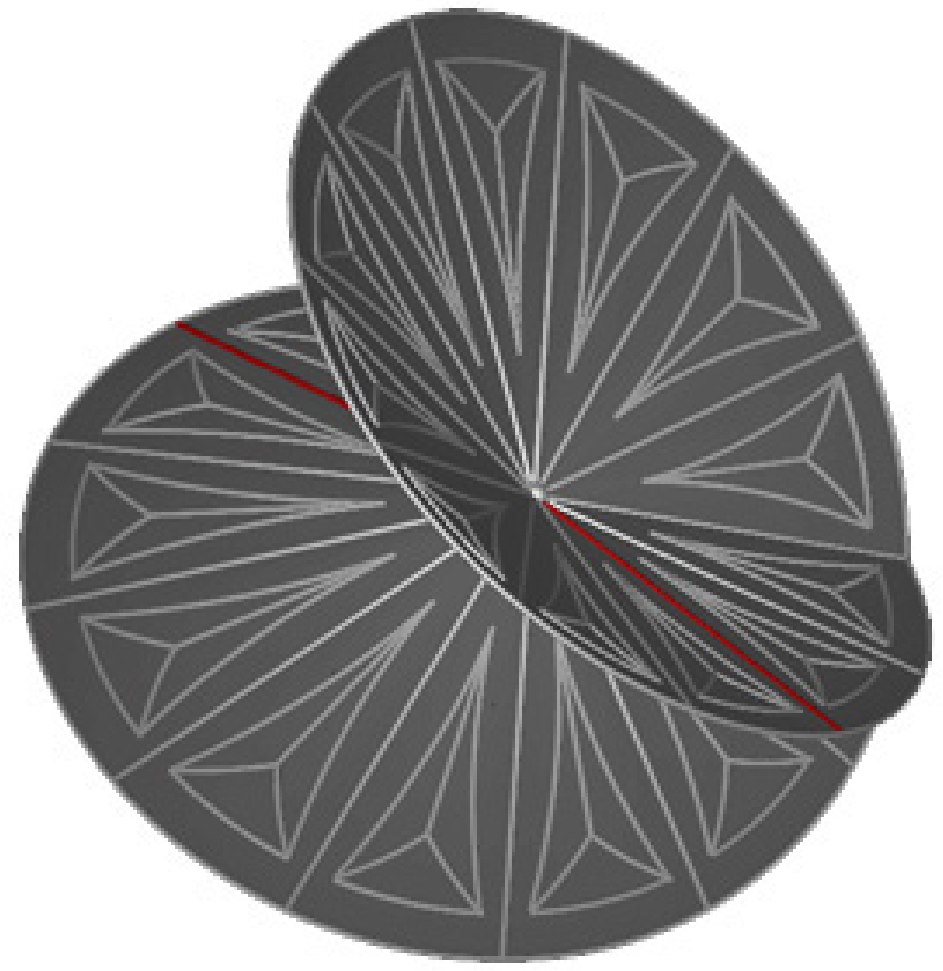,width=50mm,height=50mm}{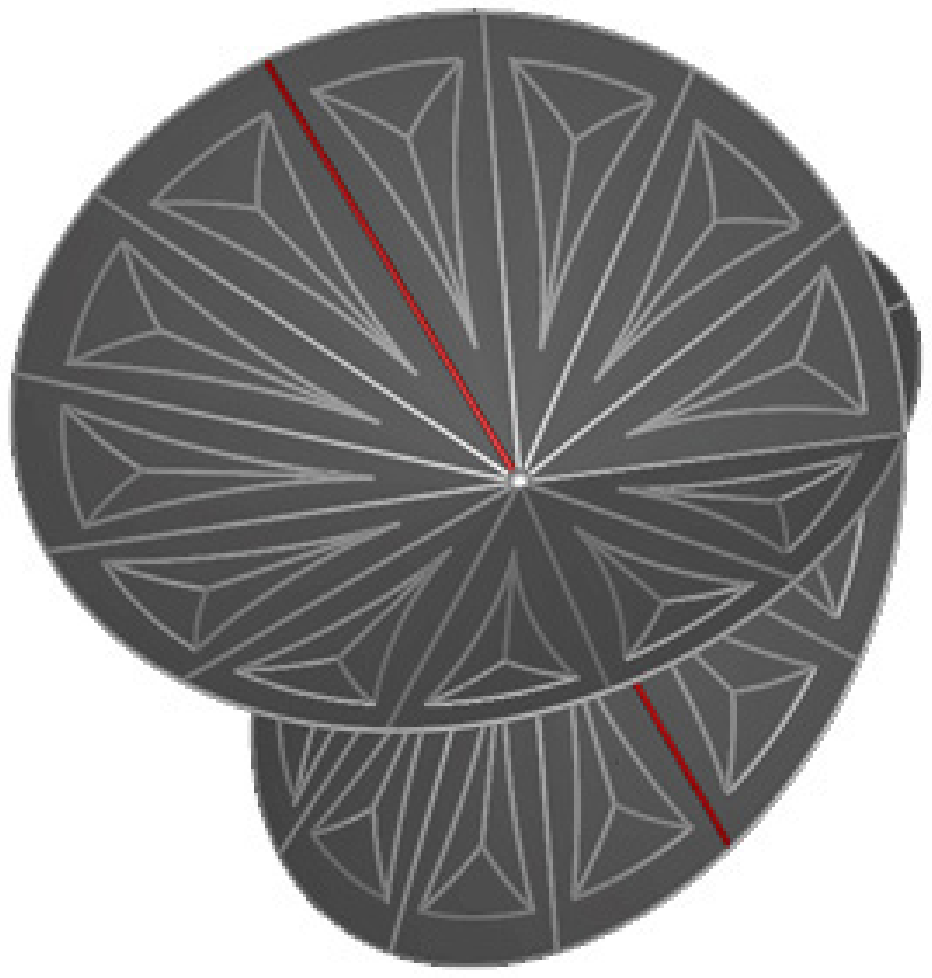,width=50mm,height=50mm}
{$W^{-}$ boson on the particle frame.
\label{GU_fig30}}
{$W^{-}$ boson's antiparticle on the particle frame (which is equivalent to the $W^{+}$ boson).
\label{GU_fig31}}

\item By the Boson Selection Rule $\mathcal{B}$ - \,$Z^{0}$, $\bar{Z}^{0}$, $W^{+}$, $\bar{W}^{+}$, $W^{-}$,
$\bar{W}^{-}$ are bosons.

\item By the Spin Rule $\mathcal{D}$, the spin of all \,$Z^{0}$, $\bar{Z}^{0}$, $W^{+}$, $\bar{W}^{+}$, $W^{-}$,
$\bar{W}^{-}$\, bosons is $1$.

\item By the Electric Charge Rule $\mathcal{E}$, the electric charge of \,$Z^{0}$, $\bar{Z}^{0}$\, is \,$0$;\,the electric charge of
\,$W^{+}$, $\bar{W}^{-}$\, is \,$+1$;\,the electric charge of \,$W^{-}$, $\bar{W}^{+}$\, is \,$-1$.

\item By the Weak Isospin Rule $\mathcal{F}$, the weak isospin of \,$Z^{0}$, $\bar{Z}^{0}$\, is \,$0$;\,the weak isospin of
\,$W^{+}$, $\bar{W}^{-}$\, is \,$+1$;\,the weak isospin of \,$W^{-}$, $\bar{W}^{+}$\, is \,$-1$.

\item By the Strong Charge Rule $\mathcal{G}$, the strong charge of each of these bosons is neutral with \,$N_{c} = 1$ for each.

\item From experimental observations, the rest masses of these bosons are \,$M_{_{\!Z}} \sim 91.1875\,GeV$\, and
\,$M_{_{\!W}} \sim 80.398\,GeV$ \cite{LEP,SLD}, which can be attributed by the Mass Rule $\mathcal{H}$ to the Higgs-Kibble
mechanism \cite{Hig1,Hig2,Hig3,Hig4,Kib,Kib1,Djo1,Bed}.

\item In agreement with the Helicity Rule $\mathcal{K}$, \,$Z^{0}$, $\bar{Z}^{0}$, $W^{+}$, $\bar{W}^{+}$, $W^{-}$,
$\bar{W}^{-}$\, can be theoretically observed with both left-handed and right-handed helicities.

\item Note that by the Equivalence Rule $\mathcal{I}$,
\begin{itemize}
\item[a)] $Z^{0}$ and $\bar{Z}^{0}$ represent the same kind of particle.
\item[b)] $W^{+}$ and $\bar{W}^{-}$ represent the same kind of particle.
\item[c)] $W^{-}$ and $\bar{W}^{+}$ represent the same kind of particle.
\end{itemize}

\item By the \textbf{CP} Transformation Rule $\mathcal{L}$ - \,$Z^{0}_{_{\!L}} \Leftrightarrow Z^{0}_{_{\!R}}$,\,
\,$\bar{Z}^{0}_{_{\!L}} \Leftrightarrow \bar{Z}^{0}_{_{\!R}}$;\,
\,$W^{+}_{_{\!L}} \Leftrightarrow W^{-}_{_{\!R}}$,\, \,$W^{-}_{_{\!L}} \Leftrightarrow W^{+}_{_{\!R}}$;\, \\
\,$\bar{W}^{+}_{_{\!L}} \Leftrightarrow \bar{W}^{-}_{_{\!R}}$,\,
\,$\bar{W}^{-}_{_{\!L}} \Leftrightarrow \bar{W}^{+}_{_{\!R}}$.\,
\\

\hskip -0.725truecm
In Table $3$ we represent the values of the related physical magnitudes of the vector bosons. \\

\begin{center}
\begin{normalsize}
\TABLE[h]{
\begin{tabular}{|l|l|l|l|l|l|l|l|}
\hline\
\hskip 0.6truecm \textbf{Bosons} & \hskip 0.1truecm \textbf{Symbol} & \hskip 0.1truecm \textbf{Associated force field}
& \hskip 0.1truecm \textbf{Charge} & \hskip 0.1truecm \textbf{Weak isospin} & \hskip 0.1truecm \textbf{Mass (GeV)}  \\
\hline
\hline %
\hskip 0.4truecm vector boson &  & \hskip 1.1truecm neutral carrier &  &  &  \\
\hskip 0.5truecm $Z^{0}$ particle & \hskip 0.6truecm $Z^{0}$ & \hskip 0.8truecm of the weak force & \hskip 0.7truecm $0$
& \hskip 1.2truecm $0$ & \hskip 0.5truecm $\sim\,91.1875 $  \\
\hline %
\hskip 0.4truecm vector boson &  & \hskip 1.1truecm neutral carrier &  &  &   \\
\hskip 0.1truecm $Z^{0}$ antiparticle & \hskip 0.5truecm $\bar{\,Z^{0}}$ & \hskip 0.8truecm of the weak force &  \hskip 0.7truecm $0$
& \hskip 1.2truecm $0$ & \hskip 0.5truecm $\sim\,91.1875 $  \\
\hline %
\hskip 0.4truecm vector boson &  & \hskip 1.0truecm positive carrier &  &  &  \\
\hskip 0.4truecm $W^{+}$ particle & \hskip 0.5truecm $W^{+}$ & \hskip 0.8truecm of the weak force & \hskip 0.5truecm $+\,1$
& \hskip 0.9truecm $+\,1$ & \hskip 0.5truecm $\sim\,80.398 $  \\
\hline %
\hskip 0.4truecm vector boson &  & \hskip 1.0truecm negative carrier &  &  &  \\
\hskip 0.1truecm $W^{+}$ antiparticle & \hskip 0.25truecm $\bar{\,\,\,\,W^{+}}$ & \hskip 0.8truecm of the weak force & \hskip 0.5truecm $-\,1$
& \hskip 0.9truecm $-\,1$ & \hskip 0.5truecm $\sim\,80.398 $  \\
\hline %
\hskip 0.4truecm vector boson &  & \hskip 1.0truecm negative carrier &  &  &  \\
\hskip 0.4truecm $W^{-}$ particle & \hskip 0.5truecm $W^{-}$ & \hskip 0.8truecm of the weak force & \hskip 0.5truecm $-\,1$
& \hskip 0.9truecm $-\,1$ & \hskip 0.5truecm $\sim\,80.398 $  \\
\hline %
\hskip 0.4truecm vector boson &  & \hskip 1.0truecm positive carrier &  &  &  \\
\hskip 0.1truecm $W^{-}$ antiparticle & \hskip 0.25truecm $\bar{\,\,\,\,W^{-}}$ & \hskip 0.8truecm of the weak force & \hskip 0.5truecm $+\,1$
& \hskip 0.9truecm $+\,1$ & \hskip 0.5truecm $\sim\,80.398 $  \\
\hline %
\end{tabular}
\caption{The vector bosons in the SM.}}
\end{normalsize}
\end{center}

\item (Yang-Mills Weak Field Equations \cite{Dhar2, Ho}) The components of the weak field are classified via the following correspondence.
\begin{itemize}
\item[a)] The neutral component (\fig{GU_fig32}):
{\setlength\arraycolsep{2pt}
\bea \label{QuBo1}
F_{0}^{(0)}\,& \rightarrow &\,\Lb +(\underline{0},\,\rho^{2})\varphi_{i}R(\beta),\,\beta + \gamma \Rb\,;
\nonumber \\
F_{1}^{(0)}\,& \rightarrow &\,\Lb +(\underline{0},\,\sigma\rho^{2})\varphi_{i}R(\beta),\,\beta + \gamma \Rb\,;
\nonumber \\
F_{2}^{(0)}\,& \rightarrow &\,\Lb -(\underline{0},\,\rho^{2})R(\gamma)\varphi_{i},\,\beta + \gamma \Rb\,;
\nonumber \\
F_{3}^{(0)}\,& \rightarrow &\,\Lb -(\underline{0},\,\sigma\rho^{2})R(\gamma)\varphi_{i},\,\beta + \gamma \Rb\,.
\eea}

\item[b)] The positive component (\fig{GU_fig33}):
{\setlength\arraycolsep{2pt}
\bea \label{QuBo2}
F_{0}^{(+)}\,& \rightarrow &\,\Lb +(\underline{3},\,\rho)\varphi_{i}R(\beta),\,\beta + \gamma \Rb\,;
\nonumber \\
F_{1}^{(+)}\,& \rightarrow &\,\Lb +(\underline{3},\,\rho^{2})\varphi_{i}R(\beta),\,\beta + \gamma \Rb\,;
\nonumber \\
F_{2}^{(+)}\,& \rightarrow &\,\Lb +(\underline{3},\,\rho)\varphi_{i}R(\gamma),\,\beta + \gamma \Rb\,;
\nonumber \\
F_{3}^{(+)}\,& \rightarrow &\,\Lb +(\underline{3},\,\rho^{2})\varphi_{i}R(\gamma),\,\beta + \gamma \Rb\,.
\eea}

\item[c)] The negative component (\fig{GU_fig34}):
{\setlength\arraycolsep{2pt}
\bea \label{QuBo3}
F_{0}^{(-)}\,& \rightarrow &\,\Lb -(\underline{3},\,\sigma\rho^{2})R(\beta)\varphi_{i},\,\beta + \gamma \Rb\,;
\nonumber \\
F_{1}^{(-)}\,& \rightarrow &\,\Lb -(\underline{3},\,\sigma\rho)R(\beta)\varphi_{i},\,\beta + \gamma \Rb\,;
\nonumber \\
F_{2}^{(-)}\,& \rightarrow &\,\Lb -(\underline{3},\,\sigma\rho^{2})R(\gamma)\varphi_{i},\,\beta + \gamma \Rb\,;
\nonumber \\
F_{3}^{(-)}\,& \rightarrow &\,\Lb -(\underline{3},\,\sigma\rho)R(\gamma)\varphi_{i},\,\beta + \gamma \Rb\,.
\eea}

\begin{center}
\FIGURE[h]{
\centerline{\epsfig{file=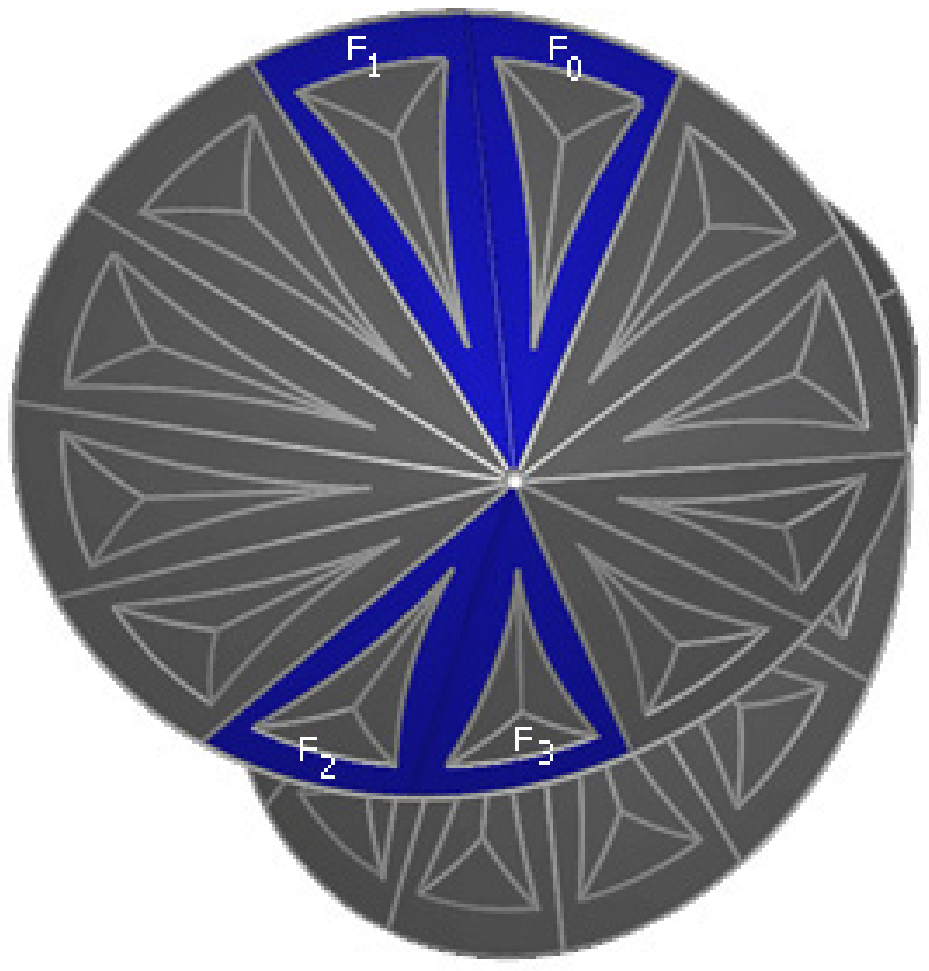,width=50mm,height=50mm}}
\caption{$Z^{0}$ boson and the neutral component of the weak field.}
\label{GU_fig32}}
\end{center}

\DOUBLEFIGURE[h]{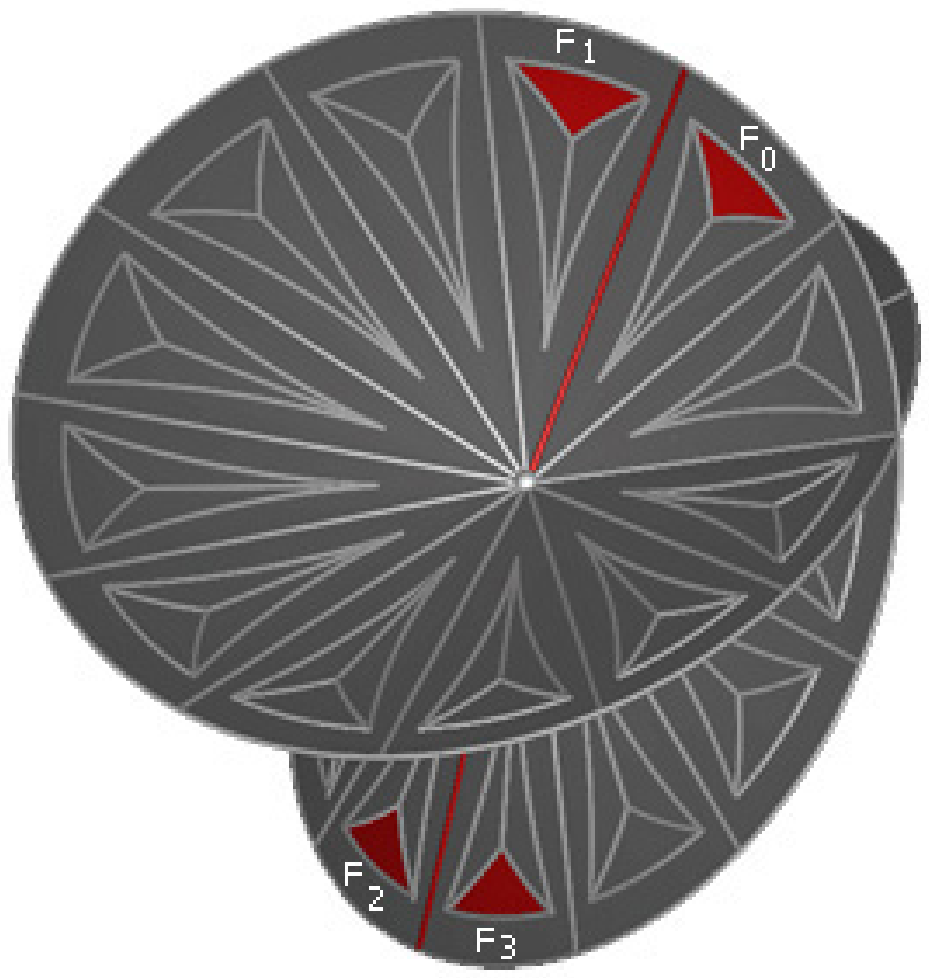,width=50mm,height=50mm}{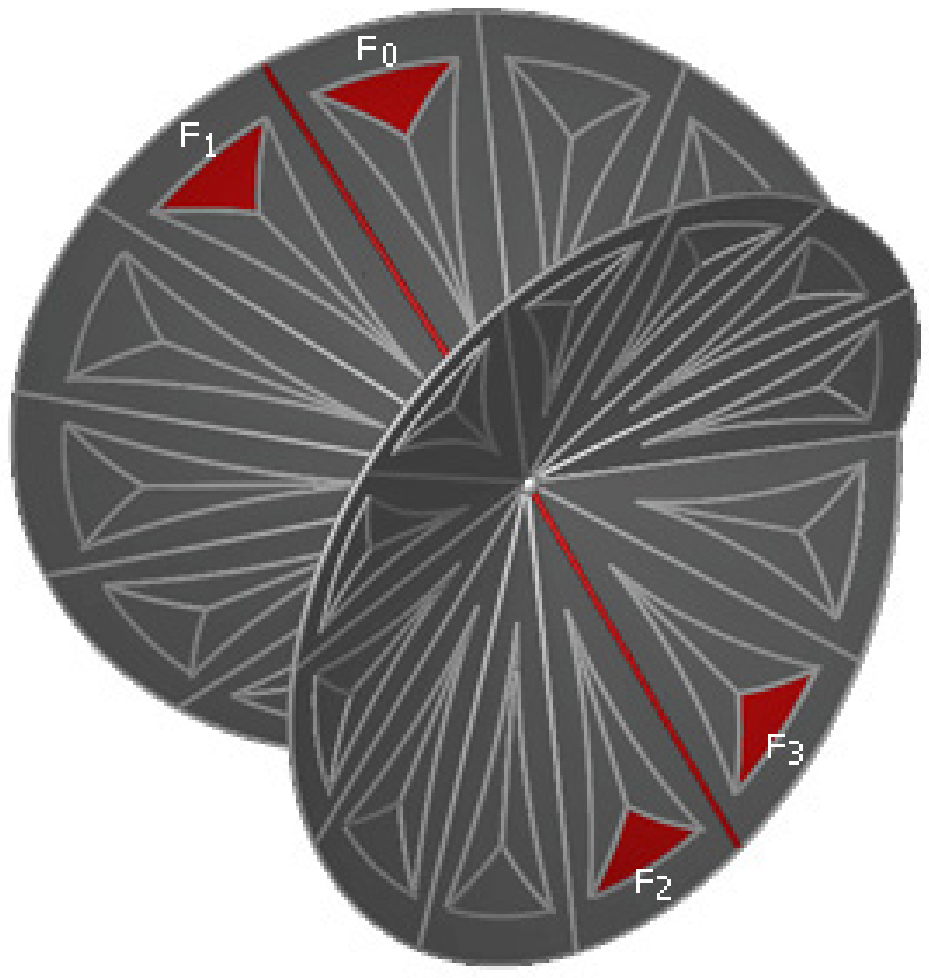,width=50mm,height=50mm}
{$W^{+}$ boson and the positive component of the weak field.
\label{GU_fig33}}
{$W^{-}$ boson and the negative component of the weak field.
\label{GU_fig34}}
\end{itemize}

Three covariant $4$-vectors
\begin{displaymath}
\Lb F_{0}^{(w)},\,F_{1}^{(w)},\,F_{2}^{(w)},\,F_{3}^{(w)} \Rb\,\,\,\,\,\,\,\,\,\,\mbox{where}\,\,\,\,\,w\,=\,0,+,-\,
\end{displaymath}
in \eq{QuBo1}, \eq{QuBo2} and \eq{QuBo3} are from the tensors \,$F_{\mu\nu}^{(0)}$, $F_{\mu\nu}^{(+)}$ and $F_{\mu\nu}^{(-)}$\,
whereby three electromagnetic-type fields are defined which, in turn, correspond to three vector bosons \,$Z^{0}$ and $W^{\pm}$.
We also have three electric and three magnetic fields corresponding to the antisymmetric tensors \,$F_{\mu\nu}^{(0)}$,
$F_{\mu\nu}^{(+)}$ and $F_{\mu\nu}^{(-)}$\, given by the following correspondence:
\vskip0.5truecm
\begin{displaymath}
\begin{tabular}{|c|c|c|c|}
\hline
$F_{00}^{(w)}$ & $F_{01}^{(w)}$ & $F_{02}^{(w)}$ & $F_{03}^{(w)}$ \\
\hline
$F_{10}^{(w)}$ & $F_{11}^{(w)}$ & $F_{12}^{(w)}$ & $F_{13}^{(w)}$ \\
\hline
$F_{20}^{(w)}$ & $F_{21}^{(w)}$ & $F_{22}^{(w)}$ & $F_{23}^{(w)}$ \\
\hline
$F_{30}^{(w)}$ & $F_{31}^{(w)}$ & $F_{32}^{(w)}$ & $F_{33}^{(w)}$ \\
\hline
\end{tabular}\,\,\,\,\,\,\,\longrightarrow\,\,\,\,\,\,\,
\begin{tabular}{|c|c|c|c|}
\hline
$0$ & $-E^{1(w)}$ & $-E^{2(w)}$ & $-E^{3(w)}$ \\
\hline
$E^{1(w)}$ & $0$ & \,\,\,\,$H^{3(w)}$ & $-H^{2(w)}$ \\
\hline
$E^{2(w)}$ & $-H^{3(w)}$ & $0$ & \,\,\,\,$H^{1(w)}$ \\
\hline
$E^{3(w)}$ & \,\,\,\,$H^{2(w)}$ & $-H^{1(w)}$ & $0$ \\
\hline
\end{tabular}\,\,\,\,\,\,\,\,\,\,\mbox{where}\,\,\,\,\,w\,=\,0,+,-\,.
\end{displaymath}
\end{enumerate}

\vskip2truecm

\subsection{The Weak Gauge Group}
The gauge group $G_{w}$ for the weak field is defined by specifying its generators as in \fig{GU_fig35}. The row and column labels
\begin{displaymath}
\begin{tabular}{|c|c|}
\hline
$\pm\rho$  \\
\hline
$\pm\sigma\rho$  \\
\hline
\end{tabular}\,\,,\,\,\,\,\,
\begin{tabular}{|c|c|}
\hline
$\pm\rho^{2}$  \\
\hline
$\pm\sigma\rho^{2}$  \\
\hline
\end{tabular}
\end{displaymath}
in \fig{GU_fig35} specify the Schr\"{o}dinger discs of the three bosons \,($Z^{0}$, $W^{\pm}$) and the weak force as defined by
\eq{QuBo1}, \eq{QuBo2} and \eq{QuBo3}, respectively. The first component of the row and column labels consists of the weak
charge; the second component specifies how the weak gauge group will be embedded in the strong gauge group \cite{Dhar2}.
\begin{center}
\FIGURE[h]{
\centerline{\epsfig{file=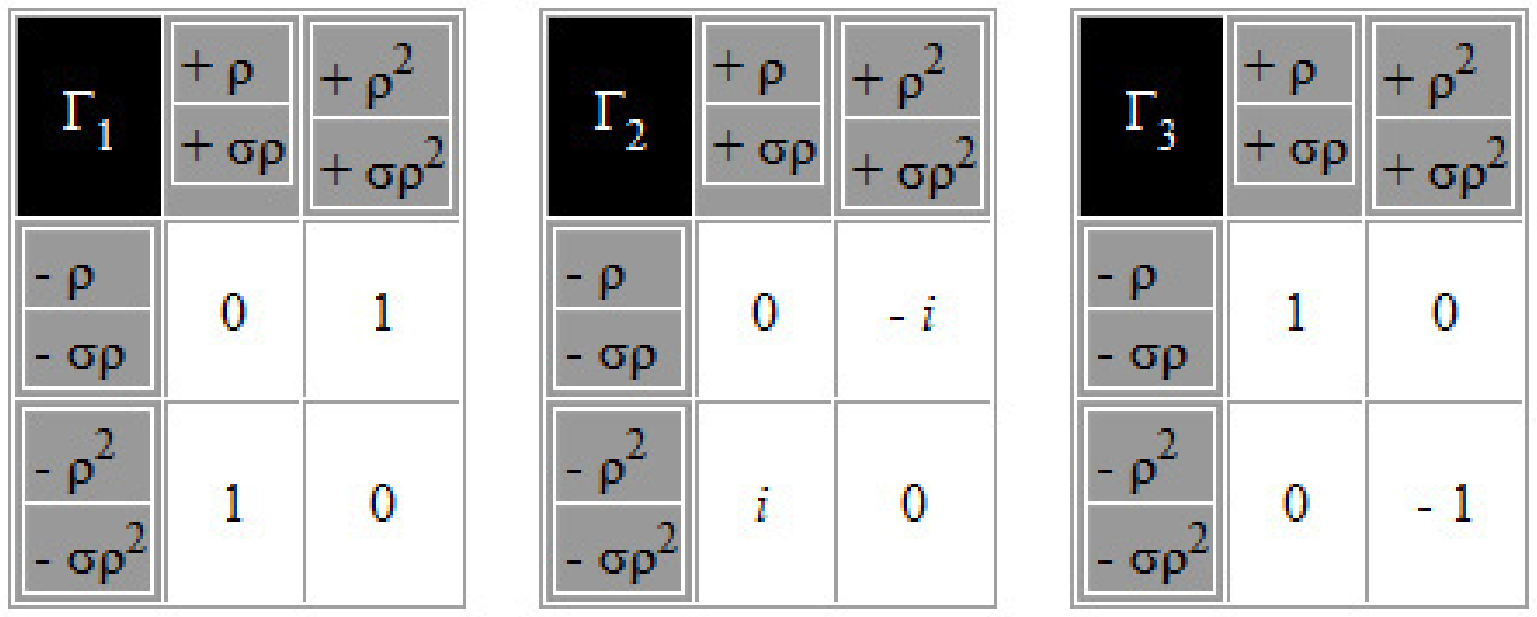,width=94mm,height=42mm}}
\caption{The generators of the weak gauge group \,$G_{w} = SU(2)$\,.}
\label{GU_fig35}}
\end{center}

The three \emph{observable} vector bosons are defined as:\, \,$[Z^{0}]\,=\,\Gamma_{1}$,\, \,$[W^{+}]\,=\,\Gamma_{2}$\, and
\,$[W^{-}]\,=\,\Gamma_{3}$.\, Each of these bosons is regarded as a superposition of the weak charge - weak anticharge
labels of its row and column multiplied by a complex number viewed on the $z$-plane. Since $t = z^{2}$, a rotation by an
angle $\theta$ around the origin of the $z$-plane corresponds to a rotation by an angle $2\theta$ around the branch point
of the \textit{t}-Riemann surface. In particular, we notice that the multiplication by \,$\pm{i}$ and $\pm{1}$ on the
$z$-plane corresponds to rotations of the $t$-Riemann surface by $\pm{180}$ and $\pm{360}$ degrees around the branch point,
respectively. This means that for the three observable vector bosons \,$[Z^{0}]$,\, \,$[W^{+}]$\, and \,$[W^{-}]$,\,
the rays defining the corresponding three vector bosons $Z^{0}$, $W^{+}$ and $W^{-}$ are permuted amongst themselves
(and not any other rays) on the particle frame. Otherwise stated, \,$[Z^{0}]$,\, \,$[W^{+}]$\, and \,$[W^{-}]$\,
correspond to superpositions of particle frames for $Z^{0}$, $W^{+}$ and $W^{-}$ (see \fig{GU_fig36}).

\begin{center}
\FIGURE[h]{
\centerline{\epsfig{file=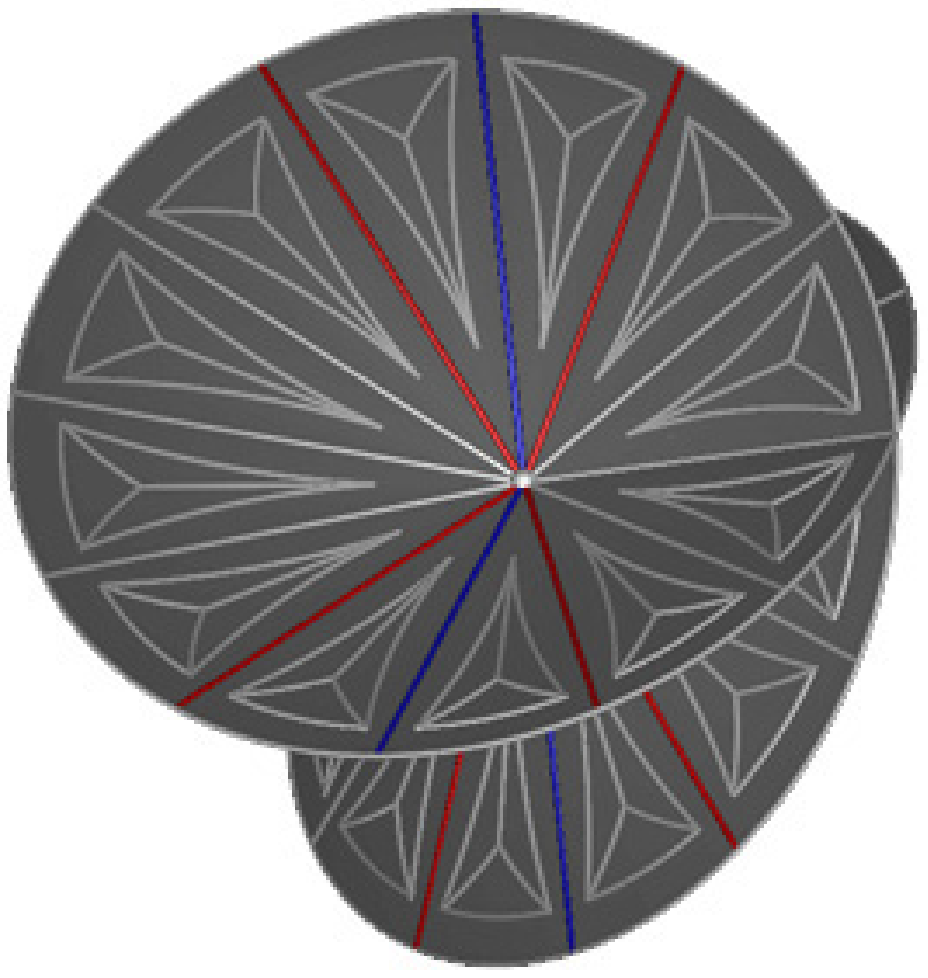,width=50mm,height=50mm}}
\caption{The observable vector bosons \,$[Z^{0}]$,\, \,$[W^{+}]$\, and \,$[W^{-}]$.}
\label{GU_fig36}}
\end{center}

The weak gauge group acts on the bosons \,$[Z^{0}]$,\, \,$[W^{+}]$\, and \,$[W^{-}]$\, by conjugation viewed on the
$z$-plane: \,$\Gamma_{i}^{-1}\Gamma_{j}\Gamma_{i}$\,\, for \,$i,j = 1,2,3$.\, Again, since \,$t = z^{2}$,\, the rotation by the angle $\theta$ around the origin of the $z$-plane corresponds to the
rotation by the angle $2\theta$ around the branch point of the \textit{t}-Riemann surface, and the multiplication by
$\pm{i}$ and $\pm{1}$ on the $z$-plane corresponds to rotations of the \textit{t}-Riemann surface by $\pm{180}$ and
$\pm{360}$ degrees around the branch point, respectively. This means that for any weak gauge transformation, the rays
defining the corresponding three vector bosons $Z^{0}$, $W^{+}$ and $W^{-}$ are permuted amongst themselves (and not any
other rays) on the particle frame.

\subsection{The Weinberg angle}
The Weinberg angle $\Theta_{W}$ is a parameter that gives a relationship between the masses of $W^{+}$, $W^{-}$ and $Z^{0}$
bosons \,$\Lb M_{_{\!Z}} \cos{\Theta_{W}} = M_{_{\!W}} \Rb$,\, as well as the ratio of the weak $Z^{0}$ mediated interaction,
called its mixing. Indeed, from \fig{GU_fig32}, \fig{GU_fig33}, \fig{GU_fig34} and \fig{GU_fig36}, it is apparent that the
components of the weak $Z^{0}$ field mix with the components of the weak $W^{\pm}$ fields, and the angle subtended by the
mixing Schr\"{o}dinger discs on the particle frame is exactly $\pi/6$ radians or $30$ degrees, as shown in \fig{GU_fig37}.
Hence, \,$\Theta_{W} = 30$\, degrees on the particle frame. This is in good agreement with the SLAC experiment \cite{Mix}
which estimates \,$\sin{{\!}^{2}(\Theta_{W})} = 0.2397$,\, i.e. \,$\Theta_{W} = 29.3137$\, degrees (this is a ``running''
value, depending on the momentum at which it is measured, with a significance of $6$ standard deviations). In our
consideration, the Weinberg angle appears to be a measure of the strength of the weak force.

\begin{center}
\FIGURE[h]{
\centerline{\epsfig{file=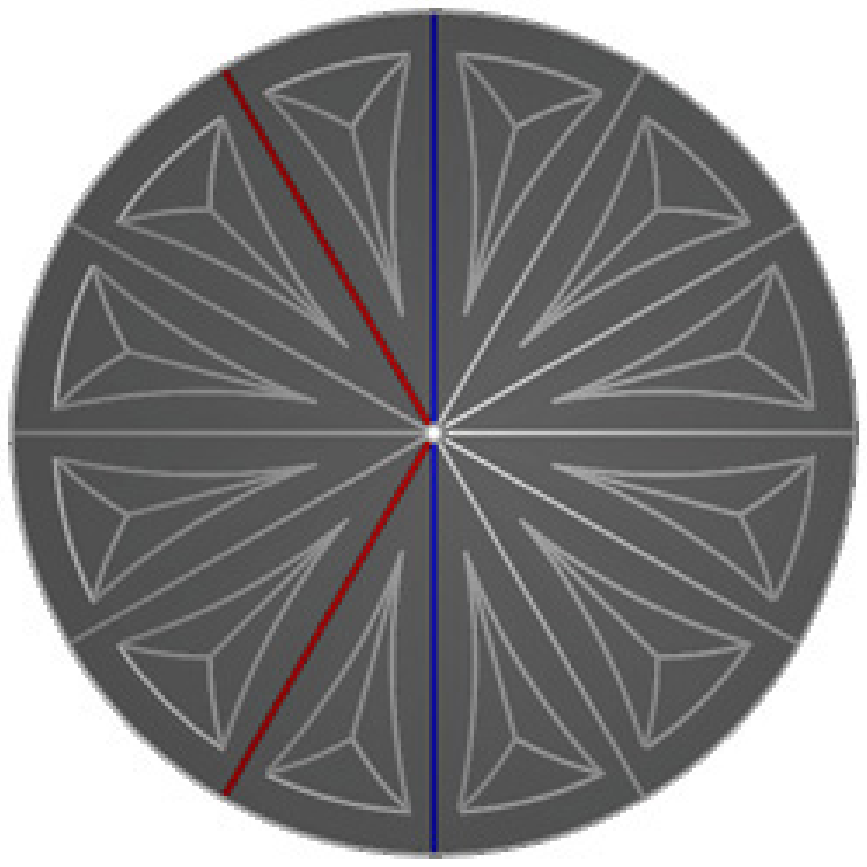,width=50mm,height=50mm}}
\caption{The components of the weak $Z^{0}$ field subtends an angle \,$\Theta_{W} = 30$\, degrees with the $W^{+}$ and $W^{-}$
components of the weak fields.}
\label{GU_fig37}}
\end{center}

\subsection{The Cabibbo angle}
The Cabibbo angle is a measure of the probability that one flavor of quark will change into other flavors under the action
of the weak force. Cabibbo's idea \cite{Cab} originated from a need to explain two observed phenomena at the time:
\begin{quote}
1) the transitions \,$u \leftrightarrow d$,\, \,$e \leftrightarrow \nu_{e}$\, and \,$\mu \leftrightarrow \nu_{\mu}$\, had
similar amplitudes (or probabilities); \\
2) the transitions with a change in strangeness \,$\Delta S = 1$\, had smaller probabilities than those with \,$\Delta S = 0$.
(Otherwise stated, the strangeness-changing coupling is weaker than the strangeness-preserving coupling.)
\end{quote}
The solution consisted of postulating weak universality to resolve the first issue, along with a mixing angle (the Cabibbo
angle $\theta_{c}$) between $d$ and $s$ quarks to resolve the second issue.
This solution was based on a suggestion of a superposition state $d\,'$ of the down quark $d$ and strange quark $s$ with the
doublet
\beq \label{QuBo4}
\left( \begin{array}{cc}
u \\
\,d\,'
\end{array} \right)\,=\,
\left( \begin{array}{c}
u \\
d\!\cdot\!\cos{\theta_{c}}\,+\,s\!\cdot\!\sin{\theta_{c}}
\end{array} \right)\,,
\eeq
where $d$ and $s$ quarks are said to form a weak interaction eigenstate. But this was only for the first generation of quarks. For the three known generations of quarks in the SM, the basic idea led to the general model of defining a
unitary rotation $V_{CKM}$\footnote{$CKM$: Cabibbo-Kobayashi-Maskawa} between the weak eigenstates of the quarks and the
flavor eigenstates
\beq \label{QuBo5}
\left( \begin{array}{c}
\,d\,' \\
\,s\,' \\
\,b\,'
\end{array} \right)\,=\,
V_{CKM}\left( \begin{array}{c}
d \\
s \\
b
\end{array} \right)\,.
\eeq \label{QuBo6}
Let us write the matrix $V_{CKM}$ explicitly as
\beq
V_{CKM}\,=\,\left( \begin{array}{ccc}
V_{ud} & V_{us}  & V_{ub}  \\
V_{cd} & V_{cs}  & V_{cb}  \\
V_{td} & V_{ts}  & V_{tb}  \\
\end{array} \right)\,,
\eeq
where $|V_{ij}|^2$ represents the probability that a quark of flavor $i$ decays to a quark of flavor $j$.
A standard parameterization of the $CKM$ matrix uses three Euler angles ($\theta_{12}$, $\theta_{23}$, $\theta_{13}$)
and one \textbf{CP}-violating phase $\delta_{13}$ \cite{ChaKe} such that
\beq  \label{QuBo7}
V_{CKM}\,=\,\left( \begin{array}{ccc}
c_{12}c_{13} & c_{13}s_{12} & \,\,\,\,s_{13}\,e^{i\delta_{13}} \\
- c_{23}s_{12} - c_{12}s_{13}s_{23}\,e^{i\delta_{13}} & c_{12}c_{23} - s_{12}s_{13}s_{23}\,e^{i\delta_{13}} & c_{13}s_{23} \\
s_{12}s_{23} - c_{12}c_{23}s_{13}\,e^{i\delta_{13}} & - c_{12}s_{23} - c_{23}s_{12}s_{13}\,e^{i\delta_{13}} & c_{13}c_{23}
\end{array} \right)\,
\eeq
where $c_{ij} = \cos{\theta_{ij}}$\, , \,$s_{ij} = \sin{\theta_{ij}}$   and $\theta_{12}$ is the Cabibbo angle. The currently best
known values for the standard parameters are \,$\theta_{12} = 13.04^{\circ} \pm 0.05^{\circ}$,\,
\,$\theta_{13} = 0.201^{\circ} \pm 0.011^{\circ}$,\, \,$\theta_{23} = 2.38^{\circ} \pm 0.06^{\circ}$\, and
\,$\delta_{13} = 1.20 \pm 0.08$ \cite{Wol}.

It is well known that the decay of hadrons by the weak interaction can be viewed as a process of decay of their constituent quarks. In the decay process, a quark of charge \,$+2/3$ $(u,\,c,\,t)$\, is always
transformed into a quark of charge \,$-1/3$ $(d,\,s,\,b)$\, and vice versa. This is because the transformation proceeds by
the exchange of the charged $W$ bosons, which must alter the charge by one unit. The general trend is that the quarks will
decay to the most massive quark possible, leading to the pattern
\,\,$t \rightarrow b \rightarrow c \rightarrow s \rightarrow u \leftrightarrow d$\,\, \cite{Roh}. Hypothetically, we can
consider nine possible quark decays of \,$d$, $s$, $b$\, quarks into \,$u$, $c$, $t$\, quarks:
\,$d \rightarrow u$,\, \,$d \rightarrow c$,\, \,$d \rightarrow t$,\, \,$s \rightarrow u$,\, \,$s \rightarrow c$,\,
\,$s \rightarrow t$,\, \,$b \rightarrow u$,\, \,$b \rightarrow c$,\, \,$b \rightarrow t$\, \footnote{We are only interested
in these nine decay modes since the $CKM$ matrix operates only on \,$d,\,s,\,b$\, quarks.}. However, from the pattern above,
there are only four out of these nine decays that can actually occur, namely: \,$d \rightarrow u$,\, \,$s \rightarrow u$,\,
\,$b \rightarrow u$,\, \,$b \rightarrow c$.\, The other five decays are not permitted by the pattern because of the violation
of mass conservation. Hence, one can conjecture that the probability of occurrence of such a quark decay mode involving the
weak interaction is $4/9$. We have already defined the Weinberg angle that measures the strength of the weak force as
\,$\Theta_{W} = \pi/6$\, on the particle frame. We can now define the Cabibbo angle by the formula
\beq \label{QuBo8}
\theta_{c} = (4/9)\Theta_{W}.
\eeq
Thus $\theta_c $ is approximately $13.33^{\circ}$ which is close to the experimentally observed value. The small difference between the theoretical and experimental values may be caused by the fact that the four quark decay probabilities are not exactly equal.

Returning to the matrix $V_{CKM}$, we notice that it must be unitary. This leads to the following two constraints
\beq \label{QuBo9}
\sum_{k}|V_{ik}|^2 = 1\,,
\eeq
for all $i$, and
\beq \label{QuBo10}
\sum_{k}V_{ik}V_{jk}^{*} = 0\,,
\eeq
for all $i$ and $j$.
The first constraint means that the sum of all couplings of any of green (charge $2/3$) quarks to all yellow (charge $1/3$)
quarks is the same for all three generations (see \fig{GU_fig38}). This relation is called weak universality (noted at the
beginning) which is interpreted in our context. Theoretically, it is a consequence of the fact that all $SU(2)$ doublets
couple with the same strength to the vector bosons of the weak interaction.
\begin{center}
\FIGURE[h]{
\centerline{\epsfig{file=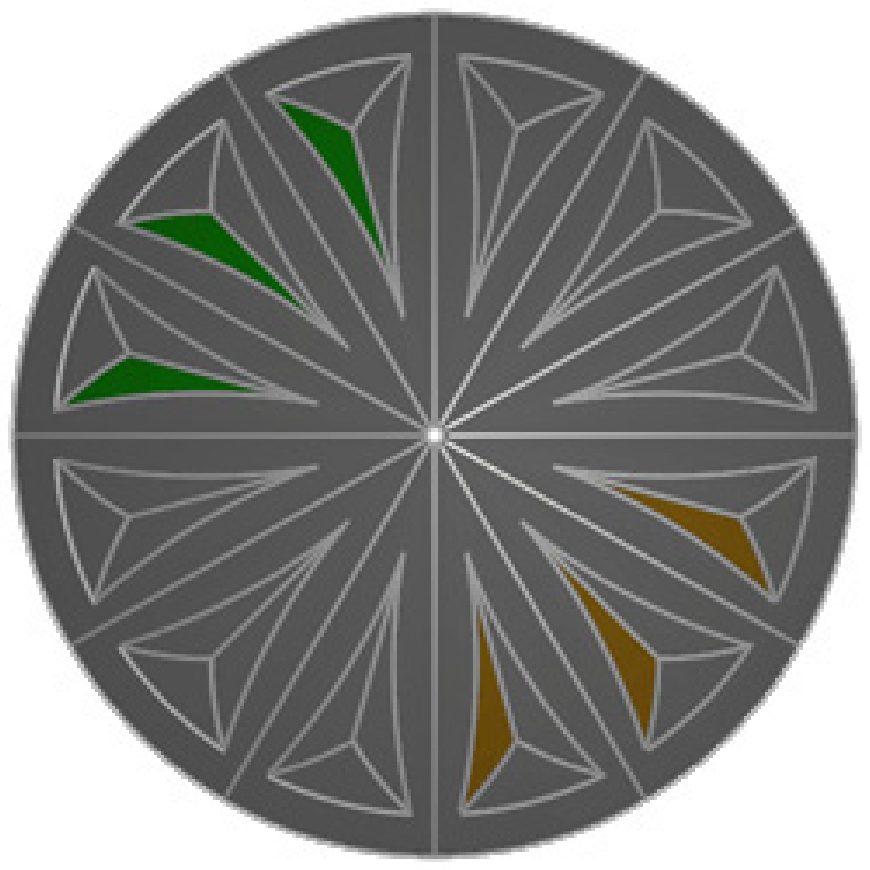,width=50mm,height=50mm}}
\caption{The superposition of three generations of quarks on the particle frame.}
\label{GU_fig38}}
\end{center}
The second constraint has a very nice geometrical interpretation. For any fixed $k$ and different $i$ and $j$, this is a
constraint on three complex numbers which states that these numbers form sides of a triangle in the complex plane. There
are six choices of $i$ and $j$ (three independent), and hence six such triangles, each of which is called a unitary
triangle. The orientation of the triangles depends on the phases of the quark fields. Their shapes can be very different,
but they all have the same area which can be related to the \textbf{CP} violating phase. The area vanishes for the specific
parameters in the SM for which there would be no \textbf{CP} violation. For our purposes, we can represent six unitary
triangles by three green and three yellow regions that exhibit the charges of the quarks on the particle frame, as
expressed by a superposition of six particle frames shown in \fig{GU_fig38}. Since these regions can never have zero area,
we must have \textbf{CP} violation in weak interactions involving the quarks.
%\vskip1truecm
%\vskip1truecm

\section{Higgs Boson mass from the Particle Frame: comparisons with some mass constraints of the SM and MSSM}
\label{sec:Higgs}
%\vskip1truecm
\subsection{Some constraints on mass values of the SM Higgs Boson}
The Higgs mass in the SM is given by: \,$m_{h_{SM}}^{2} = (1/2)\lambda\,v^{2}$\, where $\lambda$ is the Higgs self-coupling
parameter. The value of the SM Higgs mass has not been predicted yet because $\lambda$ is unknown at present. However, various
theoretical considerations place constraints on the Higgs mass. In contrast, the Higgs couplings to fermions (bosons) are
predicted by the theory to be proportional to the corresponding particle masses (squared-masses). In particular, the SM
Higgs boson is a $\textbf{CP}$-even scalar, and its couplings to gauge bosons, Higgs bosons and fermions are given by
\cite{CaHa}:
\vskip1truecm
{\setlength\arraycolsep{2pt}
\bea \label{Hig1}
g_{hf\bar{f}}\,& = &\,\FFr{m_{f}}{v}\,,\,\,\,\,\,\,\,\,\,\,\,\,\,\,\,\,\,\,\,\,\,\,\,\,\,\,\,\,\,\,\,\,\,\,\,\,\,\,\,
\,\,\,\,\,g_{hVV}\,=\,\FFr{2\,m_{V}^{2}}{v},
\,\,\,\,\,\,\,\,\,\,\,\,\,\,\,\,\,\,\,\,\,\,\,\,\,\,\,\,\,\,\,\,\,\,\,\,
g_{hhVV}\,=\,\FFr{2\,m_{V}^{2}}{v^{2}}\,,
\nonumber \\
g_{hhh}\,& = &\,\FFr{3}{2}\lambda\,v\,=\,\FFr{3\,m_{h_{SM}}^{2}}{v}\,,\,\,\,\,\,\,\,\,\,\,\,
g_{hhhh}\,=\,\FFr{3}{2}\lambda\,=\,\FFr{3\,m_{h_{SM}}^{2}}{v^{2}}\,,
\eea}
\vskip1truecm
where \,$h \equiv h_{SM}$,\, \,$V = W$ or $Z$\, and \,\,$v = 2\,M_{W}/g = 246\,GeV$\,\, (g=gluon). The magnitude $v$ is the
scalar vacuum expectation value which determines the electroweak scale, and is also expressed through the Fermi coupling:
\,$v = \Lb \sqrt{2}\,G_{F} \Rb^{-1/2} = 246\,GeV$.

Nonetheless, various experimental measurements give constraints on the mass of the SM Higgs boson. The mass values from
these constraints are believed to be in the region of its actual existence. We will show three of them consecutively.\,\,\,
a) Summary of electroweak precision measurements at $LEP1$, $LEP2$, $SLC$ and the Tevatron \cite{LEP,SLD} gives the
constraint summarized in \fig{Higgs1}.
\begin{center}
\FIGURE[h]{
\centerline{\epsfig{file=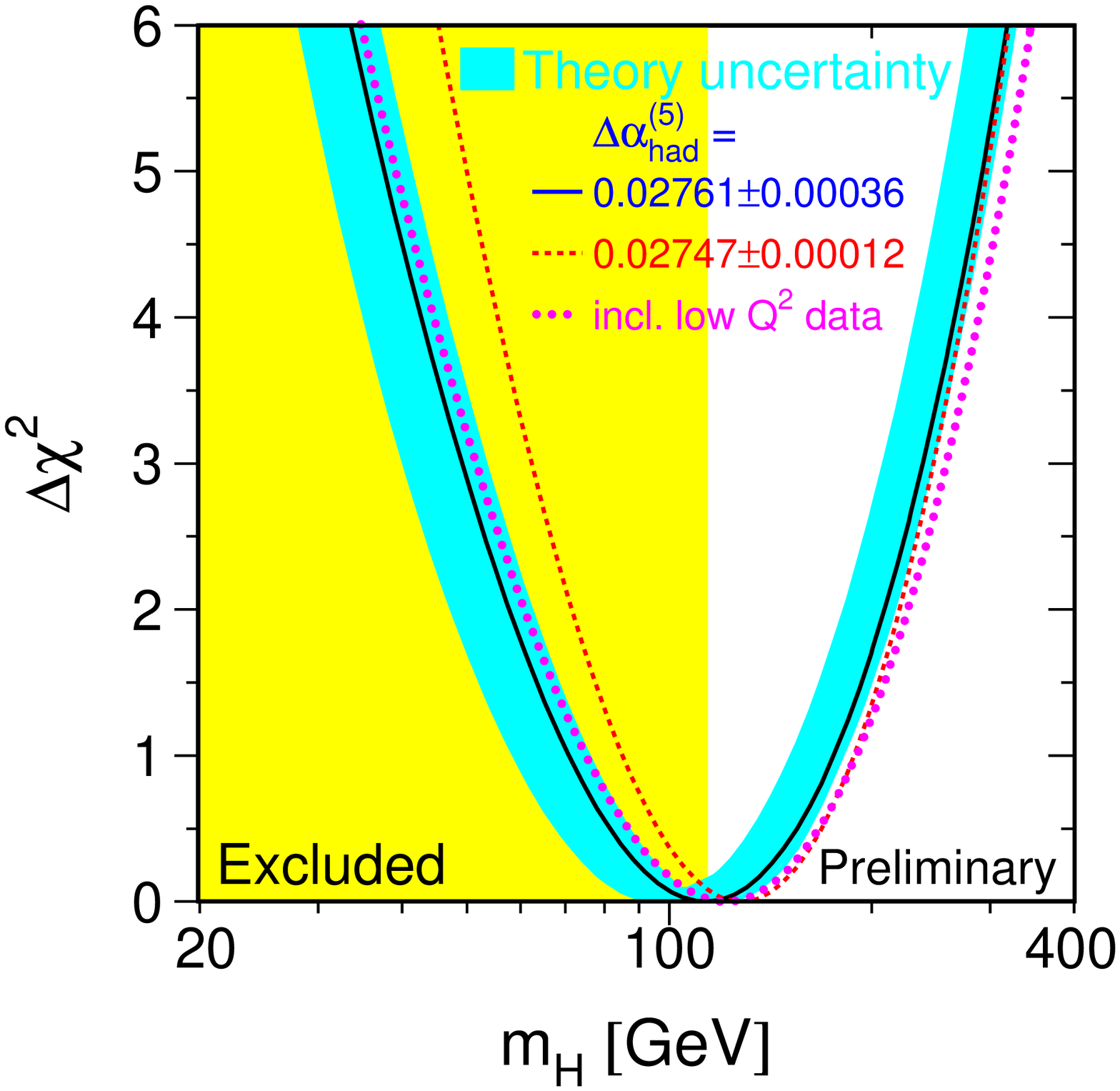,width=65mm,height=65mm}}
\caption{The $\Delta\chi^{2}$ of the fit to the electroweak precision data as a function of $M_{H}$, with the uncertainties
included in \cite{LEP,SLD}.}
\label{Higgs1}}
\end{center}
The central value of the SM Higgs boson mass was estimated to be $M_{H}\,=\,114^{+69}_{-45}\,GeV$.\,\,\, b) Detailed
simulations in \cite{Agu,Des} based on the Higgs-strahlung process (which is one of the main production mechanisms for the
SM Higgs particles in $e^{+}e^{-}$ collisions) shows that the SM Higgs boson can possibly exist at $\sim 120\,GeV$,\,
as shown in \fig{Higgs2}.
\begin{center}
\FIGURE[h]{
\centerline{\epsfig{file=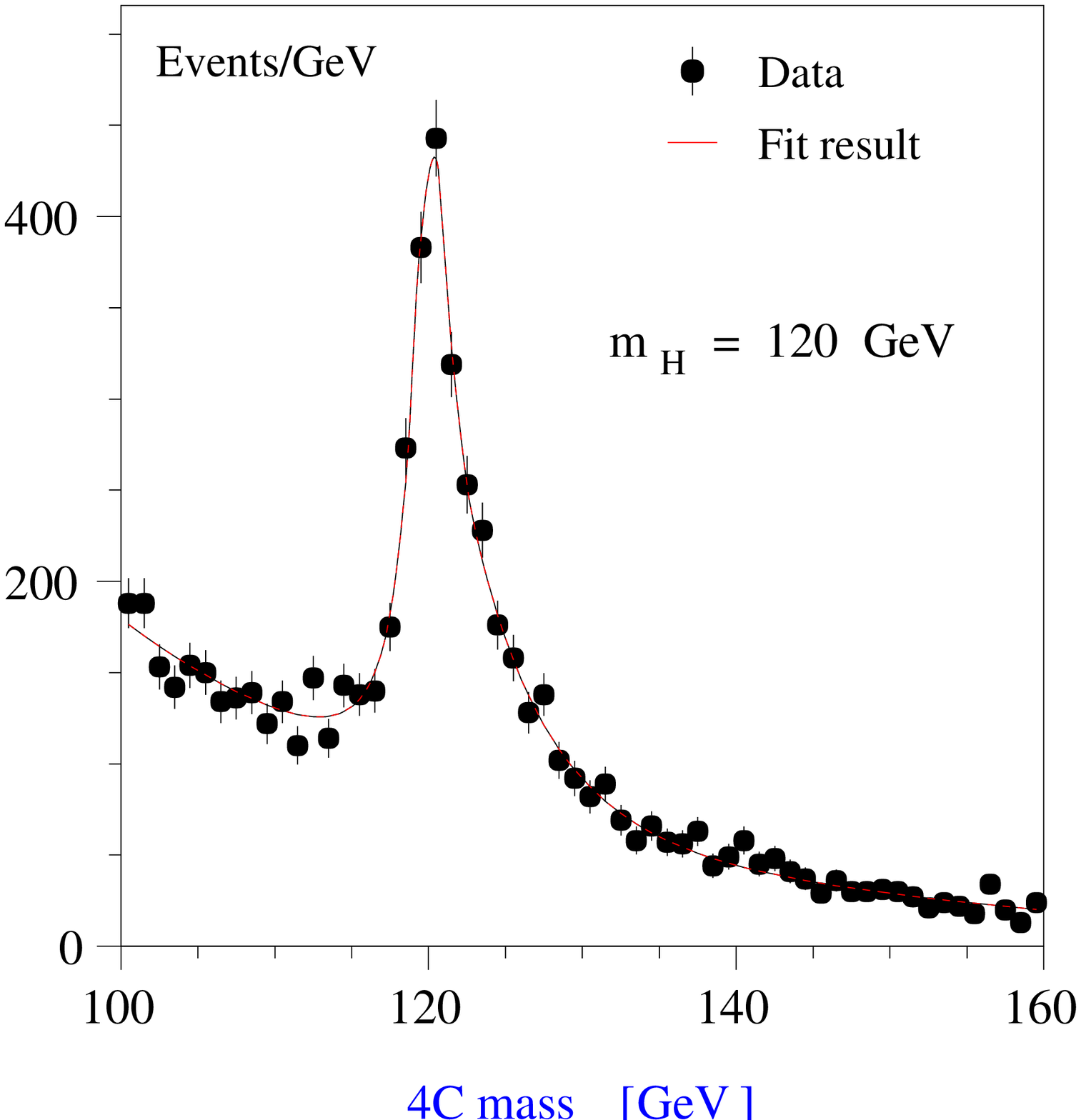,width=65mm,height=65mm}}
\caption{Possible detection for \,$M_{H} = 120\,GeV$\, of the SM Higgs boson at the proposed ILC \cite{Agu}.}
\label{Higgs2}}
\end{center}
\begin{center}
\FIGURE[h]{
\centerline{\epsfig{file=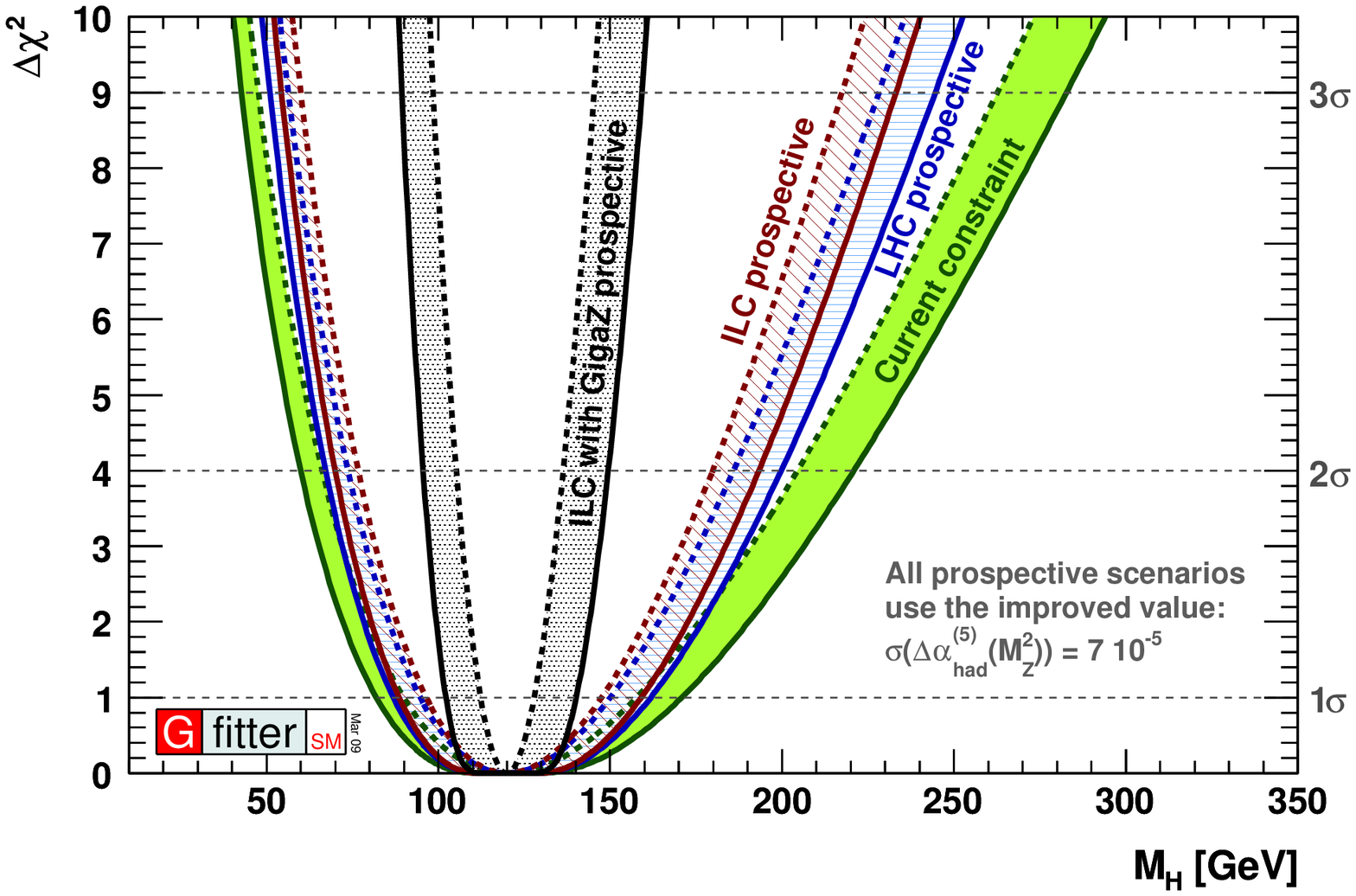,width=90mm,height=80mm}}
\caption{\textsf{Gfitter} constraints on the Higgs boson mass obtained for four future scenarios. Parabolas in
$\Delta\chi^{2}$ are shown with their theoretical error bands. From wider to narrower: current constraint, LHC expectation,
ILC expectations excluding and including the so-called Giga-Z option \cite{Fla}.}
\label{Higgs3}}
\end{center}
c) \fig{Higgs3} projects improvements in the global-fit (\textsf{Gfitter}) constraints on the mass of the SM Higgs boson in
light of measurements to be carried out at the LHC, and measurements that might be made at the proposed International Linear
Collider. From this figure we can conclude that the central value of the SM Higgs boson mass $\sim 120\,GeV$.

\subsection{Some constraints on mass values of the MSSM Higgs Bosons}
Contrary to the case of the SM, in the MSSM two Higgs doublets are required. This results in five physical Higgs bosons
instead of the single Higgs boson of the SM. These are the light and heavy $\textbf{CP}$-even Higgs bosons $-$ $h$ and $H$,\,
the $\textbf{CP}$-odd Higgs boson $-$ $A$,\, and the charged Higgs bosons $-$ $H^{\pm}$. The supersymmetric structure of the
theory imposes very strong constraints on the Higgs spectrum. Out of the six parameters which describe the MSSM Higgs sector,
$M_{h}$, $M_{H}$, $M_{A}$, $M_{H^{\pm}}$, $\beta$ and $\alpha$, only two parameters, which can be taken as \,$\tan{\beta}$\,
and $M_{A}$, are free parameters at the tree-level. These magnitudes are defined as follows \cite{Mar,Mar1,Mar2,Djo2,Djo3,CaHa}:
\beq \label{Hig2}
\tan{\beta}\,=\,\FFr{v_{2}}{v_{1}}\,=\,\FFr{v\,\sin{\beta}}{v\,\cos{\beta}}\,,\,\,\,\,\,\,\,\,\,\,\mbox{(where}\,\,\,\,\,\,\,\,\,\,
0\leq\,\beta\,\leq \pi/2\,\,\,\,\,\mbox{and}\,\,\,\,\, v^{2}\,=\,v_{1}^{2}\,+\,v_{2}^{2})
\eeq
as the ratio of two neutral Higgs field vacuum expectation values;
\beq \label{Hig3}
M_{A}^{2}\,=\,-\Lb \FFr{1}{\cos{2\beta}}(m_{{\!}_{H_{1}^{2}}}\,-\,m_{{\!}_{H_{2}^{2}}}\,+\,M_{Z}^{2} \Rb\,,
\eeq
as the mass of the $\textbf{CP}$-odd scalar boson \footnote{$m_{{}_{H_{1}}}$ and $m_{{}_{H_{2}}}$ are the so-called
soft-SUSY breaking terms for the Higgs boson masses in the Higgs potential $V_{H}$ \cite{Mar,Mar1,Mar2,Djo2,Djo3}.};
\beq \label{Hig4}
M_{H^{\pm}}^{2}\,=\,M_{A}^{2}\,+\,M_{W}^{2}\,,
\eeq
as the mass of the charged Higgs boson;
\beq \label{Hig5}
\alpha\,=\,\FFr{1}{2}\arctan\!\!{\left[\tan{2\beta}\,\FFr{M_{A}^{2}\,+\,M_{Z}^{2}}{M_{A}^{2}\,-\,M_{Z}^{2}}\right]}\,,
\,\,\,\,\,\,\,\,\,\,\mbox{(where}\,\,\,\,\,-\pi/2\leq\,\alpha\,\leq 0)
\eeq
as the angle of rotation by which the physical bosons $h$ and $H$ are obtained from the neutral components
($H_{1}^{0}$ and $H_{2}^{0}$) of the Higgs fields;
and ultimately, the masses of these $\textbf{CP}$-even Higgs bosons are given by:
\beq \label{Hig6}
M_{h,H}^{2}\,=\,\FFr{1}{2}\left[M_{A}^{2}\,+\,M_{Z}^{2}\,\mp\,\sqrt{(M_{A}^{2}\,+\,M_{Z}^{2})^{2}\,-\,4M_{A}^{2}M_{Z}^{2}\,
\cos^{2}\!{2\beta}}\right]\,.
\eeq
\vskip1truecm
We concentrate on the constraints for \,$M_{h,H}$\, bosons which are $\textbf{CP}$-even, like the SM Higgs boson,
which is also $\textbf{CP}$-even but a scalar. Various interesting theoretical approaches give constraints on the masses
of the MSSM Higgs bosons, including \,$M_{h,H}$.\, We look specifically at four results obtained in \,\cite{Djo2,Djo3} and
\,\cite{CaHa}. There one can see
\vskip1truecm
\begin{quote}
$\bullet$  Production cross sections (see \fig{Higgs4}) \cite{Djo2,Djo3} of the MSSM Higgs bosons in $e^{+}e^{-}$ collisions as
functions of the masses for \,$\tan{\beta} = 30$\, and \,$\sqrt{s} = 500\,GeV$. The cross section for $hZ$ production is
large for large values of $M_{h}$, being of \,$\mathcal{O}(100\,fb)$;\, by contrast, the cross section for $HZ$ is large
for light $H$ (implying small $M_{H}$). The largest values lie at $\sim 129\,GeV$.  \\
$\bullet$ The neutral MSSM Higgs production cross sections (see \fig{Higgs5} and \fig{Higgs6}) \cite{CaHa} at the Tevatron
\,$(\sqrt{s} = 2\,TeV)$\, for gluon fusion \,$gg \rightarrow \phi$;\, vector-boson fusion
\,$qq \rightarrow qqV^{*}V^{*} \rightarrow qqh,qqH$;\, vector-boson bremsstrahlung
\,$q\bar{q} \rightarrow V^{*} \rightarrow hV/HV$\, ($V^{*}$ is a virtual vector boson, and \,$V = W$ or $Z$);\, and the
associated production \,$gg,q\bar{q} \rightarrow \phi b\bar{b}/\phi t\bar{t}$\, including all known QCD corrections, where
\,$\phi = h,H$ or $A$ \footnote{The notations in \fig{Higgs4} can be understood in the same way.}. \\
$\bullet$ Branching ratios of the MSSM Higgs bosons $h$ and $H$, with \,$\tan{\beta} = 30$\, \cite{CaHa}. There is a clear
picture of the various decay modes. The range of $M_{H}$ shown corresponds to \,$90\,GeV < M_{A} < 130\,GeV$,\, whereas the
range of $M_{h}$ shown corresponds to \,$128\,GeV < M_{A} < 1\,TeV$.\, However, for the central values of $h$ and $H$ bosons'
masses we see that \,$M_{h}^{max} \simeq 125.9\,GeV$\, and \,$M_{H}^{min} \simeq 126.1\,GeV$\, (see \fig{Higgs7}).
\end{quote}

\begin{center}
\DOUBLEFIGURE[h]{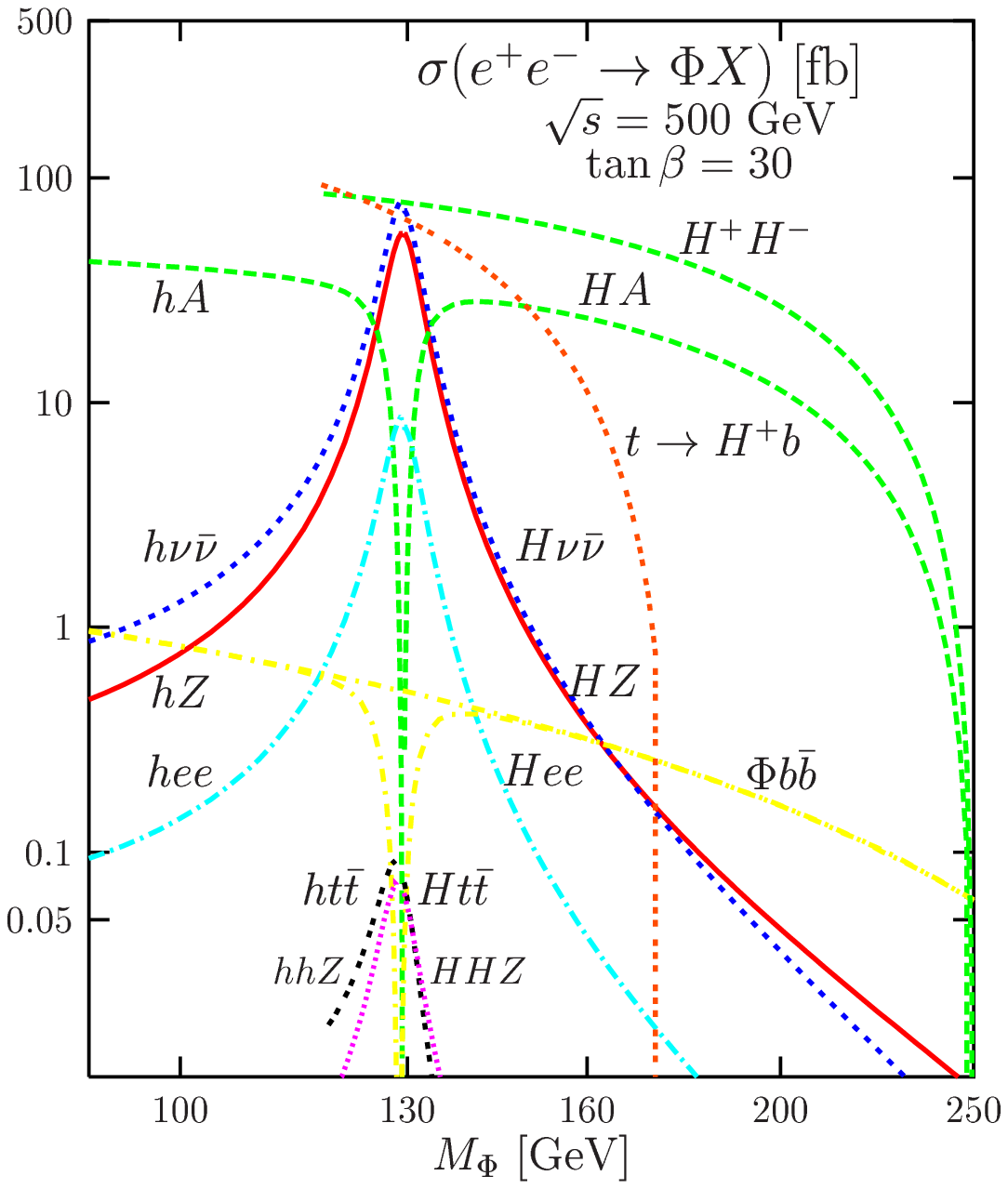,width=75mm,height=75mm}{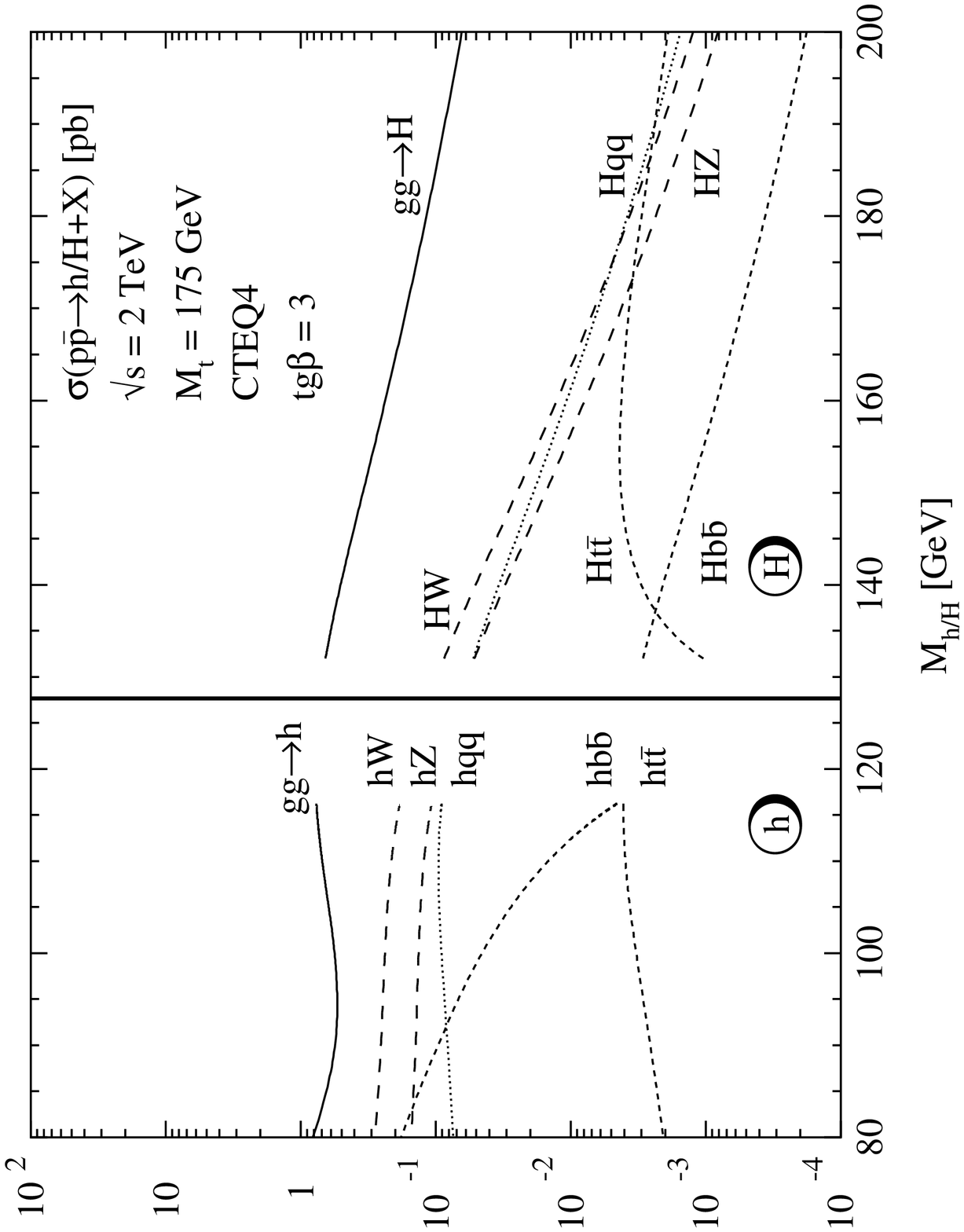,width=75mm,height=75mm,angle=-90}
{The production scenario for \,$\tan{\beta} = 30$\, and \,$\sqrt{s} = 500\,GeV$\, from \cite{Djo2,Djo3}.\label{Higgs4}}
{The production scenario for \,$\tan{\beta} = 3$\, from \cite{CaHa}. \label{Higgs5}}
\end{center}
\begin{center}
\DOUBLEFIGURE[h]{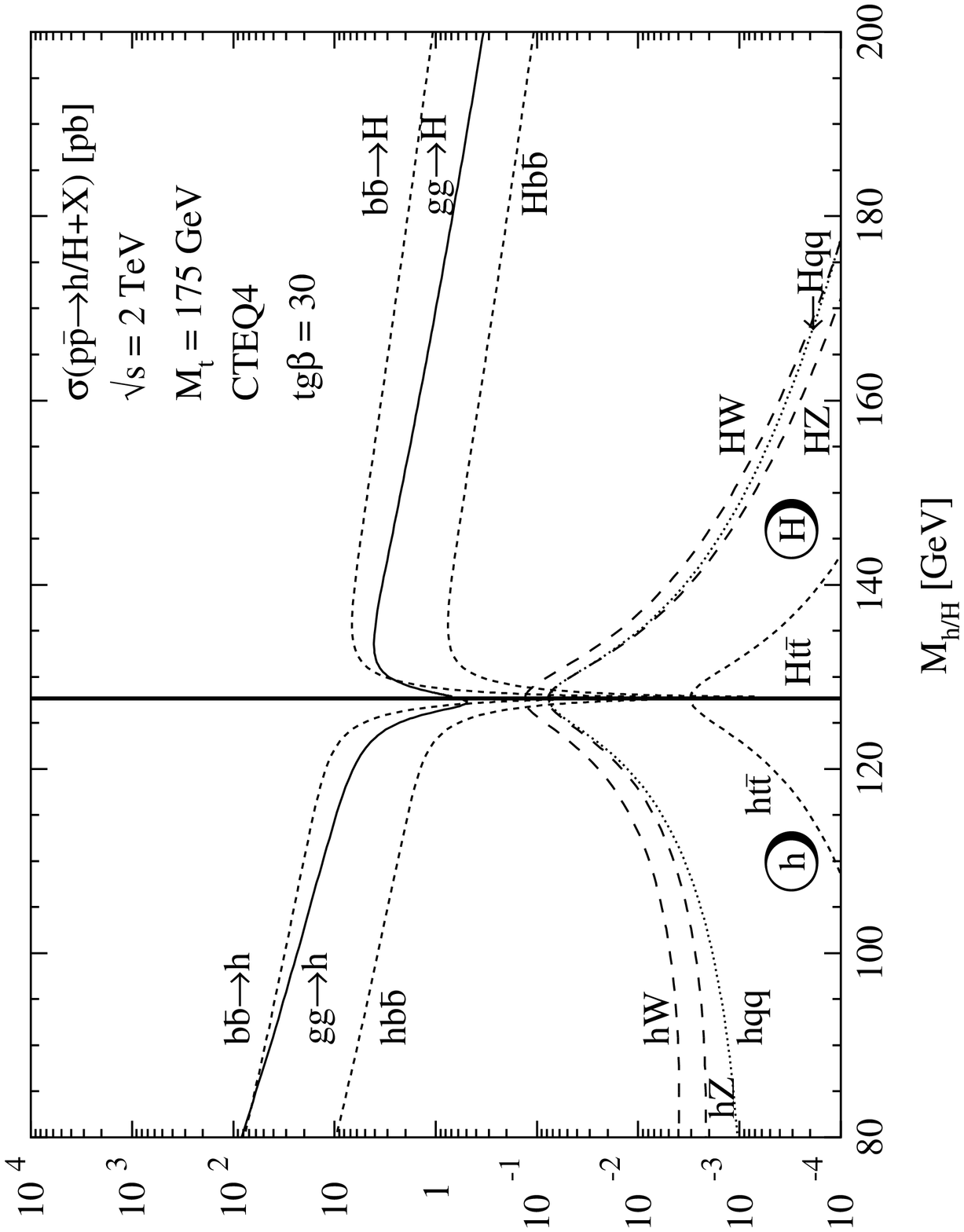,width=75mm,height=75mm,angle=-90}{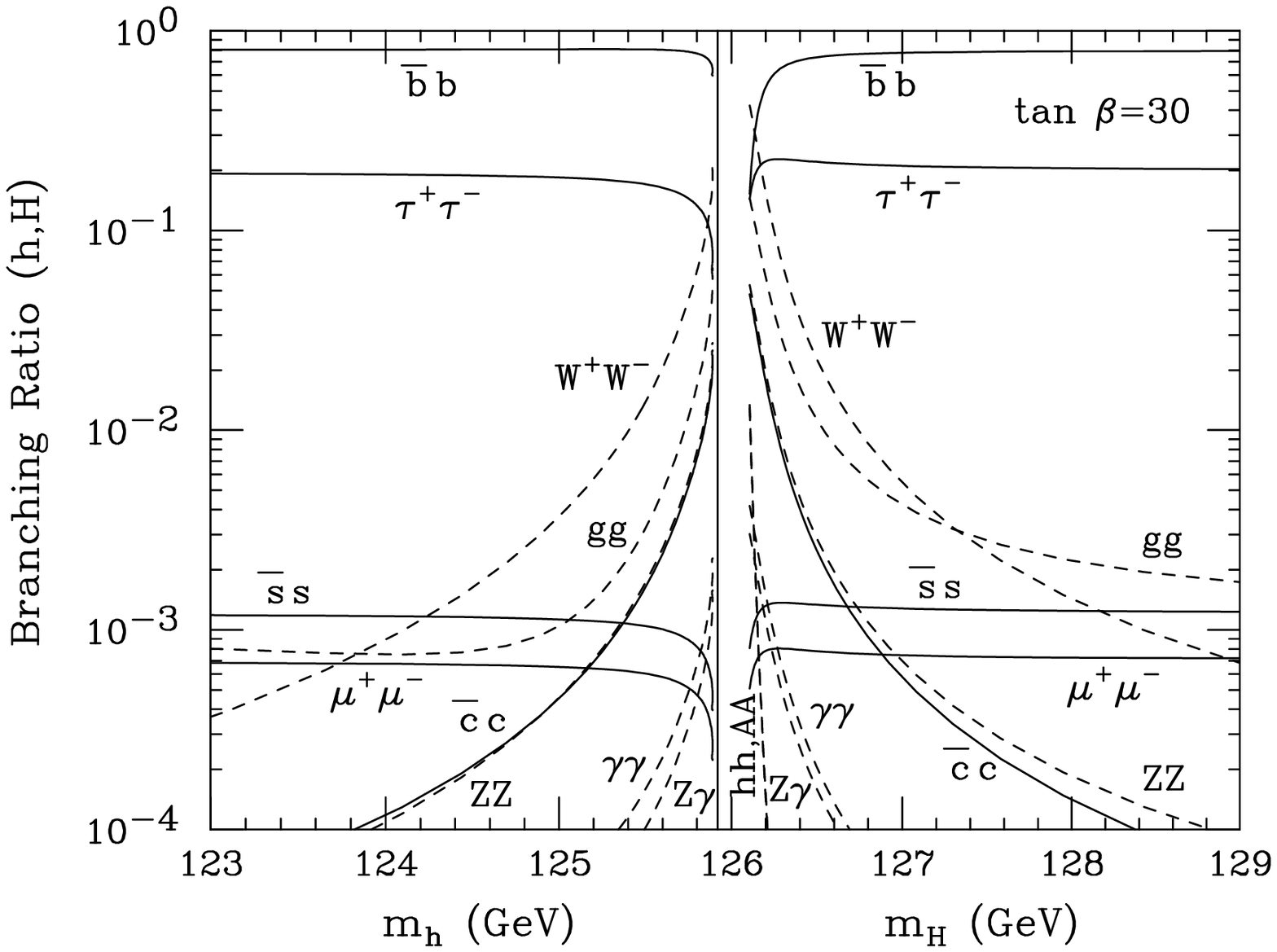,width=75mm,height=56mm}
{The production scenario for \,$\tan{\beta} = 30$\, from \cite{CaHa}.
\label{Higgs6}}
{Branching ratios of the MSSM Higgs bosons $h$ and $H$, with \,$\tan{\beta} = 30$\, from \cite{CaHa}.
\label{Higgs7}}
\end{center}

\subsection{Higgs Boson mass from the Particle Frame}
The Higgs particle attributes mass to all particles in the SM, including itself. If we select the unique permutation \,$\psi_{Higgs} = R(\beta)\varphi_{i}$\, associated with the massive Higgs particle then
it must have \,$\beta = 1$,\, according to the Mass Rule $\mathcal{H}$. We also select $\gamma = 1$. Then the particle
frame corresponds to the \textit{t}-Riemann surface with this choice of $\varphi_{i}$, $\beta$ and $\gamma$.

By the Higgs Selection Rule $\mathcal{C}$, the Higgs particle is given as the intersection of all $24$ discs of the
\textit{t}-Riemann surface. We may regard the Higgs particle as the intersection of the discs \,$1,...,12$\, of the upper
sheet (the origin of the upper sheet), and the Higgs antiparticle as the intersection of the discs \,$13,...,24$\, of the
lower sheet (the origin of the lower sheet). However, the Higgs particle and antiparticle are identified as the branch point
$(0,\,\beta + \gamma)$ of the t-Riemann surface. These $24$ discs together represent the Schr\"{o}dinger discs corresponding
to the Higgs field, and the branch point represents the Higgs particle on the particle frame, according to \fig{FCT_fig7} and
\fig{Higgs8}. So that by the Higgs Selection Rule $\mathcal{C}$, the Higgs particle is a scalar boson.
By the Spin Rule $\mathcal{D}$, the spin of the Higgs boson is $0$; by the Electric Charge Rule $\mathcal{E}$, its electric
charge is $0$, by the Weak Isospin Rule $\mathcal{F}$, its weak isospin is $0$; and by the Strong Charge Rule $\mathcal{G}$,
its strong charge is neutral with \,$N_{c} = 1$.\, The Higgs boson has not been observed yet, but it is an inevitable
consequence of the Higgs-Kibble mechanism.

\begin{center}
\FIGURE[h]{
\centerline{\epsfig{file=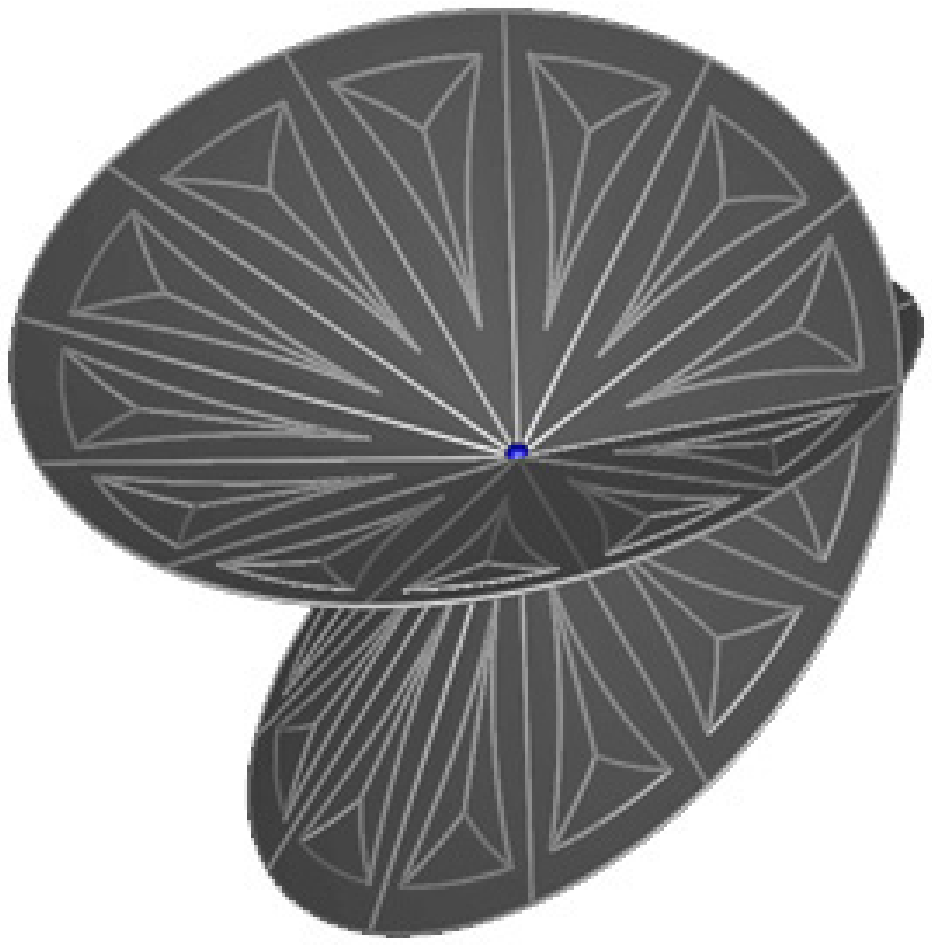,width=60mm,height=60mm}}
\caption{The Higgs boson on the particle frame.}
\label{Higgs8}}
\end{center}

Furthermore, by the Higgs Selection Rule $\mathcal{C}$ and Mass Rule $\mathcal{H}$ for the massive Higgs boson, we assume
that the permutation $\varphi_j$ has the rest mass equal to half of the sum of the rest masses of all other bosons defined on the particle frame \footnote{Only the rest masses of the three vector bosons $Z^{0}$, $W^{+}$ and $W^{-}$ contribute to the sum. All the other bosons are massless.}.
The mechanism is the following. Since the Higgs particle/antiparticle will be identified (as a Cooper pair), their combined
energy would then be the sum of the masses of all other bosons defined on the particle frame. We can have all types of bosons superposed on a single particle frame, and the single Cooper pair of the Higgs particle/antiparticle must be able to attribute energy/rest mass to
all types of bosons on this particle frame, by the Higgs-Kibble mechanism. The particle frames of the bosons can be superposed
at a point in space-time because they follow the Bose-Einstein statistics. Hence, this Cooper pair must have \emph{at least}
enough energy to attribute the sum of the rest masses of all types of bosons defined on the particle frame. On the other hand, the most important property of Bose condensation is that the Cooper pair of the Higgs particle/antiparticle must have minimum energy, so
it can have \emph{at most} the energy required to attribute the sum of the rest masses of all types of bosons defined on the particle frame. This must be the lowest energy state possible for the Higgs boson when it undergoes Bose condensation. Summarizing all these facts, we obtain the formula for the Higgs boson mass defined on the particle frame as

{\setlength\arraycolsep{2pt}
\bea \label{Hig02}
M_{{\!}_{H^{0}}}\,& = &\, \Lb M_{_{\!Z}}\,+\,M_{_{\!W^{+}}}\,+\,M_{_{\!W^{-}}} \Rb/2\,=\,
\nonumber \\
\,& = &\,(91.1875\,GeV\,+\,80.398\,GeV\,+\,80.398\,GeV)/2\,=\,125.992\,GeV\,\simeq\,126\,GeV.
\eea}

In conclusion, we would like to note that we have calculated the Higgs boson mass from formula $5.7$ by taking into account the whole topological and algebraic structure of the \textit{t}-Riemann surface, the particle frame and the proof of the four color theorem. We find that our result is in close agreement with the values of the SM and MSSM Higgs boson masses depicted in \fig{Higgs1}, \fig{Higgs2}, \fig{Higgs3}, \fig{Higgs4}, \fig{Higgs5}, \fig{Higgs6} and \fig{Higgs7}, although those results were derived using a different approach. We would like to especially stress the exact agreement of our result with the central mass values in
\fig{Higgs7}: \,$M_{h}^{max} \simeq 125.9\,GeV$\, and \,$M_{H}^{min} \simeq 126.1\,GeV$\, for \,$\tan{\beta} = 30$,\, in spite of the fact that our result is for the SM Higgs $H^{0}$ boson.

\vskip 2.0truecm
\begin{center}
{\Large{\bf{Appendix}}} \\

\vskip 0.5truecm
{\large{\bf{An explicit construction of the Steiner system \,$\mathbf{S(5, 8, 24)}$\\* from the proof of the Four Color
Theorem}}} \\
\end{center}

A \emph{posteriori} the proof of the four color theorem (lemmas $1$-$23$), we know that the Steiner system
\,$S(N\!+\!1, 2N, 6N)$\, has been constructed and \,$N = 4$.\, Then we can return to the beginning and use the lemmas
$1$-$23$ to construct the Steiner system \,$S(5, 8, 24)$\, explicitly. By LEMMA $4$, the $24$ points of the system
\,$S(5, 8, 24)$\, are the elements of the underlying set \,$\textit{\textbf{Z}}_{4}]S_{3}$\, as shown in Table $1$.
Then, using the definitions in the proof of the four color theorem and LEMMA $23$, we construct all the blocks of the system
\,$S(5, 8, 24)$.\, We use the computer program ``Steiner.exe'' to explicitly calculate the distinct blocks, making use of the
definition in \eq{Theor49}

\begin{displaymath}
\left\{\begin{array}{ll}
Fix\!\downarrow\!\!(H) \downarrow\!\!
\left(\begin{array}{ccc}
\psi            \!\! & \\
\,\,\psi^{\mu}  \!\! & \\
\end{array} \right)
\left. \begin{array}{ccc}
\mid  \hskip 0.024truecm & \\
\mid  \hskip 0.024truecm & \\
\end{array} \right.
& \textrm{\!\!\!\!\!\!\!\!$\mid$}
\,\,\left(\begin{array}{ccc}
\psi            \!\!& \\
\,\,\psi^{\mu}  \!\!& \\
\end{array} \right)
\in
G
\end{array} \right\}\,.
\end{displaymath}

It turns out that there are 759 distinct blocks in total. To view all the blocks, download the file ``appendix.zip" from the following hyperlink:

\begin{center}
\href{http://www.dharwadker.org/khachatryan/higgs/appendix.zip}{http://www.dharwadker.org/khachatryan/higgs/appendix.zip}
\end{center}

Unzip the file ``appendix.zip" and open the program ``Steiner.exe'' to explore the complete structure of the Steiner system \,$S(5, 8, 24)$. First check that the three properties in the definition of the Steiner system \,$S(5, 8, 24)$\, are indeed satisfied
(see Sec $2.1.1$):
\begin{quote}
$\bullet$ There are $24$ points. \\
$\bullet$ Each block consists of $8$ points. \\
$\bullet$ Any set of $5$ points is contained in a unique block.
\end{quote}

\vskip 0truecm
\hskip -0.8truecm
{\bf{Examples}}

In the program `Steiner.exe'', the user may select any number of points and the program explicitly finds all the blocks that
contain the selected points. The user may also explore some of the beautiful combinatorial properties of the system
\,$S(5, 8, 24)$,\, such as:
\begin{quote}
$\bullet$ Any $5$ points are contained in a unique block (by definition). \\
$\bullet$ Any $4$ points are contained in exactly $5$ blocks. \\
$\bullet$ Any $3$ points are contained in exactly $21$ blocks. \\
$\bullet$ Any distinct pair of points are contained in exactly $77$ blocks. \\
$\bullet$ Any single point is contained in exactly $253$ blocks.
\end{quote}
Let us explicitly calculate a couple of examples:
\begin{quote}
$\bullet$ The $5$ points \,$\left\{(\underline{0},1),(\underline{1},1),(\underline{2},1),(\underline{3},1),
(\underline{0},\sigma)\right\}$\, are contained in the unique block
\begin{displaymath}
{\bf{B}_{1}} = \,\left\{(\underline{0},1),(\underline{1},1),(\underline{2},1),(\underline{3},1),
(\underline{0},\sigma),(\underline{1},\sigma),(\underline{2},\sigma),(\underline{3},\sigma)\right\} = Fix\!\downarrow\!\!(H)
\end{displaymath}
by LEMMA 21. \\
$\bullet$ The $5$ points \,$\left\{(\underline{0},1),(\underline{1},1),(\underline{2},1),(\underline{3},\rho^{2}),
(\underline{2},\sigma\rho^{2})\right\}$\, are contained in the unique block
\begin{displaymath}
{\bf{B}_{5}} = \,\left\{(\underline{0},1),(\underline{1},1),(\underline{2},1),(\underline{3},1),
(\underline{1},\rho^{2}),(\underline{2},\rho^{2}),(\underline{3},\rho^{2}),(\underline{2},\sigma\rho^{2})\right\}.
\end{displaymath}.
\end{quote}

In this way, the reader can calculate many other examples to explore the combinatorial structure of the Steiner system \,$S(5, 8, 24)$.

%\section{Conclusions}

%\section*{Acknowledgments}

%\input{Higgs-ref.tex}


\begin{thebibliography}{1}

\bibitem{Dhar1}
A.~Dharwadker, A New Proof of the Four Colour Theorem (2000), \href{http://www.dharwadker.org/}{$http://www.dharwadker.org/$}
%``A New Proof of the Four Colour Theorem''


\bibitem{Dhar2}
A.~Dharwadker, Grand Unification of the Standard Model with
Quantum Gravity (2008), \href{http://www.dharwadker.org/standard_model/}{$http://www.dharwadker.org/standard{}_{-}model/$}
%``Grand Unification of the Standard Model with
%``Quantum Gravity''


\bibitem{Cour}
R.~Courant and H.~Robbins, What is Mathematics?, Oxford University Press (1996).
%``What is Mathematics?''


\bibitem{ApHak2}
K.~Appel and W.~Haken, Every Planar Map is Four-Colorable, Bull. Amer. Math. Soc. 82, (1976).
%``Every Planar Map is Four-Colorable''

\bibitem{DharPir1}
A.~Dharwadker and S.~Pirzada, Graph Theory, Orient Longman and
Universities Press of India (2008).
%``Graph Theory''


\bibitem{DharPir2}
A.~Dharwadker and S.~Pirzada, Applications of Graph Theory, Journal of the Korean Society for Industrial and Applied Mathematics 11 (4) (2007).
%``Applications of Graph Theory''

\bibitem{Di}
R.~Diestel, Graph Theory, Springer-Verlag (2005).
%``Graph Theory''


\bibitem{Aal}
T.~Aaltonen et al., CDF Collaboration, Phys. Rev. Lett. 102
(2009) 021802, arXiv: 0809.3930 [hep-ex].
%``''


\bibitem{CDF_D0_1}
The CDF Collaboration, Search for $H\,\rightarrow\,WW$ production at CDF using
$3.0\,fb^{-1}$ of Data, CDF conference note $9500$.
%``Search for $H\,\rightarrow\,WW$ production at CDF using
%$3.0\,fb^{-1}$ of Data''
%``''

\bibitem{CDF_D0_2}
The D0 Collaboration, Search for Higgs boson production in dilepton plus missing
transverse energy final states with $3.0\,-\,4.2\,fb^{-1}$ of
$p\bar{p}$ collisions at $\sqrt{s}\,=\,1.96\,TeV$'', D0 conference
note 5871.
%``Search for Higgs boson production in dilepton plus missing
%transverse energy final states with $3.0\,-\,4.2\,fb^{-1}$ of
%$p\bar{p}$ collisions at $\sqrt{s}\,=\,1.96\,TeV$''
%``''



\bibitem{Ber}
G.~Bernardi et al., Combined CDF and Dzero Upper Limits on Standard Model Higgs Boson Production at High Mass (2008), arXiv: 0808.0534 [hep-ex].
%``''


\bibitem{Tev}
Tevatron New Phenomena, Higgs Working Group for the CDF collaboration
and D0 collaboration (2009), arXiv: 0903.4001 [hep-ex].
%``Tevatron New Phenomena Higgs Working Group and CDF Collaboration
%and D0 collaboration''

\bibitem{Gold1}
J.~Goldstone, A.~Salam and S.~Weinberg, Broken Symmetries, Phys. Rev. 127 (1962) 965; 
%``Broken Symmetries''


\bibitem{Gold2}
J. Goldstone, Field Theories with Superconductor Solutions, Nuov. Cim. 19 (1961) 154.
%``Field Theories with Superconductor Solutions''


\bibitem{Hig1}
P.~Higgs, Spontaneous Symmetry Breakdown without Massless Bosons, Phys. Rev. 145 (1966) 1156.
%``Spontaneous Symmetry Breakdown without Massless Bosons''

\bibitem{Hig2}
P.~Higgs, Broken Symmetries, Massless Particles and Gauge Fields, Phys. Lett. 12 (1964) 132.
%``Broken Symmetries, Massless Particles and Gauge Fields''

\bibitem{Hig3}
P.~Higgs, Broken Symmetries and the Masses of Gauge Bosons , Phys. Rev. Lett. 13 (1964) 508.
%``Broken Symmetries and the Masses of Gauge Bosons''

\bibitem{Hig4}
F.~Englert and R.~Brout, Broken Symmetry and the Mass of Gauge Vector Mesons, Phys. Rev. Lett. 13 (1964) 321.
%``Broken Symmetry and the Mass of Gauge Vector Mesons''


\bibitem{Kib}
T.W.~Kibble, Symmetry Breaking in Non-Abelian Gauge Theories, Phys. Rev. 155 (1967) 1554.
%``Symmetry Breaking in Non-Abelian Gauge Theories''

\bibitem{Kib1}
G.~Guralnik, C.~Hagen and T.W.~Kibble, Global Conservation Laws and Massless Particles, Phys. Rev. Lett. 13 (1964)
585.
%``Global Conservation Laws and Massless Particles''

\bibitem{Djo1}
A.~Djouadi, The Anatomy of Electro-Weak Symmetry Breaking. I: The Higgs boson in the Standard Model, Phys. Rept. 457 (2008) 1, hep-ph/0503172.
%``The Anatomy of electro-weak symmetry breaking. I:
%The Higgs boson in the standard model''


\bibitem{Bed}
V.~A.~Bednyakov, N.~D.~Giokaris and A.~V.~Bednyakov, On Higgs mass generation mechanism in the Standard Model,
hep-ph/0703280.
%``On Higgs mass generation mechanism in the Standard Model''

\bibitem{Noj}
M.~Nojiri, T.~Plehn and G.~Polesello, Physics Beyond the Standard Model: Supersymmetry, arXiv: 0802.3672 [hep-ph].
%``Physics Beyond the Standard Model: Supersymmetry''


\bibitem{Mar}
S.~Martin, The Supersymmetry primer, in Perspectives Supersymmetry, World Scientific, Singapore (1998) hep-ph/9709356. 
%```The Supersymmetry primer'', in Perspectives Supersymmetry''


\bibitem{Mar1}
H.~Haber and G.~Kane, The search for supersymmetry: probing physics beyond the standard model, Phys. Rept. 117 (1985) 75. 
%``The search for supersymmetry''


\bibitem{Mar2}
H.~Nilles, Supersymmetry, Supergravity and Particle Physics, Phys. Rept. 110 (1984) 1.
%``Supersymmetry, Supergravity and Particle Physics''


\bibitem{Djo2}
A.~Djouadi, The Anatomy of Electro-Weak Symmetry Breaking. II: The Higgs bosons in the Minimal Supersymmetric Model, Phys.Rept. 459 (2008) 1, hep-ph/0503173.
%``The Anatomy of electro-weak symmetry breaking. II.
%The Higgs bosons in the minimal supersymmetric model''


\bibitem{Djo3}
A.~Djouadi, The Higgs sector of supersymmetric theories and the implications for high-energy colliders, Eur. Phys. J. C 59 (2009) 389, arXiv: 0810.2439 [hep-ph].
%``The Higgs sector of supersymmetric theories
%and the implications for high-energy colliders''



\bibitem{CaHa}
M.~Carena and H.~Haber, Higgs boson theory and phenomenology, Prog. Part. Nucl. Phys. 50 (2003) 63, hep-ph/0208209.
%``Higgs boson theory and phenomenology''


\bibitem{Low}
I.~Low and S.~Shalgar, Implications of the Higgs Discovery in the
MSSM Golden Region, arXiv: 0901.0266 [hep-ph].
%``Implications of the Higgs Discovery in the
%MSSM Golden Region''


\bibitem{HaHe}
T.~Hahn, S.~Heinemeyer, W.~Hollik, H.~Rzehak and G.~Weiglein, Higgs Masses and More in the Complex MSSM with FeynHiggs,
 arXiv: 0710.4891 [hep-ph].
%``Higgs Masses and More in the Complex MSSM
%with FeynHiggs''

\bibitem{PDG}
Particle Data Group, Review of Particle Physics, Phys. Lett. B 667 (2008) 1.
%``Review of Particle Physics''

\bibitem{Stil}
J.~Stillwell, Classical Topology and Combinatorial Group Theory,
Springer-Verlag (1993).
%``Classical Topology and Combinatorial Group Theory''


\bibitem{Ass}
J.H.~Van Lint and R.M.~Wilson,  A Course in Combinatorics, Cambridge University Press (1992).
%``Design Theory''


\bibitem{Dhar3}
A.~Dharwadker, The Witt Design - The Steiner system S(5,8,24)
explicitly computed (2002), \href{http://www.dharwadker.org/witt.html}{$http://www.dharwadker.org/witt.html$} 
%``The Witt Design - The Steiner system S(5,8,24)
%explicitly computed by Ashay Dharwadker''


\bibitem{Tits}
J.~Tits, Sur les syst\`{e}mes de Steiner associ\'{e}s aux trois grands groupes de Mathieu, Rend. Math. e Appl. (5)23 (1964) 166.
%``Sur les systemes de Steiner associes aux trois grands groupes de
%Mathieu''


\bibitem{Lan}
S.~Lang, Algebra, Springer-Verlag (2002). \\
%``Algebra''


\bibitem{DharSmi}
A.~Dharwadker and J.D.H.~Smith, Split Extensions and Representations of Moufang Loops, Comm. in Alg. 23(11) (1995) 4245.
%``Split Extensions and Representations of Moufang Loops''


\bibitem{Eilen}
S.~Eilenberg, Extensions of General Algebras, Ann. Soc. Polon. Math. 21 (1948).
%``Extensions of General Algebras`''


\bibitem{Rot}
J.~Rotman, An Introduction to the Theory of Groups, Springer-Verlag (1995).
%``An Introduction to the Theory of Groups''


\bibitem{Hall}
P.~Hall, On Representatives of Subsets, J. London Math. Soc. 10 (1935) 26.
%``On Representatives of Subsets''


\bibitem{Ahl}
L.V.~Ahlfors, Complex Analysis, McGraw-Hill Book Company (1979). 
%``Complex Analysis''


\bibitem{Ri}
A.~Dharwadker,  Riemann Surfaces, Electronic Geometry Models, Model 2002.05.001 (2003), 
\href{http://www.eg-models.de/models/Surfaces/Riemann_Surfaces/2002.05.001/}{$http://www.eg{-}models.de/models/Surfaces/Riemann{}_{-}Surfaces/2002.05.001/$} 


\bibitem{Kami}
Official Super-Kamiokande Press Release, Evidence for massive neutrinos (1998), \href{http://neutrino.phys.washington.edu/~superk/sk_release.html}{$http://neutrino.phys.washington.edu/{\sim}superk/sk{}_{-}release.html$} 

\bibitem{Coop3}
J. R. Schrieffer, Theory of Superconductivity, Addison-Wesley (1988).
%`` ``Cooper Pairs''. Book''

\bibitem{Coop2}
T. Keilmann and J. J. Garcia-Ripoll, Dynamical Creation of Bosonic Cooper-Like Pairs, Phys. Rev. Lett. 100 (2008) 110406.
%`` ``Cooper Pairs''. Paper''


\bibitem{Coop1}
L.~Cooper, Bound electron pairs in a degenerate Fermi gas, Phys. Rev. 104(4) (1956) 1189.
%``Bound electron pairs in a degenerate Fermi gas''


\bibitem{Ho}
G.~'t Hooft, In search of the ultimate building blocks, Cambridge University Press (1997).
%``In search of the ultimate building blocks''


\bibitem{Schro}
D.~Schroeder, Thermal Physics, Addison Wesley Longman (2000).
%`` Thermal Physics''


\bibitem{LEP}
The LEP Collaborations ALEPH, DELPHI, L3 and OPAL and the LEP Electroweak Working Group,
hep-ex/0612034.


\bibitem{SLD}
The ALEPH, DELPHI, L3, OPAL and SLD Collaborations, the LEP
Electroweak Working Group and the SLD Electroweak and Heavy
Flavour Groups, Phys. Rept. 427 (2006) 257, \href{http://www.cern.ch/LEPEWWG/}{$http://www.cern.ch/LEPEWWG/$} 
%``CERN''


\bibitem{Mix}
The SLAC E158 Collaboration, Precision Measurement of the Weak Mixing Angle in M$\oslash$ller Scattering, Phys. Rev. Lett. 95:081601 (2005) 20, hep-ex/0504049.
%``Precision Measurement of the Weak Mixing Angle in Moller
%Scattering, by SLAC E158 Collaboration''


\bibitem{Cab}
N.~Cabibbo, Unitary Symmetry and Leptonic Decays, Phys. Rev. Lett. 10 (12) (1963) 531.
%``Unitary Symmetry and Leptonic Decays''


\bibitem{ChaKe}
L.~Chau and W.-Y.~Keung, Comments on the Parametrization of the Kobayashi-Maskawa Matrix, Phys. Rev. Lett. 53 (1984) 1802.
%``''


\bibitem{Wol}
Particle Data Group Review, Values obtained from values of Wolfenstein parameters (2008).
%``''


\bibitem{Roh}
J.~Rohlf, Modern Physics from a to $Z^{0}$, Wiley (1994).
%``Modern Physics from a to Z^{0}''


\bibitem{Agu}
J.~A.~Aguilar-Saavedra et al., TESLA Technical Design Report, Part
$3$, Report DESY-$2001$-$011$, hep-ph/0106315.
%``''


\bibitem{Des}
K.~Desch et al., Report of the Higgs Working Group for the
extended ECFA-DESY study, Amsterdam (2003), hep-ph/0311092.
%``''


\bibitem{Fla}
H.~Flaecher et al., Gfitter - Revisiting the Global Electroweak
Fit of the Standard Model and Beyond, arXiv: 0811.0009 [hep-ph],
\href{http://www.cern.ch/gfitter/}{$http://www.cern.ch/gfitter/$} 
%``Gfitter - Revisiting the Global Electroweak
%Fit of the Standard Model and Beyond''






\end{thebibliography}
\end{document}